\documentclass[a4paper, fleqn, usenatbib]{mnras}

\usepackage{newtxtext,newtxmath}

\usepackage[T1]{fontenc}
\usepackage{ae,aecompl}
\usepackage{booktabs}  

\usepackage{graphicx}   
\usepackage{amsmath}    
\usepackage{gensymb}
\usepackage[flushleft]{threeparttable}
\usepackage{multirow}

\usepackage{tabularx}
\newcolumntype{Y}{>{\centering\arraybackslash}X}

\usepackage{pdflscape}
\usepackage{CJK}

\graphicspath{{figs/}}

\newcommand{\angstrom}{\textup{\AA}}

\def\lsim{\mathrel{\rlap{\lower4pt\hbox{\hskip1pt$\sim$}}
    \raise1pt\hbox{$<$}}}                
\def\gsim{\mathrel{\rlap{\lower4pt\hbox{\hskip1pt$\sim$}}
    \raise1pt\hbox{$>$}}}                

\title[Radio AGN Selection and Characterization in Three LSST DDFs]{Radio AGN Selection and Characterization in Three Deep-Drilling Fields of
the Vera C. Rubin Observatory Legacy Survey of Space and Time}

\author[S. Zhu et al.]{Shifu Zhu,$^{1,2}$\thanks{E-mail: SFZAstro@gmail.com (PSU)}
W. N. Brandt,$^{1,2,3}$
Fan Zou,$^{1,2}$
Bin Luo,$^{4,5}$
Qingling Ni,$^6$
Yongquan Xue,$^{7,8}$
and Wei Yan$^{1,2}$
\\
$^1$Department of Astronomy \& Astrophysics, The Pennsylvania State University, University Park, PA 16802, USA\\
$^2$Institute for Gravitation and the Cosmos, The Pennsylvania State University, University Park, PA 16802, USA\\
$^3$Department of Physics, 104 Davey Lab, The Pennsylvania State University, University Park, PA 16802, USA\\
$^4$School of Astronomy and Space Science, Nanjing University, Nanjing, Jiangsu 210093, China\\
$^5$Key Laboratory of Modern Astronomy and Astrophysics (Nanjing University), Ministry of Education, Nanjing 210093, China\\
$^6$Institute for Astronomy, University of Edinburgh, Royal Observatory, Edinburgh EH9 3HJ, UK\\
$^7$CAS Key Laboratory for Research in Galaxies and Cosmology, Department of Astronomy, University of Science and Technology of China, Hefei 230026, China\\
$^8$School of Astronomy and Space Sciences, University of Science and Technology of China, Hefei 230026, China}
\date{Accepted XXX. Received YYY; in original form ZZZ}

\pubyear{2020}

\begin{document}
\label{firstpage}
\pagerange{\pageref{firstpage}--\pageref{lastpage}}
\maketitle
\begin{abstract}
    The Australia Telescope Large Area Survey (ATLAS) and the VLA survey in the XMM-LSS/VIDEO deep field provide deep ($\approx $15~$\micro$Jy~beam$^{-1}$) 
    and high-resolution ($\approx$ 4.5--8 arcsec) radio coverage of the three XMM-SERVS 
    fields (W-CDF-S, ELAIS-S1, and XMM-LSS). 
    These data cover a total sky area of 11.3 deg$^2$ and contain $\approx 11000$ radio components. 
    Furthermore, about 3~deg$^2$ of the XMM-LSS field also has deeper MIGHTEE data that achieve a median RMS of 5.6 $\micro$Jy beam$^{-1}$
    and detect more than 20000 radio sources. We analyze all these radio data and find source counterparts at other wavebands utilizing deep 
    optical and IR surveys. The nature of these radio sources is studied using radio-band properties (spectral slope and morphology), 
    and the IR-radio correlation. 
    Radio AGNs are selected and compared with those selected using other methods (e.g. X-ray).
    We found 1656 new AGNs that were not selected using X-ray and/or MIR methods.
    We constrain the FIR-to-UV SEDs of radio AGNs using {\sc cigale}
    and investigate the dependence of radio AGN fraction upon
    galaxy stellar mass and star-formation rate.
\end{abstract}




\section{Introduction}
\label{sec:intro}
During 2025--2035, the Vera C. Rubin Observatory will perform the most ambitious 
optical time-domain imaging survey yet,
the Legacy Survey of Space and Time (LSST; \citealt{ivezic2019}).
Specifically, Rubin will repeatedly survey
the southern sky through six filters ($ugrizy$) that cover 320--1060~nm.
Billions of galaxies are expected to be mapped and cataloged by the LSST.
In particular, Rubin will visit five 10--20-deg$^2$ deep-drilling fields (DDFs) more frequently than
typical regions in the wide survey and reach a much better depth (e.g. \citealt{brandt2018, scolnic2018}).
The five DDFs are COSMOS (Cosmic Evolution Survey), 
W-CDF-S (Wide Chandra Deep Field-South),
ELAIS-S1 (European Large-Area ISO Survey-S1),
XMM-LSS (XMM-Large Scale Structure), and EDF-S (Euclid Deep Field-South).
These DDFs have been selected because of their rich multi-wavelength datasets, 
including (but not limited to) the MeerKAT International GHz Tiered Extragalactic 
Exploration (MIGHTEE; \citealt{jarvis2016, heywood2022}) survey in the radio,
the Spitzer DeepDrill (\citealt{lacy2021}) survey and the VISTA Deep Extragalactic Observation (VIDEO; \citealt{jarvis2013}) survey
in the infrared, and the XMM-Spitzer Extragalactic Representative Volume Survey (XMM-SERVS; \citealt{chen2018, ni2021}) in the X-ray.

Active galactic nuclei (AGNs) are one of the driving science topics in these DDFs (e.g. \citealt{brandt2018}).
The LSST survey will provide a large high-quality DDF data set that supports
studies of, e.g., 
long-term AGN variability selection, 
AGN continuum variability, 
the continuum reverberation mapping of accretion disks, 
and stellar tidal disruption events. 
To support such studies, it is critical to obtain the most complete AGN samples possible using selection methods across the entire electromagnetic spectrum (e.g. \citealt{padovani2017}) in these DDFs.
Among the five DDFs, the COSMOS field is the most well-studied on deg$^2$ scales,
and COSMOS AGN samples selected by various methods are available (e.g. \citealt{marchesi2016, chang2017, smolcic2017}).
The EDF-S field presently has relatively poorer multiwavelength data compared with the other DDFs.
Therefore, we focus on improving AGN samples in the remaining three DDFs: W-CDF-S, ELAIS-S1, and XMM-LSS.
These three DDFs are also the XMM-SERVS fields (e.g. \citealt{chen2018, ni2021}).

Different AGN selection methods are often complementary to each other, and the most complete AGN samples are obtained only using all methods together.
In the W-CDF-S, ELAIS-S1, and XMM-LSS fields,
AGNs have been selected using MIR colors and spectral energy distributions (SEDs; e.g. \citealt{zou2022}), 
optical-NIR colors (e.g. \citealt{wolf2004, berta2006, nakos2009}), optical variability (e.g. \citealt{poulain2020}), 
and X-ray methods (e.g. \citealt{chen2018, ni2021}).
A fraction of AGNs launch powerful relativistic jets,
and strong radio emission is a defining feature of these extragalactic jets.
A significant number of AGNs can thus be found in the radio band as well.
Many of these jetted AGNs, traditionally also called radio-loud AGNs 
(RL AGNs),\footnote{We use ``RL AGNs'' and ``radio AGNs'' 
interchangeably in this paper
to refer to those objects for which the radio emission is mainly produced by the active nuclei.}
cannot be easily selected using X-ray or MIR methods, especially using moderate-depth X-ray or MIR data (e.g. \citealt{hickox2009}).

In this work, we aim to deliver useful catalogs of radio AGNs in the W-CDF-S, ELAIS-S1, and XMM-LSS fields.
Some radio AGNs have been selected in these three DDFs (e.g. \citealt{tasse2008, mao2010, mao2012}).
They generally use radio data that are often shallower than RMS $=30 \micro$Jy~beam$^{-1}$.
Several deep radio surveys have been completed in the three DDFs,
including the Australia Telescope Large Area Survey (ATLAS) in
the W-CDF-S and ELAIS-S1 fields (\citealt{norris2006, hales2014, franzen2015}),
the VLA survey in the XMM-LSS field (\citealt{heywood2020}), and 
the MIGHTEE survey that is also in the XMM-LSS field but covers a smaller sky area (\citealt{heywood2022}).
We use these surveys to select radio AGNs to a depth of RMS $\approx6$--$15\micro$Jy~beam$^{-1}$ at 1.4~GHz.
The basic properties of the surveys are shown in Table~\ref{tab:atlascat}.
These surveys are sufficiently deep to select most of the prime radio AGNs because 
star forming-related processes in the host galaxies dominate
the fainter radio sky (e.g. \citealt{padovani2016}).

Selecting AGNs using radio data is not as straightforward as using some other wavebands.
Radio data often have a low resolution (compared with most optical/infrared data) to reach a high sensitivity, 
and the morphology of radio jets can be complex.
It is challenging to find reliable optical/IR counterparts for a large number of radio sources.
We thus perform careful multiwavelength matching using the superb available co-located survey datasets before AGN selection.
We select radio AGNs using the outstanding radio morphology of their 
extended jets and lobes, flat-spectrum radio core emission,
and excess radio flux that cannot be explained by star formation.
The radio-excess method is the workhorse of RL-AGN selection.
We find more than 1700 AGNs from the radio band that have not been found from X-ray or MIR methods.
In addition to the radio AGN catalogs, we also give a few basic illustrative science results.
We investigate the X-ray properties of these radio AGNs using new sensitivive {\it XMM-Newton} data from XMM-SERVS.
We also study the FIR-to-UV SEDs of radio AGNs and constrain the stellar masses and star-formation rates of their host galaxies.
Further science explorations of these RL AGNs will be in future papers.

In \S~\ref{sec:xmatch}, we describe in detail the radio data of the 
three DDFs we use in this paper.
In \S~\ref{sec:x}, we find the optical/IR counterparts of radio sources
using probabilistic (i.e. Bayesian) cross-matching and visual-inspection methods.
We investigate the multiwavelength properties 
of radio sources, select RL AGNs,
and investigate the \mbox{X-ray} properties of these AGNs in \S~\ref{sec:multi_wav}.
We investigate the SEDs of our RL AGNs against {\sc cigale} models in \S~5.
We summarize our results in \S~6.

We adopt a flat-$\Lambda$CDM cosmology with $H_0=70$~km~s$^{-1}$~Mpc$^{-1}$ and $\Omega_\mathrm{m}=0.3$ in this paper.
Spectral index is defined as $\alpha$ in $f_\nu\propto v^{\alpha}$.
We use AB magnitudes unless otherwise unmentioned.
All sky coordinates are in J2000 frame.

\section{Radio data}
\label{sec:xmatch}
A few improvements to the radio catalogs are made to aid the following analyses, as described in the next two subsections.


\subsection{Extracting ATLAS sources using {\sc Aegean}}
\label{sec:aegean}
\begin{figure}
\centering
\includegraphics[width=0.48\textwidth, clip]{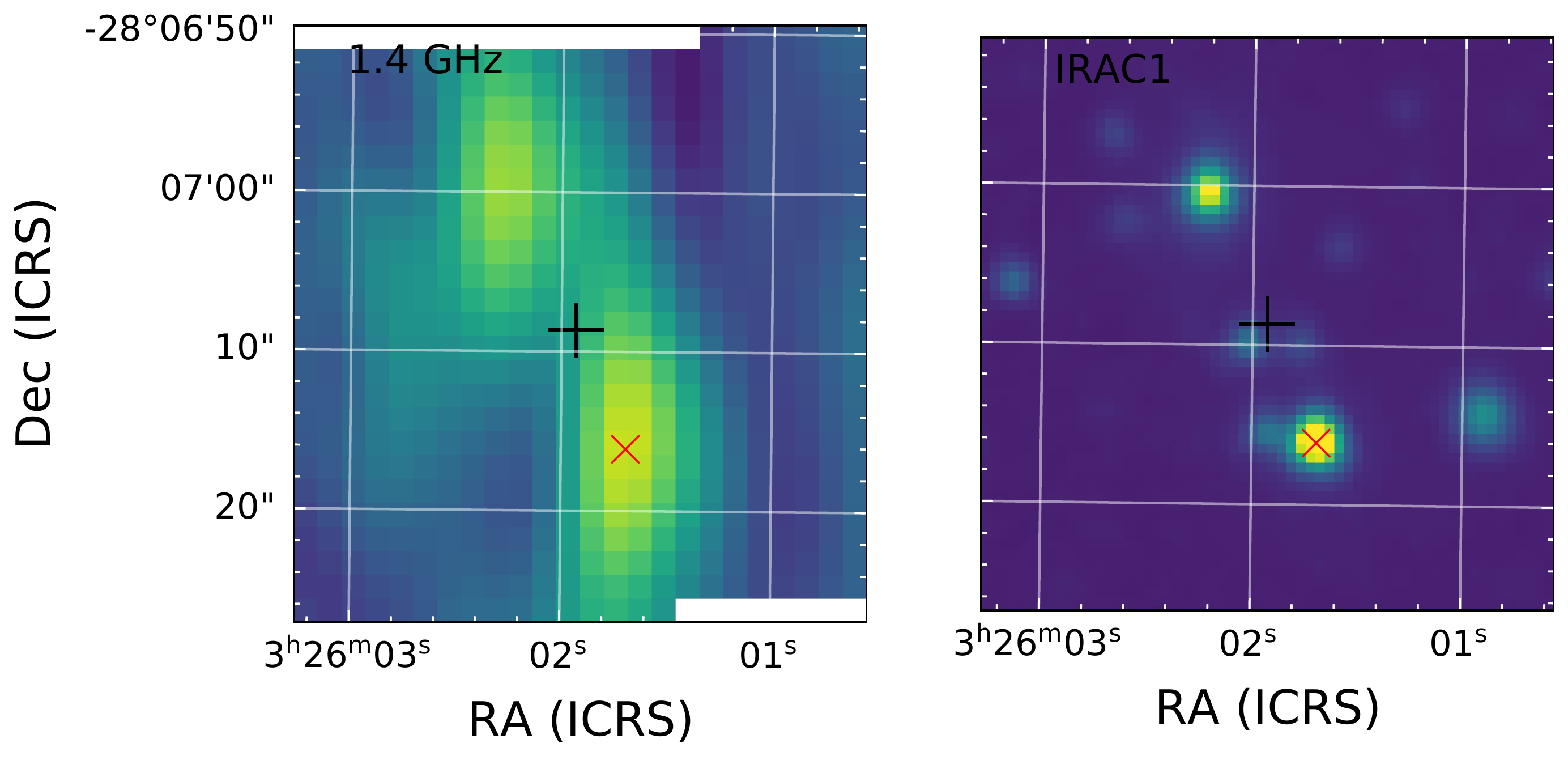}
    \caption{One example of a confused ATLAS component (ID=CI0008) that is distinguished by {\sc aegean}.
    Left: The 1.4 GHz image ($0.6\times0.6$ arcmin) centered at the position in the ATLAS DR3 catalog (black plus sign). The red cross is the position of a radio source found by {\sc aegean}.
    Note that we only keep the one with the highest SNR if multiple radio sources are found by {\sc aegean}.
    Right: The corresponding IRAC1/SERVS image (this region is outside the footprint of the VIDEO survey) clearly shows that the original ATLAS component is associated with two IR counterparts.}
\label{fig:aegean}
\end{figure}

We performed our own radio-source extraction for ATLAS for two reasons:
1. The ATLAS DR3 catalog (\citealt{franzen2015}) only reports the flux-weighted positions of radio sources, 
which often deviate from the brightest pixels, making automatic multi-wavelength cross-matching a difficult task.
2. We find that some of the single components in the \citet{franzen2015} catalog are apparently composed of two radio sources that can be separated (see Fig.~\ref{fig:aegean} for an example).

The science mosaics and rms maps for the W-CDF-S and ELAIS-S1 fields were obtained from the authors of \citet{franzen2015}.
We create 1$\times$1 arcmin$^2$ cutouts for each radio component in the ATLAS DR3 catalog and use {\sc Aegean} (\citealt{hancock2018}) to extract radio sources.
Since we do not intend to produce a new catalog release for the ATLAS data but rather to make the minimum necessary corrections to the existing catalog,
we decide to keep the number of total radio components unchanged.
Furthermore, we make no changes to the radio components that belong to multi-component radio sources with visually assigned hosts (see \S~\ref{sec:vishost}).
From the resulting list of radio sources for each cutout, we only keep the one with the highest peak flux density (i.e. the brightest).
If the kept radio source is closer to another original ATLAS source instead of the one from which we create the cutout,
we treated this source extraction as a failure due to bright-source contamination.
We also visually inspected the results and found another 68 and 48 failure cases due to 
bright-source contamination in the W-CDF-S and ELAIS-S1 fields, respectively.
In total, we obtain {\sc Aegean}-produced radio flux densities and positions for 2781/3034 and 1924/2084 radio components in the W-CDF-S and ELAIS-S1 fields, respectively.

\begin{figure}
\centering
\includegraphics[width=0.45\textwidth, clip]{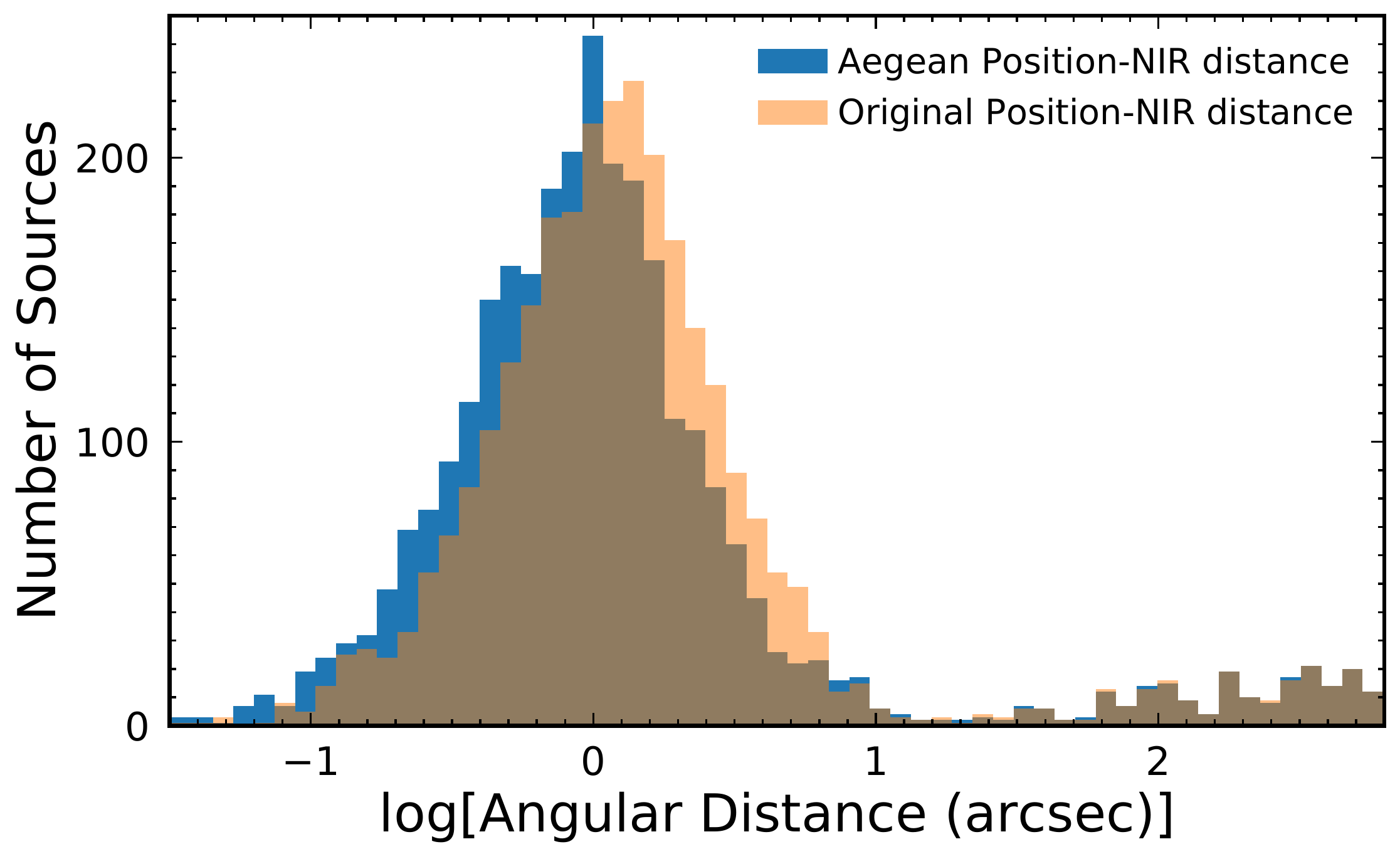}
\caption{Distance to nearest NIR neighbours for the {\sc Aegean}-produced positions (blue) 
    and original flux-weighted positions (orange) in the W-CDF-S field.
    The {\sc Aegean} positions are closer to the NIR positions than the original positions.}
\label{fig:pos_corr}
\end{figure}

To illustrate the improvements of the radio positions, 
we calculate the distance to nearest NIR neighbours for the {\sc Aegean} and original positions in the W-CDF-S field, 
which are compared in Fig.~\ref{fig:pos_corr}.
A large fraction of the nearest neighbours are the NIR counterparts of these radio sources, though not all of them are.
The {\sc Aegean}-produced positions are generally closer to their nearest NIR neighbours than the original radio positions, 
making it easier to distinguish between real associations and false matching in \S~\ref{sec:lrx}.

\begin{figure}
\centering
\includegraphics[width=0.48\textwidth, clip]{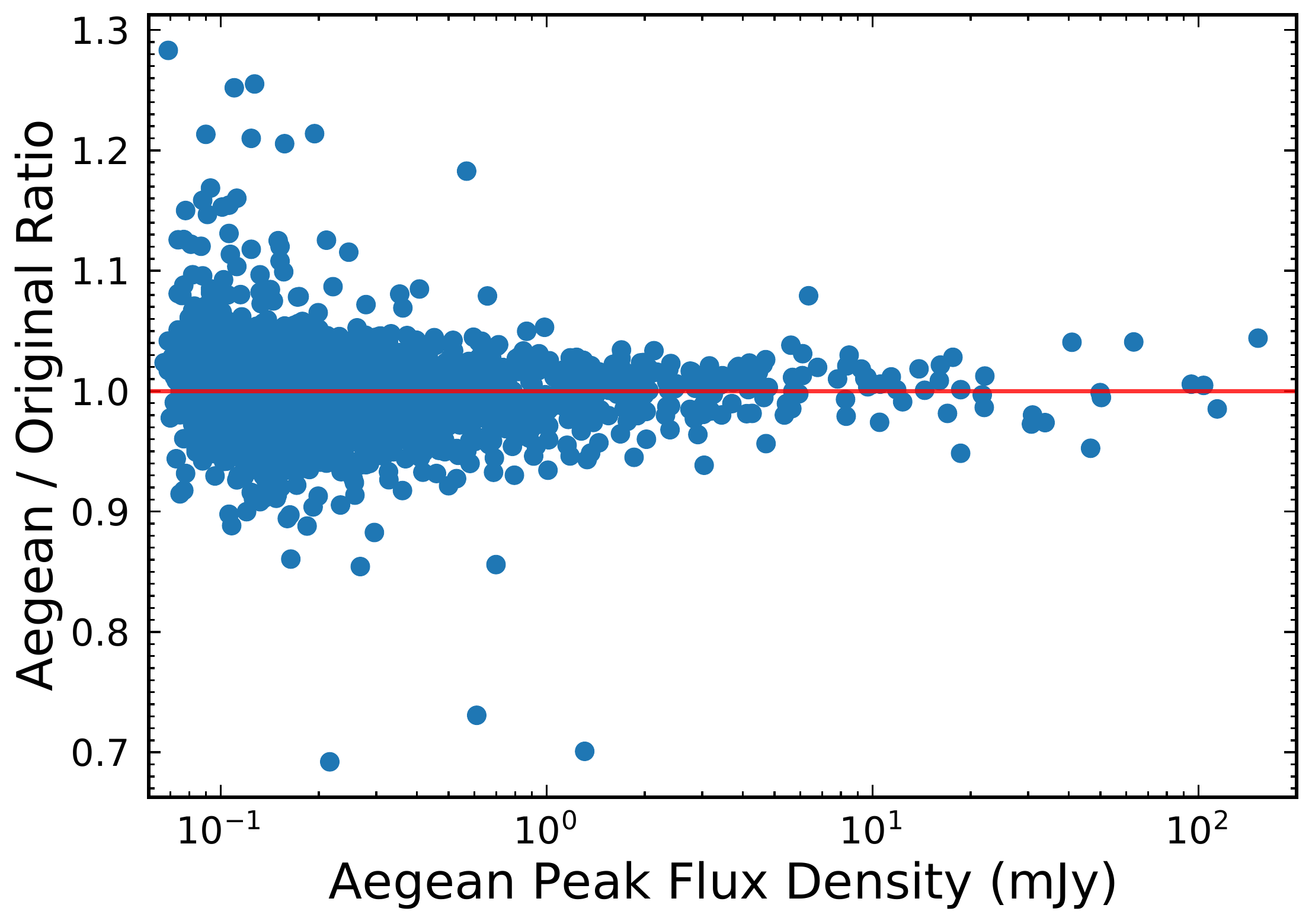}
\caption{A comparison between the peak flux densities derived 
    by {\sc Aegean} and those in the original ATLAS DR3 catalog 
    in the W-CDF-S field.
The median ratio is 1.0002 in this plot.}
\label{fig:aegeanflux}
\end{figure}

Different from the situation for positions,
we would like to keep the {\sc Aegean}-produced flux densities consistent with those in the original catalog.
Since the median {\sc Aegean}-to-original flux density ratio is $\approx0.95$,
we applied a correction of 1.05 to the {\sc Aegean}-produced fluxes.
We compare the {\sc Aegean}-produced peak flux densities (after correction) to those in the original catalog for 
the 2635 sources consistently classified as unresolved point sources in the W-CDF-S field in Fig.~\ref{fig:aegeanflux}.

\begin{table*}
\centering
\caption{Summary of deep radio surveys in the three studied XMM-SERVS fields.}
\label{tab:atlascat}
\begin{threeparttable}[b]
\begin{tabularx}{\linewidth}{@{}Y@{}}
\begin{tabular}{ccccc}
\hline
\hline
    Survey &  Area (deg$^2$) & Sensitivity ($\micro$Jy beam$^{-1}$) & Angular Resolution (arcsec) & No. of Radio Components \\
\hline
    ATLAS/W-CDF-S & 3.6 & 14 & 16$\times$7 & 3034  \\
    ATLAS/ELAIS-S1 & 2.5 & 17 & 12$\times$8 & 2084  \\
    VLA/XMM-LSS & 5.0 & 16 & 4.5 & 5762  \\
    MIGHTEE/XMM-LSS\tnote{a} & 3.5 & 5.6 & 8.2 & 20274 \\
\hline
\end{tabular}
\end{tabularx}
\begin{tablenotes}
\item[a] We use the Level-1 Early Science catalog, which is based on the low-resolution/high-sensitivity image with a robust weighting value of 0.0 (\citealt{heywood2022}).
\end{tablenotes}
\end{threeparttable}
\end{table*}

\begin{table*}
\centering
    \caption{
    Summary of the cross matching of the radio catalogs with DES, VIDEO, and SERVS catalogs. 
    Column (1): Radio catalog. 
    Column (2): Median positional error of the radio sources in units of arcsec.
    Column (3): Optical/IR catalog. 
    Column (4): Number of radio sources in the footprint of the corresponding catalog and not in the masked region due to bright stars.
    Column (5): Total number of candidates with $r<r_{\rm cut}$.
    Column (6): The expected fraction of radio sources that have a genuine counterpart in the optical/IR catalog.
    Column (7): Completeness of the matching. The fraction of radio sources with $R_{\max}>0.5$.
    Column (8): Purity of the matching. The fraction of counterparts (i.e. $R_{\max}>0.5$) that are genuine.
    Column (9): False-negative fraction. The fraction of the radio sources that have a genuine counterpart but with $R_{\max}\le0.5$.
    Column (10), (11), and (12): Predicted values of Column (7) , (8), and (9). We report here the averaged values of all radio sources for cross matching. See Appendix~\ref{sec:appendix1}.
    }
\label{tab:xmatch}
\begin{threeparttable}[b]
\begin{tabularx}{\linewidth}{@{}Y@{}}
\begin{tabular}{lcccccccccccccc}
\hline
\hline
    Radio Cat. & $\sigma$&  OIR Cat. &  $N_{\rm r}$ & $N_{\rm c}$ &  $Q$ & $\mathcal{C}$ & $\mathcal{P}$ & FN & ${\mathcal{\overline C}}$ & $\mathcal{\overline P}$ & $\mathrm{\overline{FN}}$ \\
          &             arcsec &  & &     &               &               &    & & \multicolumn{3}{c}{(Predicted)} \\
    (1)   &   (2)       &  (3) & (4) & (5) & (6) & (7) & (8) & (9) & (10) & (11) &(12) \\
\hline
     ATLAS/W-CDF-S
    & 1.54 & DES &3010& 5254 &0.74 & 0.76 & 0.93 & 0.042 & 0.76 & 0.92& 0.050 & \\
    &      & VIDEO & 2777 & 6630 &0.87 & 0.87 & 0.93 & 0.066 &0.89 & 0.93& 0.051 \\
    &    & SERVS & 3010& 4440 &0.87 & 0.89 & 0.97 & 0.011 & 0.89 & 0.97 & 0.021\\
\hline
    ATLAS/ELAIS-S1
    &  1.26  &DES & 2066& 2973 &0.75 & 0.79 & 0.94 & 0.042 & 0.77 &0.93 & 0.038 \\
    &   &VIDEO & 1240 & 2762 &0.87 & 0.88 & 0.93 & 0.063 & 0.89& 0.94& 0.045 \\
    &   &SERVS & 2066& 2515 &0.87 & 0.88 & 0.98 & 0.016 & 0.89& 0.97 & 0.015 \\
\hline
    VLA/XMM-LSS
    & 0.85 &DES& 5689 & 6247&0.76 & 0.77 & 0.97 & 0.015 &  0.77& 0.96 & 0.019 \\
    &  &VIDEO& 4843 & 8052 &0.92 & 0.93 & 0.97 & 0.027 & 0.93& 0.96 & 0.027 \\
    &  &SERVS & 5494 & 5355 &0.92 & 0.92 & 0.99 & 0.004 & 0.89& 0.99 & 0.004 \\
\hline
    MIGHTEE/XMM-LSS
    & 0.73 & DES& 20089 & 21538&0.67 & 0.69 & 0.95 & 0.033 & 0.67 & 0.94& 0.028 \\
    &  &VIDEO &19840& 32254 &0.87 & 0.88 & 0.95 & 0.041 & 0.86 & 0.95 & 0.034 \\
    &  &SERVS &20059& 21869 &0.87 & 0.88 & 0.98 & 0.014 & 0.86& 0.98 & 0.012 \\
\hline
\end{tabular}
\end{tabularx}
\end{threeparttable}
\end{table*}

\begin{figure}
\centering
\includegraphics[width=0.48\textwidth, clip]{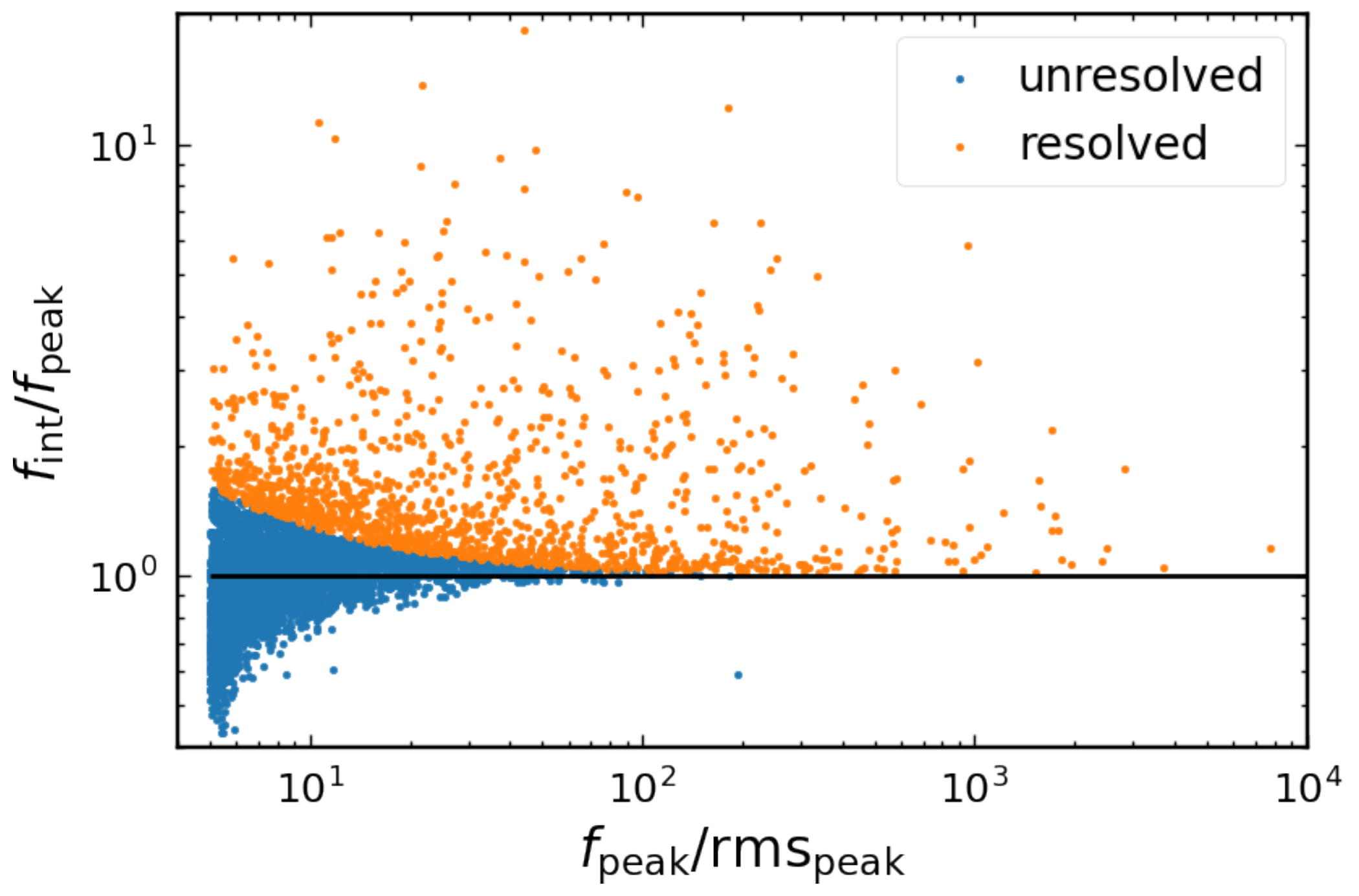}
    \caption{The resolved radio components in the VLA/XMM-LSS catalog are flagged using Eq.~\ref{fig:resolved_vla_xmmlss}.
    }
\label{fig:resolved_vla_xmmlss}
\end{figure}

\begin{figure}
\centering
\includegraphics[width=0.45\textwidth, clip]{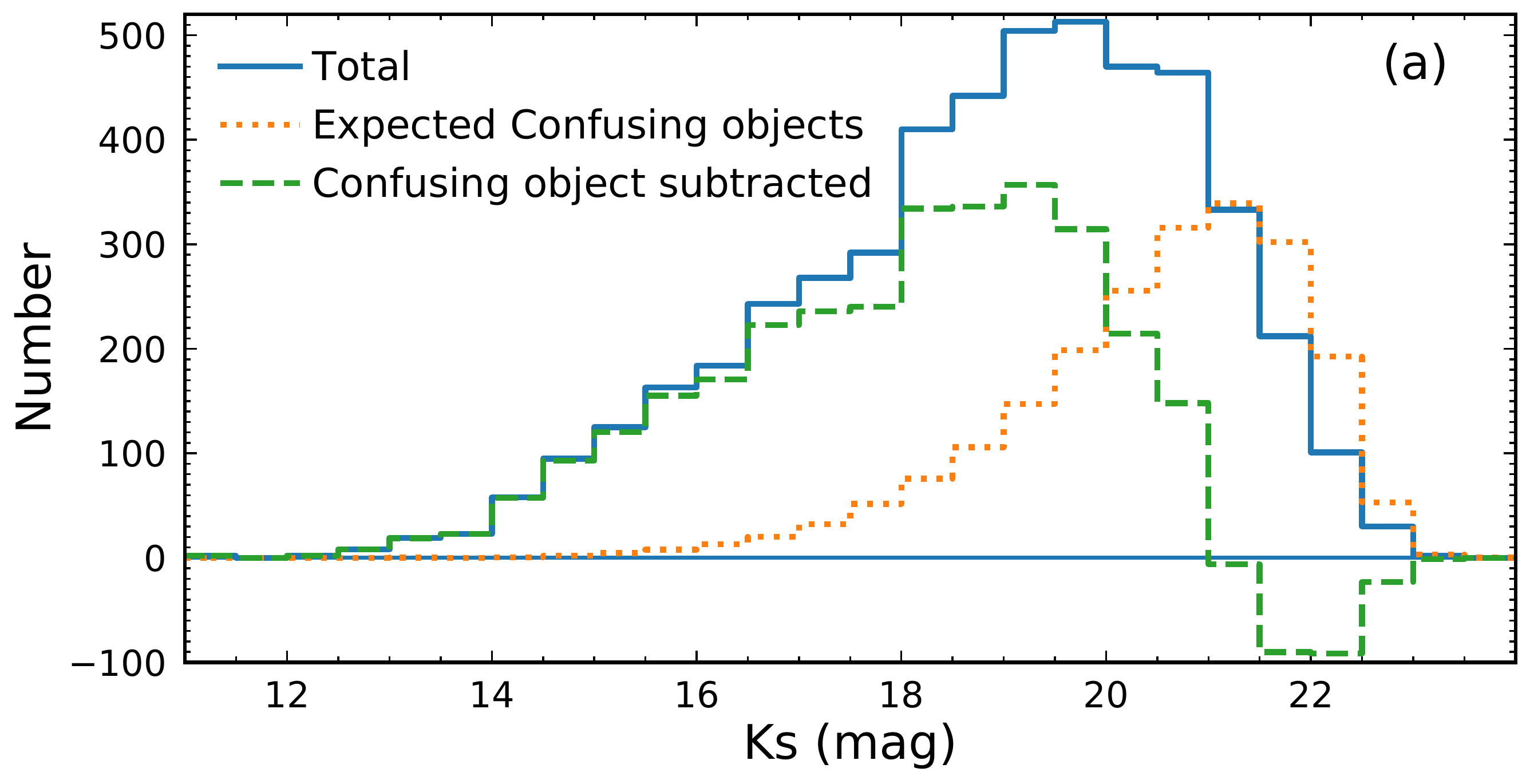}
\includegraphics[width=0.45\textwidth, clip]{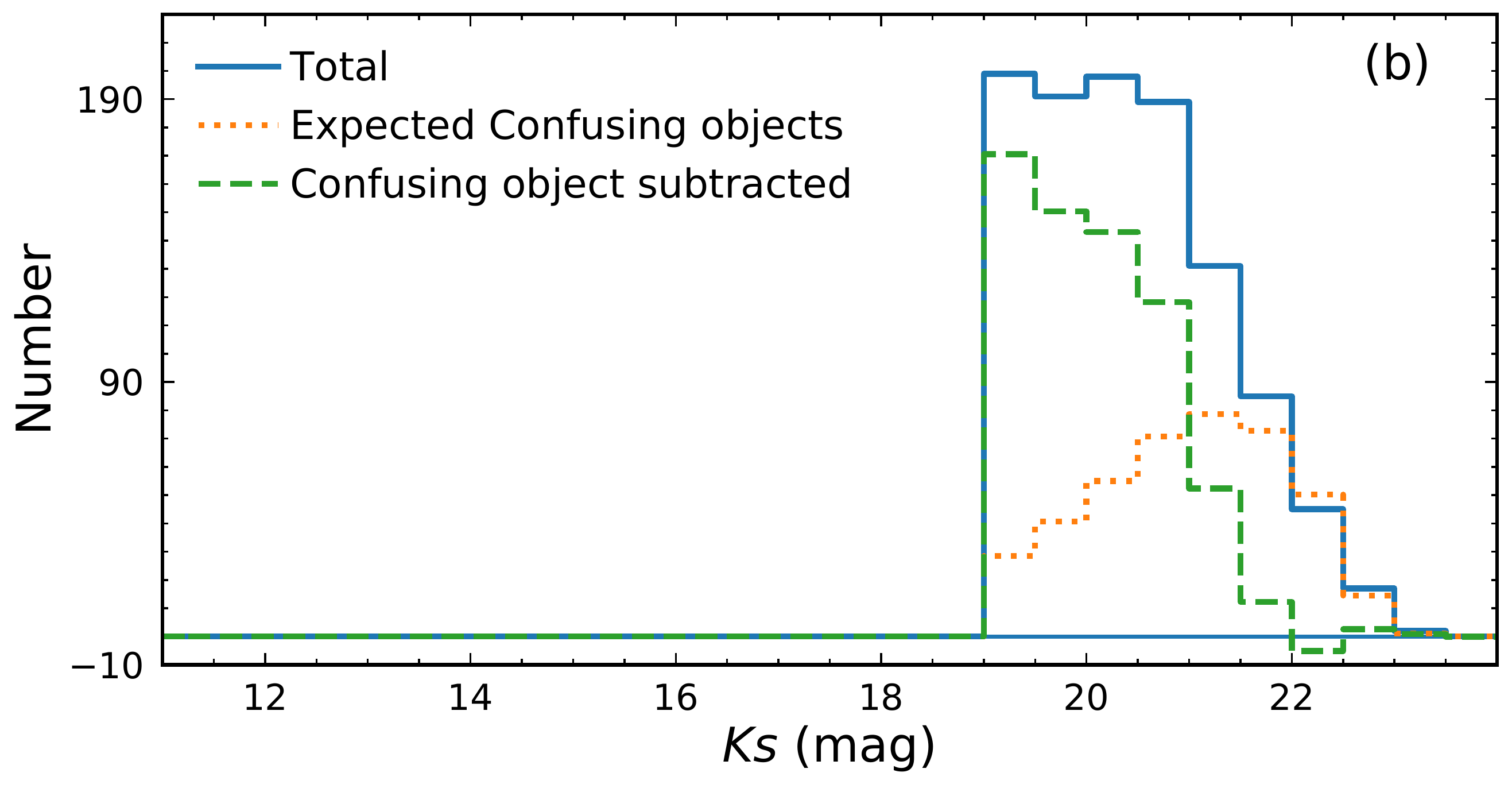}
\includegraphics[width=0.45\textwidth, clip]{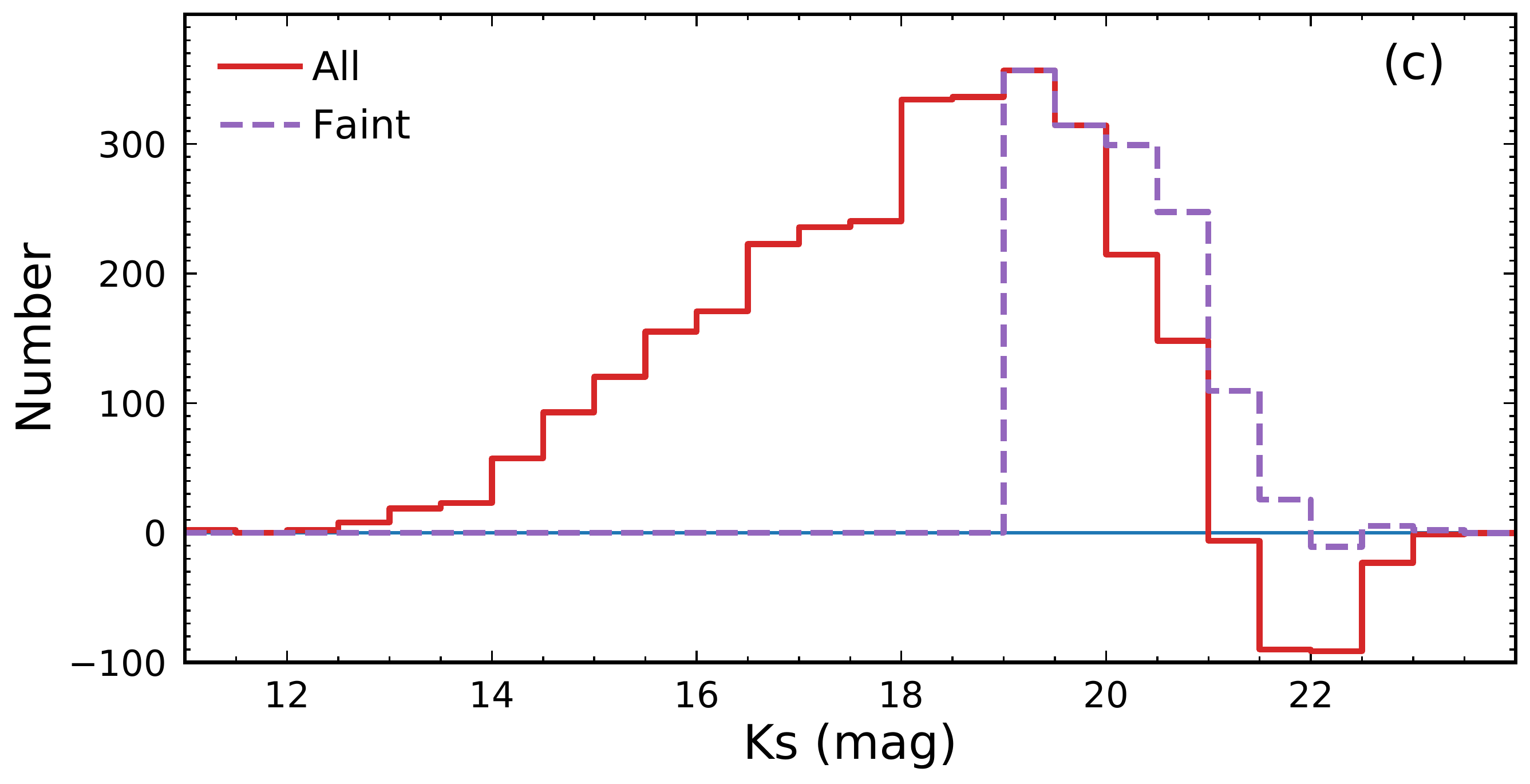}
\includegraphics[width=0.45\textwidth, clip]{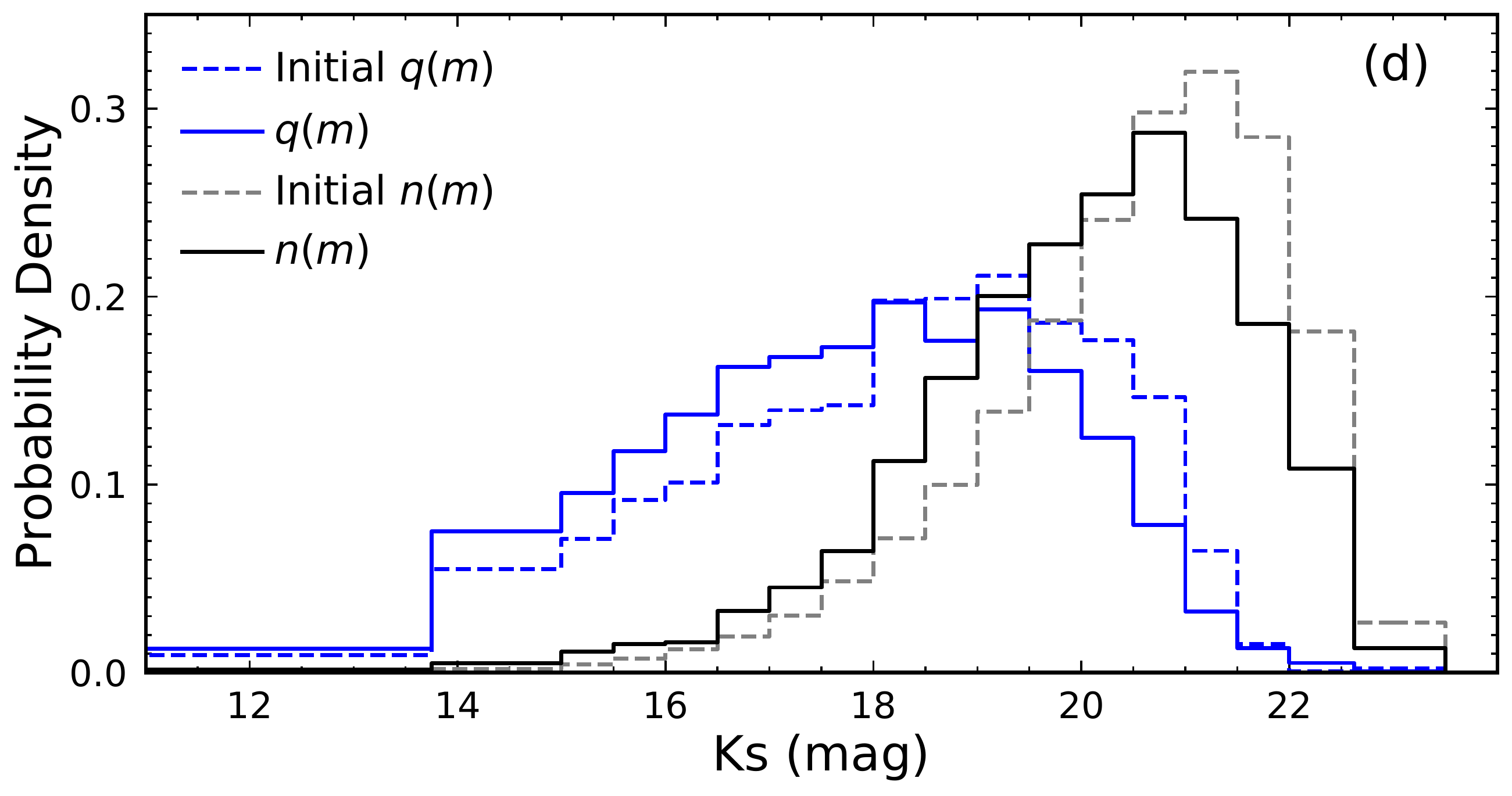}
\caption{Estimating the $Ks$-band magnitude distributions of genuine counterparts and confusing objects, $q(m)$ and $n(m)$, 
    for the ATLAS sources in the W-CDF-S field.
    (a): The magnitude distributions for all NIR sources $<5$ arcsec away from the radio sources (blue solid),
    the expected confusing objects (orange dotted), and the counterparts of the radio sources (green dashed).
    The green dashed histogram becomes negative above $21.5$ mag 
    because the counterparts of radio sources are bright NIR sources, near which the sensitivity is degraded.
    (b): Same as (a) but with all objects brighter than 19 mag masked.
    (c): The combined confusing object-subtracted distributions from (a) (at $<19$ mag) and (b) (at $\ge19$ mag).
    The distribution from $(b)$ is scaled up by a factor of about 2 to match that from (a) in the first 2 bins.
    (d): A comparison between initial estimations of $q(m)$ and $n(m)$ 
    and those after the iteration, which are the magnitude distributions used in the cross matching. 
    See \S~\ref{sec:source_stack}.
    }
\label{fig:ks_prior}
\end{figure}

\begin{figure*}
\centering
\includegraphics[width=0.95\textwidth, clip]{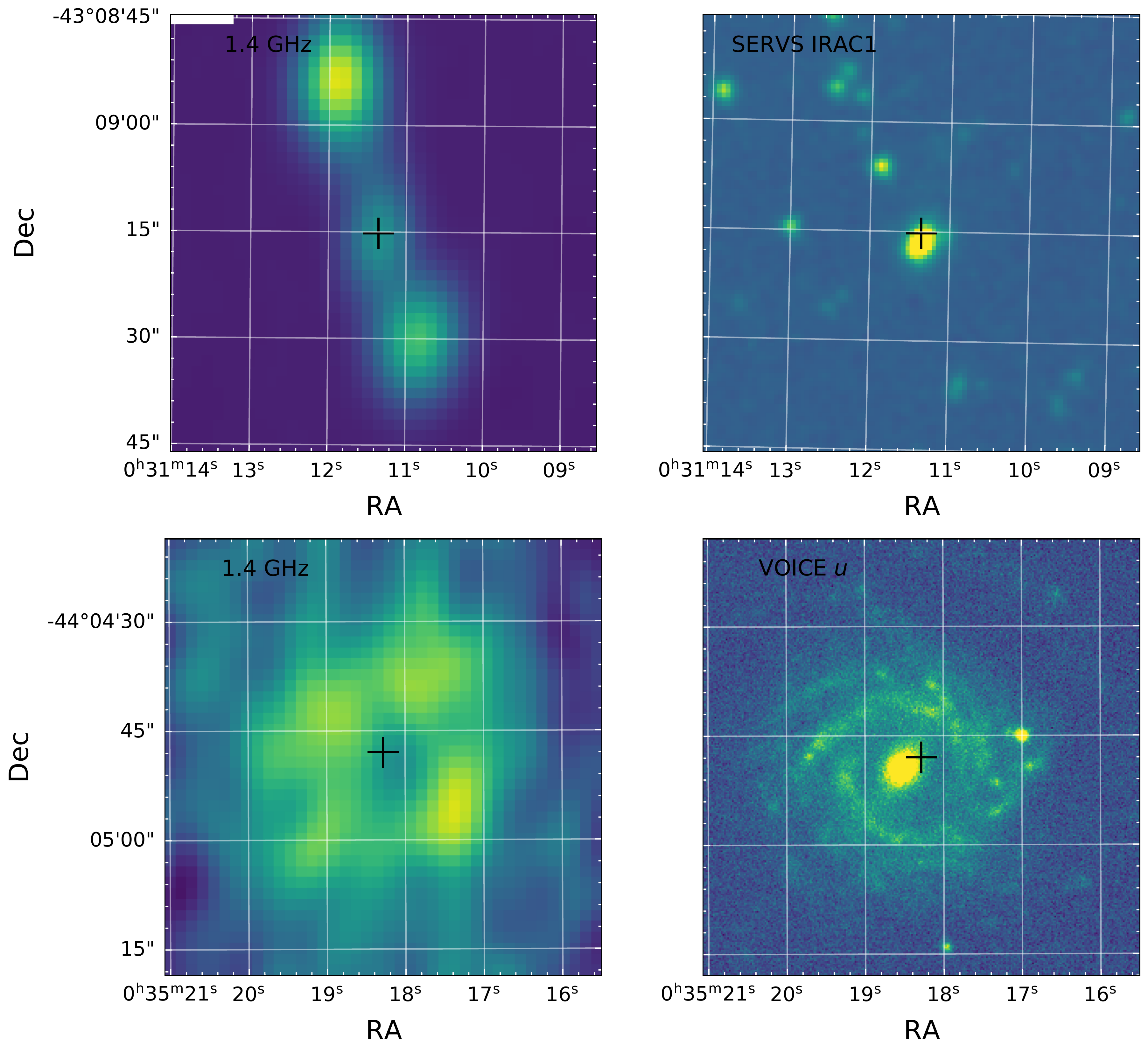}
    \caption{Examples of visually identified radio AGN (upper panels) and SF galaxies (lower panels).
    The black cross in each panel indicates the cataloged position of the central radio source.
    The top-panel example is not covered by the VIDEO data and we therefore show the IRAC1-band image.
    We show the VOICE $u$-band image for the bottom-panel example since the $u$ band seems to trace better the star-forming regions.}
\label{fig:morph_ext_example}
\end{figure*}


\subsection{Positional uncertainties of the MIGHTEE/XMM-LSS and VLA/XMM-LSS sources}
\label{sec:poserr}

The positional uncertainties (i.e. error estimation of RA and Dec of radio sources, $\sigma_\alpha$ and $\sigma_\delta$) of the MIGHTEE/XMM-LSS catalog only include statistical errors (\citealt{heywood2022}),
which decrease with the signal-to-noise ratio (SNR).
We estimate the systematic error using the 1194 radio sources with SNR $>50$,
which have a mean $\sqrt{\sigma_\alpha^2+\sigma_\delta^2}$ of only 0.05~arcsec. In comparison, the radio sources with SNR~$\le50$ have a mean positional uncertainty of 0.71~arcsec.
We match these high-SNR radio sources with the VIDEO catalog using a distance cut of 1~arcsec, which results in 803 pairs.
These MIGHTEE/VIDEO pairs have a $1\sigma$ scatter of $0.23$ and $0.21$ arcsec in RA and Dec, respectively, 
after subtracting the statistical component and positional uncertainties of VIDEO sources (assumed to be 0.1 arcsec).
We add $0.22$~arcsec to both $\sigma_\alpha$ and $\sigma_\delta$ in quadrature since the MIGHTEE/XMM-LSS survey has a circular restoring beam.

The VLA/XMM-LSS catalog does not contain positional uncertainties that can be used to cross match with optical/IR catalogs (\citealt{heywood2020}).
We estimate positional uncertainties combining statistical and systematic errors (e.g. \citealt{franzen2015}) using
\begin{equation}
    \label{eq:poserr}
    \sigma_\alpha,\sigma_\delta = \Theta\sqrt{b^2 + \Big(\frac{1}{1.4\mathrm{SNR}}\Big)^2},
\end{equation}
where $\Theta=4.5$~arcsec is the FWHM of the circular restoring beam (\citealt{heywood2020}).
To estimate $b$ in Eq.~\ref{eq:poserr}, we match the VLA (restricted to SNR $>50$) and VIDEO catalogs with a distance cut of 2~arcsec, 
which results in 301 matched radio sources. 
Again, we calculate the distance scatter and subtract the positional uncertainties of VIDEO sources (assumed to be 0.1 arcsec).
The resulting $b$ is 0.100 and 0.118 for RA and Dec, respectively.
Since the VLA/XMM-LSS survey has a circular beam, we adopt $b=0.11$ in Eq.~\ref{eq:poserr}.

\subsection{Resolved components in the VLA/XMM-LSS catalog}
We preferentially use the peak flux density if the radio source is unresolved, 
and the integrated flux density if resolved.
The source-extraction method (i.e. the {\sc profound} software) of the VLA/XMM-LSS catalog 
is different from those used in other radio data, and resolved components are not flagged.
We distinguish resolved and unresolved radio components using the ratio of integrated flux density ($f_\mathrm{int}$) to peak brightness ($f_\mathrm{peak}$).
We treat the radio component as resolved if $f_\mathrm{int}-f_\mathrm{peak}>3\times \mathrm{RMS}_\mathrm{peak}$, i.e.
\begin{equation}
    \label{eq:resolved_vla_xmmlss}
    \frac{f_\mathrm{int}}{f_\mathrm{peak}}>1+3\times\bigg(\frac{f_\mathrm{peak}}{\mathrm{RMS_\mathrm{peak}}}\bigg)^{-1},
\end{equation}
where $\mathrm{RMS}_\mathrm{peak}$ is the RMS value at the peak position of the radio source.
The results are shown in Fig.~\ref{fig:resolved_vla_xmmlss}.

\section{Cross Matching}
\label{sec:x}
We require the precise positions of the host galaxies of the radio sources for redshift measurements and assessment of multiwavelength properties. 
Optical/IR counterparts to a high completeness are presently unavailable 
for the faint radio catalogs we use in this paper (e.g. \citealt{franzen2015, heywood2022}).
\citet{fan2015} and \citet{fan2020}, with a focus on developing cross-matching algorithms for radio sources with extended morphology,
matched the 784 radio sources from the ATLAS DR1 (\citealt{norris2006}) 
and IR sources from the SWIRE survey (\citealt{lonsdale2003}) in the W-CDF-S field.
\citet{weston2018} utilized the likelihood-ratio technique to match the 
ATLAS DR3 catalog (\citealt{franzen2015}) with the Spitzer Data Fusion catalog (\citealt{vaccari2015}) and 
found $2222/3078$ (72\%) and $1626/2113$ (77\%) matches for the ATLAS sources in the W-CDF-S and ELAIS-S1 fields, respectively.
Objects associated with multiple radio components were not identified in \citet{weston2018}.

We consider the DES DR2 catalog in the optical/NIR (\citealt{abbott2021}),
VIDEO DR5 catalog in the NIR (\citealt{jarvis2013}),
and DeepDrill/SERVS IRAC1 catalog in the MIR (\citealt{lacy2021}) for cross-matching (see the next section).
The VIDEO survey provides the most suitable data set for cross-matching in terms of angular resolution and depth
but does not cover all radio sources.
On the other hand, the DES and DeepDrill/SERVS data straddle the VIDEO bands, have near complete coverage of all the radio survey area,
and have their own strengths in terms of angular resolution and detection fraction, respectively.
The coordinates of these optical/IR sources are corrected to the Gaia reference frame using the method of the 
HELP project (\citealt{shirley2019}).

\subsection{Probabilistic matching algorithm}
\label{sec:lrx}
The most widely used method to find the counterparts in a catalog 
for a list of source coordinates is to select their nearest neighbours (NN) with a distance cut (e.g. \citealt{barbier1972}).
We calculate the normalized distance between radio and optical/IR sources
\begin{align}
    r^2 &= \Big(\frac{\Delta\mathrm{RA}}{\sigma_\alpha}\Big)^2 + \Big(\frac{\Delta\mathrm{Dec}}{\sigma_\delta}\Big)^2,
\end{align}
where $\sigma_\alpha$ and $\sigma_\delta$ are the radio position uncertainties. We did not add the 
uncertainties in the optical/IR positions since those of the radio positions dominate.
Genuine counterparts and confusing 
objects (optical/IR sources that are not the counterparts of radio sources) follow the Rayleigh distribution, 
$f(r) = re^{-r^2/2}$, and the linear distribution, $g(r) = 2\lambda r$, respectively,
where $\lambda = \pi\sigma_\alpha\sigma_\delta \rho$ is the expected number of confusing objects per
error ellipse and $\rho$ the number density of confusing objects.
Therefore, the likelihood ratio $f(r)/g(r) = e^{-r^2/2}/(2\lambda)$ shows that
the nearest neighbour (with the smallest $r$) is the maximum-likelihood-ratio candidate 
(we term MLRC hereafter).

The NN method above becomes progressively less accurate with increasing numbers of confusing objects per error ellipse, $\lambda$.
To improve the matching accuracy, we also incorporate the photometric properties of the candidates.
We select all optical/IR sources with $r< r_{\rm cut}=6$ as candidates,
which is sufficient since $1-\int_0^6re^{-r^2/2}dr=1.5\times10^{-8}$.
For each candidate, we calculate the likelihood ratio
\begin{align}
    \label{eq:lr}
    L &= \frac{Q}{1-Q} \frac{e^{-r^2/2}}{2\lambda}\frac{q(m)}{n(m)},
\end{align}
where $Q$ is the fraction of the radio sources that have counterparts in the optical/IR catalogs.
Furthermore, $q(m)$ and $n(m)$ represent the normalized magnitude distributions for 
the genuine counterparts and confusing objects, respectively.
We estimate them in \S~\ref{sec:source_stack}.

Assuming that there are $N$ candidates with $r<r_{\rm cut}$ for a radio source,
the cross matching is to make an inference about labels of these candidates, 
`genuine counterpart' or `confusing object'.
The posterior probability that the $i$-th candidate is the genuine counterpart with the rest being confusing objects is
\begin{align}
    \label{eq:r}
    R_i &= \frac{L_i}{\sum_j^N L_j + 1},
\end{align}
which is called ``reliability'' in \citet{sutherland1992}.
The probability that the genuine counterpart is one of the $N$ candidates
is $R=\sum_i^N R_i$. Note that $\langle R\rangle = Q$.

In principle, we can treat $R_i$ as a weight and take 
all candidates into subsequent analyses, 
which is not always computationally feasible.
Very often, only the best candidates are selected as counterparts.
We select the candidate with the maximum likelihood ratio $L_{\max}$ (i.e. MLRC),
which has a posterior probability $R_{\max}$.
We apply a cut on $L_{\max}$ that suppresses the incidence of false positives 
and maximizes the number of matched genuine counterparts.
We show in Appendix~\ref{sec:appendix1} that the optimal cut on $L_{\max}$ is not unique for heterogeneous positional errors
but always corresponds to $R_{\max}>0.5$.
We therefore select MLRCs with $R_{\max}>0.5$ as counterparts,
and the radio sources with $R_{\max}\le0.5$ are flagged unmatched.

Our method utilizes positional error and photometric distributions of genuine counterparts and confusing objects.
There is no threshold tuning in our well-defined procedure.
Furthermore, it is straightforward to assess the performance of the matching results without 
simulations that are computationally intensive and probably less accurate (see Appendix~\ref{sec:appendix1}).
Note that this method assumes that the 
radio components are co-spatial with their optical/IR counterparts.
In \S~\ref{sec:vishost}, we describe our visual inspection 
for cases where the radio emission might extend outside of the host galaxies.

\subsection{The magnitude distribution}
\label{sec:source_stack}
The magnitude distributions, $q(m)$ and $n(m)$, are estimated utilizing the overdensity
caused by the genuine counterparts of radio sources (e.g. \citealt{ciliegi2003, chen2018}).
We first select all optical/IR sources that are 
within a radius of 5~arcsec and centered at all radio 
sources that are not affected by bright stars. 
This radius is chosen to be large enough to enclose the counterparts.
For example, we plot the resulting $Ks$-band magnitude distribution for the ATLAS/W-CDF-S sources in Fig.~\ref{fig:ks_prior} (a).
The optical/IR sources that are in the annular regions of inner 
and outer radii of 10 and 30 arcsec are selected as well,
which are used to assess the magnitude distribution of confusing sources.
The expected distribution of confusing sources within a circular region of 5 arcsec
is calculated and subtracted from the previous distribution in Fig.~\ref{fig:ks_prior} (a).
The resulting distribution then represents the counterparts of radio sources.
Notably the distribution becomes negative above 21.5 mag, 
due to the fact that the genuine counterparts 
are brighter than average VIDEO sources, and the sensitivity around bright sources is reduced (e.g. \citealt{brusa2007, smolcic2017}).
Generally, $n(m)$ and $\lambda$ estimated from regions far from the positions of radio sources are biased (e.g. \citealt{naylor2013}), 
which causes $q(m)$ also to be biased.\footnote{\label{ft:lambda}
The density of confusing objects in units of per error ellipse, $\lambda$, is affected by multiple factors in a complex manner.
The degraded sensitivity around bright objects tends to reduce $\lambda$,
and if the host galaxies of radio sources live in denser-than-average environments, the clustering effect increases $\lambda$.
Furthermore, since telescopes have finite spatial-resolving power, source confusion tends to reduce $\lambda$ (e.g. \citealt{pineau2011}), 
which affects the MIR data more severely (up to a bias factor of $\approx3$) than the optical and NIR data.}

To have a more accurate estimation of the faint part of $q(m)$, 
we first masked out all objects close to a NIR source brighter than 19 mag with a 10-arcsec circle 
and then repeat the steps in the previous paragraph.
Here, the 19 mag cut is chosen because it is close to the peak of the magnitude distribution in Fig.~\ref{fig:ks_prior} (a).
The results are shown in Fig.~\ref{fig:ks_prior} (b), which is a better estimate of the faint part of $q(m)$.
We then combine the distributions resulting from Fig.~\ref{fig:ks_prior} (a) and (b). 
The distribution from (b) is scaled up to correct for the loss of area by masking out bright sources, 
and the scaling factor is determined such that the two bins between 19 and 20 mag are consistent 
with those from panel (a). Note that the distributions in Fig.~\ref{fig:ks_prior} (a), (b), and (c) have a bin size of 0.5 mag for illustration purposes.  
In our calculation, these 0.5-mag bins are merged such that the resulting bins each have more than 20 objects to suppress fluctuations.

Finally, we perform a self-calibration for $q(m)$, $n(m)$, and $\lambda$ using an iterative
method (e.g. \citealt{luo2010}).
At this step, we do not assign counterparts to radio sources but utilize all candidates
within $r<r_{\rm cut}$ and their weights, $R_i$.\footnote{Our approach is different from 
the \texttt{auto} method of the {\sc nway} software (\citealt{salvato2018}),
which constructs $q(m)$ and $n(m)$ using safe matches and safe non-matches that result from cutting on the posterior 
probability of distance-based matching, and $Q$ and $\lambda$ are fixed.}
The initial $q(m)$, $n(m)$, and $\lambda$ are used for first cross matching, and we also put in a guessed $Q=0.9$.
After calculating the reliability, $R_i$, 
we estimate $Q=\sum_i^{N_{\rm c}} R_i/N_{\rm r}$, where $N_{\rm r}$ is 
the total number of radio sources and $N_{\rm c}$ the number 
of all candidates of all radio sources.
The new $\lambda$ is the initial value multiplied by $(N_{\rm c}-\sum_i^{N_{\rm c}}R_i)/(r_{\rm cut}^2\sum_j^{N_{\rm r}}\lambda_j)$,
and $q(m)$ and $n(m)$ are derived again using the magnitudes of all candidates weighted by $R_i$ and $1-R_i$, respectively.
These new estimations of $q(m)$, $n(m)$, $Q$, and $\lambda$ are then used in the next iteration.
Generally, the results converge rapidly and are stable after less than 5 iterations.
We show the comparison between the initial and final magnitude distributions in Fig.~\ref{fig:ks_prior}.
The iteration improves both the distributions of genuine counterparts and confusing objects.
The confusing objects around radio sources are brighter than those from blank sky as expected.

Note that the properties of the radio surveys in Table~\ref{tab:atlascat} are not uniform.
Furthermore, the quality of the optical/IR data can also vary between fields.
These factors will affect the photometric distributions and $Q$.
Our method is self-contained and relies only on the two surveys (radio and optical/IR) for 
matching.
We calculate $q(m)$, $n(m)$, $Q$, and $\lambda$ for any pair of radio-optical/IR catalogs under consideration for cross matching.

\subsection{Automatic matching results}
\subsubsection{Counterpart catalogs and internal quality assessment}
\label{sec:xm_results}
Utilizing the algorithm and photometric properties in \S~\ref{sec:lrx} and \S~\ref{sec:source_stack}, 
we performed automatic cross matching between radio and optical/IR catalogs.
Even though only MLRCs are selected as counterparts, we provide the full results in Appendix~\ref{sec:full_lrx}, 
where the reliability of each candidate is given.

We assess the matching results in several different respects.
The number of radio sources with $R_{\max}>0.5$ is denoted $N_{\rm M}$, 
and there are $N_{\rm r}$ radio sources in total under consideration for matching.
We calculate the completeness $\mathcal{C}={N_{\rm M}}/N_{\rm r}$ 
and purity $\mathcal{P}=\sum_i^{N_{\rm M}}R_i / N_{\rm M}$, which is the fraction of counterparts that are genuine.
We also calculate 
the fraction of radio sources that are detected in the optical/IR catalog but are flagged unmatched (i.e. false negatives),
FN $=(Q-\mathcal{CP})/Q$. We show the results in Table~\ref{tab:xmatch}.
Notably, our cross matching results have very high purity (0.93--0.99) and low false-negative fraction ($<0.07$).
With high purity and low false-negative fraction, $\mathcal{C}$ is ultimately limited by $Q$.
Note that $Q$ is not only affected by the depth of optical/IR observations.
A fraction of the radio sources represent extended jet/lobe emission that do not spatially coincide with their host galaxies,
and they do not have counterparts in the optical/IR catalog, which reduces $Q$.
Furthermore, confused radio sources do not have unique optical/IR counterparts and reduce $Q$ as well.
The VLA/XMM-LSS survey with the smallest beam size (see Table~\ref{tab:atlascat})
has larger $Q$ values than those of the ATLAS survey in Table~\ref{tab:xmatch}, 
even though their sensitivities are similar.
We also calculate the predicted completeness, purity, and false-negative fraction using prior 
knowledge (see Appendix~\ref{sec:appendix1}) to check for consistency; 
the predicted values are consistent excellently with those calculated using the matching results.

\subsubsection{Assessing the matching results  in the VLA/E-CDF-S field}
\label{sec:vla_ecdfs}
The VLA/E-CDF-S survey (\citealt{miller2013}) covers a small sky
region (0.324 deg$^2$) inside the W-CDF-S field;
it has a factor of 2 deeper sensitivity and a factor of about 5 smaller 
beam size compared to those of the ATLAS survey.
The typical RA and Dec errors are $\approx0.2$ arcsec and $\approx0.3$ arcsec, respectively.
We utilize the deep and high-resolution VLA/\mbox{E-CDF-S} data 
to assess the performance of our automatic cross-matching.
Note that even though the VLA/E-CDF-S data are nominally deeper,
it is possible that some extended radio emission with low surface brightness
is lost in the VLA image but recovered in the ATLAS image due to their different beam sizes.

There are 243 ATLAS radio components within the region
covered by the VLA/\mbox{E-CDF-S} survey with an rms noise cut of $\le$12~$\micro$Jy.
We first matched the resulting 243 ATLAS sources with the
VLA source list and found their nearest neighbours.
We visually inspected the ATLAS and VLA postage-stamp
images (0.5 arcmin $\times$ 0.5 arcmin) to ensure that the associations are real.
Two (CI1614 and CI1685) out of the 243 ATLAS radio sources have no counterparts in the 
VLA source catalog, but they are apparently present in the VLA image.
Therefore, these two components are genuine radio sources
and we do not exclude them from the list.

Then, we match the precise VLA positions to the optical/IR catalogs.
Note that we exclude 6 radio components that are close 
to bright stars---one radio source that is just outside the star mask region, and 5 radio sources that are formally within.
For the 150 VLA sources that are consistent 
with point sources,\footnote{We include the sources either having EXTENDEC\_FLAG == 0 or FLUX\_CHOICE\_FLAG == P.}
we performed position-based likelihood-ratio matching with the VIDEO catalog (see \S~\ref{sec:lrx}).
The resulting $R_{\max}$ is larger than 0.7 for 145 objects, and their NIR counterparts are found (the maximum VLA-VIDEO separation is 1.07 arcsec).
The remaining 5 objects with $R_{\max}<0.002$ are flagged unmatched, 
and their counterparts are not found in the DES or SERVS images either.
For the 82 VLA extended sources, visual inspection is performed.
Postage-stamp images in the DES $i$, VIDEO $Ks$, and SERVS IRAC1 bands were also produced 
alongside the radio postage-stamp images to verify associations.
In total, we find optical/IR counterparts for 216/237 radio sources.
Among the 21 unmatched radio sources, 11 are extended jets/lobes of radio galaxies,
and the remaining 10 are due to the optical/IR survey limits.

Among these 237 ATLAS sources,
203 are flagged matched in the ATLAS-VIDEO (completeness 0.86) 
matching of \S~\ref{sec:xm_results},
among which 192 are identical to the counterparts found 
with the aid of the VLA/E-CDF-S (purity 0.95), 
which is consistent with the estimates in Table~\ref{tab:xmatch}.
We inspected the 24 cases where automatic cross matching in \S~\ref{sec:xm_results} did not successfully recover the counterparts and found that 
those ATLAS sources are usually confused. Strictly speaking,
there are no unique optical/IR counterparts for these confused objects unless one of the sub-components dominates the emission.

\subsubsection{Assessing cross matching using {\sc nway}}
\label{sec:nway}

The {\sc Nway} software (\citealt{salvato2018}) performs probabilistic (i.e. Bayesian) cross matching among more than two catalogs,
utilizing positional uncertainties, source number densities, and photometric properties;
one of the catalogs is chosen as primary (i.e. the radio catalog in our case),
and combinations of objects in the remaining secondary catalogs are returned as counterparts.
All catalogs in a probabilistic cross-matching method have equal status,
and the model should describe the associations of all objects under consideration,
which include those where the object from the primary catalog does not participate.
The number of hypotheses increases rapidly with the number of catalogs, which can be appreciated from the special 
case considered by \citet{pineau2017} where each secondary catalog contains only one candidate.
Splitting the catalogs into one primary and multiple secondaries, {\sc Nway} has used an approximation to 
reduce the model complexity (see Appendix~B1 of \citealt{salvato2018}).

We cross match the ATLAS catalog with the DES, VIDEO,
and SERVS catalogs in the W-CDF-S field to assess if {\sc Nway} 
performs better than the method in \S~\ref{sec:lrx}.
Again, we utilize the \mbox{VLA/E-CDF-S} data as the reference. 
We set the completeness parameter at 0.87 and 
use the magnitude distributions obtained in \S~\ref{sec:source_stack}.
We filter the results from {\sc nway} using {\texttt match\_flag==1} and ${p_i>0.1}$.
If we use $p_\mathrm{any}>0.5$ as a criterion to select counterparts,
Using {\sc Nway} results in 188 optical/IR counterparts, among which 166 are also found by our methods.
We are interested in if {\sc Nway} finds more genuine counterparts that are flagged unmatched in \S~\ref{sec:xm_results}.
However, most of the remaining 22 counterparts are not genuine but false positives of {\sc Nway}.
There are four objects that are flagged unmatched in \S~\ref{sec:xm_results} but where {\sc Nway}
finds counterparts with $p_\mathrm{any}>0.85$.
Two of these objects are confused objects where the radio emission is produced by two pairs of close galaxies.
For these cases, it is better to have the radio sources flagged as unmatched rather than assigning incorrect radio flux densities to galaxies.
The other two are extended radio components that are not co-spatial with their host galaxies.
For these radio sources, we mainly rely upon visual inspection in the next subsection.
We conclude that {\sc Nway} does not significantly improve the cross-matching results,
and we rely upon the results of \S~\ref{sec:xm_results} in this paper.


\subsection{Visual inspection}
\label{sec:vishost}
The automatic cross-matching method in the previous subsections is 
designed for radio sources with single components that are co-spatial with their optical/IR counterparts, 
which do not include the jet and lobe emission of radio galaxies.
We perform visual inspection to identify radio sources with multiple sub-components and find their host galaxies.
Since these sub-components do not have counterparts in the optical/IR catalogs and are often not point radio sources, 
we select radio sources that are flagged unmatched in \S~\ref{sec:xm_results} and/or are flagged as resolved for visual inspection.

For each radio source, we created cutouts 
with an initial size of 1 $\times$ 1 arcmin$^2$ in the bands of radio, VIDEO $Ks$, SERVS IRAC1, and DES $i$.
When there are multiple radio images of different resolutions available, we use all radio images.
For example, since the MIGHTEE/XMM-LSS field is completely within the VLA/XMM-LSS field, for each MIGHTEE source for inspection,
we create three radio cutouts from the MIGHTEE high-sensitivity/low-resolution image (8.2 arcsec), 
the MIGHTEE low-sensitivity/high-resolution image (5 arcsec), and the VLA image (4.5 arcsec).
If the initial cutouts are too small to include all radio components associated with a radio galaxy, 
we increase the cutout size during inspection.

In the top panels of Fig.~\ref{fig:morph_ext_example}, we show an example of a radio galaxy with a lobe-core-lobe morphology found by our visual inspection.
In addition to AGNs, star-forming processes in very nearby galaxies also produce resolved radio sources and radio sources with multiple sub-components.
The bottom panels show one case where the extended radio emission is apparently associated with a star-forming spiral disk instead of the nucleus of a galaxy.

During inspection, we also flagged those radio sources that appear confused and have multiple optical/IR counterparts.
The flags are contained in Table~\ref{tab:full_lrx}.
These confused radio sources are removed from the analyses in the rest of this paper.
It is easiest to find confused radio sources in the MIGHTEE/XMM-LSS catalog thanks to the high-resolution image.

Point (unresolved) sources that have counterparts found in \S~\ref{sec:xm_results} are not visually inspected.
Source confusion is not expected to be significant for such radio sources since
the criterion of $R_\mathrm{\max}>0.5$ ensures that the MLRC dominates the likelihood ratio (see Appendix~\ref{sec:appendix1}).


We provide the final cross-matching results in Table~\ref{tab:atAll} and Table~\ref{tab:lssAll}, 
merging the results from visual inspection and the automatic method.
The radio sources appeared confused are excluded.
In total, counterparts are found for 2915/3034, 1855/2084, 5462/5762, and 18974/20274 radio sources in the ATLAS/W-CDF-S, ATLAS/ELAIS-S1, VLA/XMM-LSS, and MIGHTEE/XMM-LSS surveys.
In Table~\ref{tab:lssAll}, we have combined the results of the MIGHTEE and VLA surveys in the XMM-LSS field. 

\begin{table*}
\centering
\caption{Optical/IR counterparts of ATLAS/W-CDF-S and ATLAS/ELAIS-S1 radio sources. Only the top 5 rows are shown.
    The full table is available as online supplementary material.
    Column (1): The name of the optical/IR counterpart in the format of Jhhmmss.ssddmmss.s.
    Note that a unique optical/IR counterpart might be associated with multiple radio sources.
    Column (2): The ID of the ATLAS source.
    Column (3): The ID of the {\sc Tractor} catalog (\citealt{zou2021tractor, nyland2023}).
    Column (4): The method of the matching result, $1=$ visualization and $0=$ automatic.
    Column (5): The largest extent of radio galaxies.
    We have filled empty values with $-99$.
    }
\label{tab:atAll}
\begin{threeparttable}[b]
\begin{tabularx}{\linewidth}{@{}Y@{}}
\begin{tabular}{lcccc}
\hline
\hline
    Name &  ATLASID & TRACTORID & VIS & EXT \\
    &  & &  & arcmin\\
    (1)&  (2)& (3)& (4)& (5)\\
\hline
    J032553.86$-$280455.0 & CI0001 &      $-$99 & 0 & $-$99 \\
    J032554.05$-$284155.1 & CI0002 &      $-$99 & 0 & $-$99 \\
    J032555.30$-$283744.5 & CI0003 &      $-$99 & 0 & $-$99 \\
    J032557.32$-$280308.9 & CI0004 &      $-$99 & 0 & $-$99 \\
    J032558.27$-$281152.3 & CI0005 &      $-$99 & 0 & $-$99 \\
\hline
\end{tabular}
\end{tabularx}
\end{threeparttable}
\end{table*}

\begin{table*}
\centering
\caption{Optical/IR counterparts of MIGHTEE/XMM-LSS and VLA/XMM-LSS radio sources. Only the top 5 rows are shown.
    The full table is available as online supplementary material.
    Column (1): The name of the optical/IR counterparts in the format of Jhhmmss.ssddmmss.s.
    Note that an unique optical/IR counterpart might be associated with multiple radio sources.
    Column (2): The ID of the MIGHTEE source.
    Column (3): The ID of the VLA source.
    Column (4): The ID of the {\sc Tractor} catalog (\citealt{nyland2023}).
    Column (5): The method of the matching result, $1=$ visualization and $0=$ automatic.
    Column (6): The largest extent of radio galaxies.
    }
\label{tab:lssAll}
\begin{threeparttable}[b]
\begin{tabularx}{\linewidth}{@{}Y@{}}
\begin{tabular}{lccccc}
\hline
\hline
    Name &  MTID & VLAID & TRACTORID & VIS & EXT \\
    &  & &  & & arcmin\\
    (1)&  (2)& (3)& (4)& (5)& (6)\\
\hline
    J021401.02$-$043245.4 & $-$99 & J021400.99$-$043244.3 & $-$99 & 0 & $-$99 \\
    J021401.24$-$042924.3 & $-$99 & J021401.36$-$042923.3 & $-$99 & 0 & $-$99 \\
    J021408.85$-$051844.3 & $-$99 & J021408.85$-$051843.9 & $-$99 & 1 & $-$99 \\
    J021415.46$-$043039.4 & $-$99 & J021415.45$-$043038.4 & $-$99 & 1 & $-$99 \\
    J021415.65$-$044137.0 & $-$99 & J021415.65$-$044136.0 & $-$99 & 1 & $-$99 \\
\hline
\end{tabular}
\end{tabularx}
\end{threeparttable}
\end{table*}

\begin{table}
\centering
\caption{Definition of main samples of extragalactic radio sources}
\label{tab:mainsample}
\begin{threeparttable}[b]
\begin{tabularx}{\linewidth}{@{}Y@{}}
\begin{tabular}{lccc}
\hline
\hline
    Step &  W-CDF-S& ELAIS-S1 & XMM-LSS \\
\hline
    Cross-matching & 2762 & 1851 & 18261 \\
    VIDEO Coverage & 2421 & 1057 & 17115 \\
    Bright Star Mask & 2408 & 1051 & 17036 \\
    Remove Stars & 2380 & 1041 & 16985 \\
\hline 
\end{tabular}
\end{tabularx}
\end{threeparttable}
\end{table}

\begin{landscape}
\begin{table}
\centering
    \caption{Radio AGNs selection and {\sc cigale} modeling results. 
    Only the top 5 rows are given.
    The full table is available as online supplementary material.
    Column (1): Field name. 
    Column (2): Object name.
    Column (3): Redshift.
    Column (4): Redshift type.
    Column (5)(6): Lower and upper limit of Column (3) if the redshift type in Column (4) is photometric.
    Column (7): Three flags that indicate if the object is selected as a radio AGN based on morphology, radio slope, and radio excess.
    Column (8): The flag for X-ray AGNs, and $-1$, 1, and 0 represent X-ray unmatched, X-ray AGNs, and X-ray galaxies, respectively.
    Column (9): The flag for MIR AGNs.
    Column (10)(11): The flux density and error at 1.4 GHz (20 cm).
    Column (12)(13): The flux density and error at 24 $\mu$m.
    Column (14)(15)(16): The stellar mass, SFR, and $\chi_\nu^2$ of the {\sc cigale} fits with a galaxy-only model.
    Column (17)(18)(19): The stellar mass, SFR, and $\chi_\nu^2$ of the {\sc cigale} fits with a galaxy$+$agn model.
    Column (20): The number of detected UV-to-FIR flux densities used in the {\sc cigale} modeling.}
\label{tab:ragn_table}
\begin{threeparttable}[b]
\begin{tabularx}{\linewidth}{@{}Y@{}}
\begin{tabular}{lcccccccccccc}
\hline
\hline
    Field &  Name & $z$ & $z$-type & $z_\mathrm{phot, lo}$& $z_\mathrm{phot, up}$ & Radio AGN & X-ray AGN &MIR AGN & $S_\mathrm{1.4GHz}$ & $S_\mathrm{err, 1.4GHz}$ & $S_\mathrm{24\mu m}$ & $S_\mathrm{err, 24\mu m}$ \\
          &             &  & &     &               &               &    & & mJy & mJy &mJy & mJy\\
    (1)   &   (2)       &  (3) & (4) & (5) & (6) & (7) & (8) & (9) & (10) & (11) &(12) & (13) \\ 
\hline
    W-CDF-S &J032646.45$-$284952.7  & 0.853 &  phot &  0.762 & 0.903&  000 &  $-$1 &  0&0.333&0.041&0.8098&0.0149 \\
    W-CDF-S &J032647.93$-$283142.0  & 0.967 &  phot &  0.862 & 1.058&  001 &  $-$1 &  0&0.631&0.040&0.0597&0.0165 \\
    W-CDF-S &J032648.14$-$284329.9  & 0.781 &  phot &  0.668 & 0.905&  001 &  $-$1 &  0&0.297&0.034&0.0878&0.0145 \\
    W-CDF-S &J032648.20$-$275747.2  & 1.152 &  phot &  1.034 & 1.163&  001 &  $-$1 &  0&0.895&0.049&0.2128&0.0175 \\
    W-CDF-S &J032648.36$-$280003.8  & 1.705 &  phot &  1.629 & 1.979&  001 &  $-$1 &  0&0.180&0.021&0.0768&0.0196 \\
\hline 
\hline
    $M_{*,\rm gal}$ & SFR$_\mathrm{gal}$ & $\chi^2_\mathrm{\nu,gal}$ & $M_\mathrm{*,agn}$ & SFR$_\mathrm{agn}$ & $\chi^2_\mathrm{\nu,agn}$ & $n_\mathrm{det}$ \\ 
       $10^9M_\odot$   & $M_\odot$ yr$^{-1}$   & &  $10^9M_\odot$   & $M_\odot$ yr$^{-1}$  & \\  
    (14)   &   (15)       &  (16) & (17) & (18) & (19) & (20) \\
\hline 
    99.4&  17.9&  7.3&  50.7&  26.8& 7.9 & 18\\
    112.4 & 0.90&	0.90 & 154.4 & 0.21 & 0.79 & 11 \\
    164.7 & 0.17 & 0.31& 145.3 & 0.13& 0.40 & 10 \\
    6.2 & 67.3 & 2.0 & 6.9 & 39.1 & 2.2 & 10 \\
    166.7 & 2.0 & 1.0 & 154.9 & 0.89 & 0.94 & 8 \\
\hline 
\end{tabular}
\end{tabularx}
\end{threeparttable}
\end{table}
\end{landscape}

\begin{figure*}
\centering
\includegraphics[width=0.35\textwidth, clip]{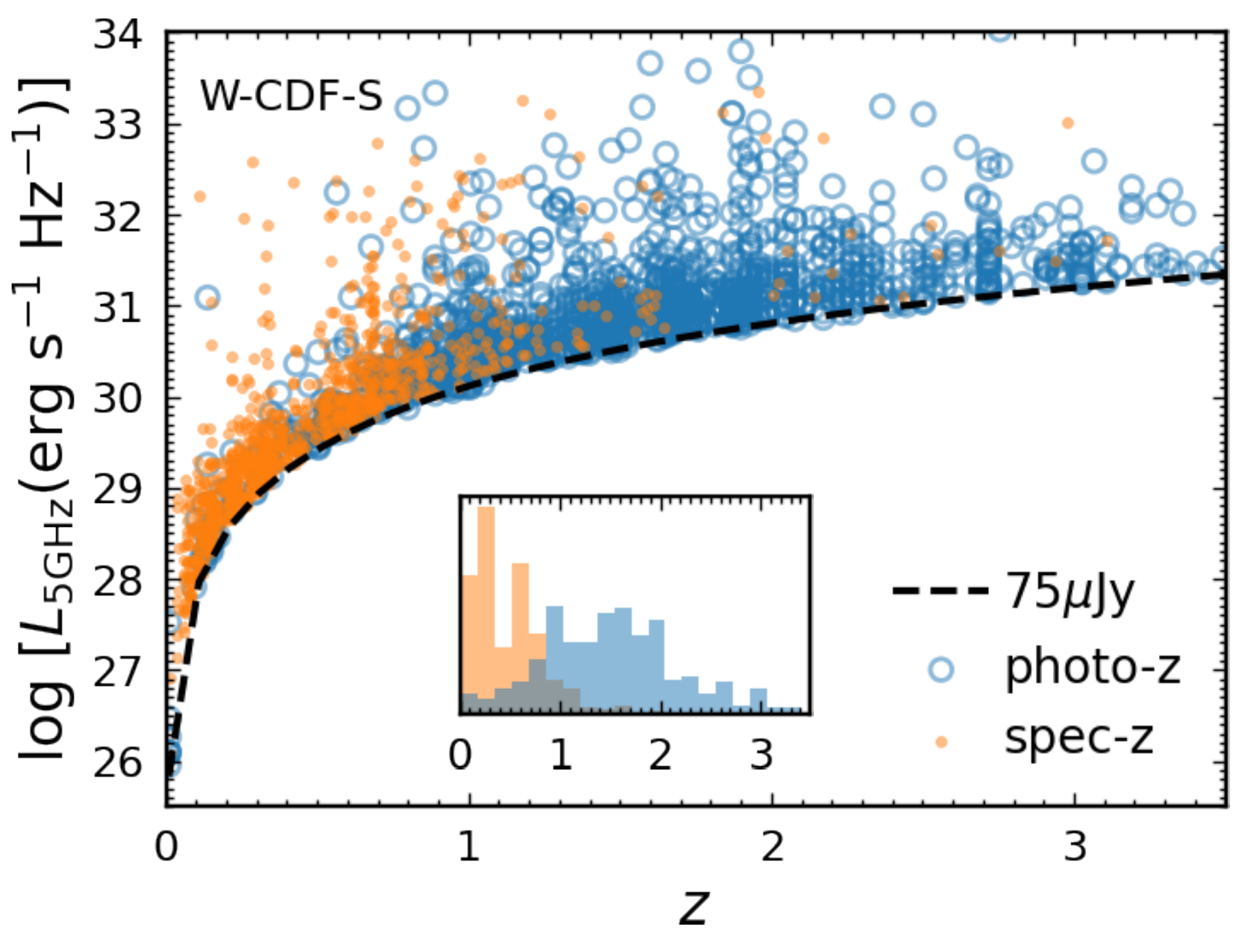}
\includegraphics[width=0.31\textwidth, clip]{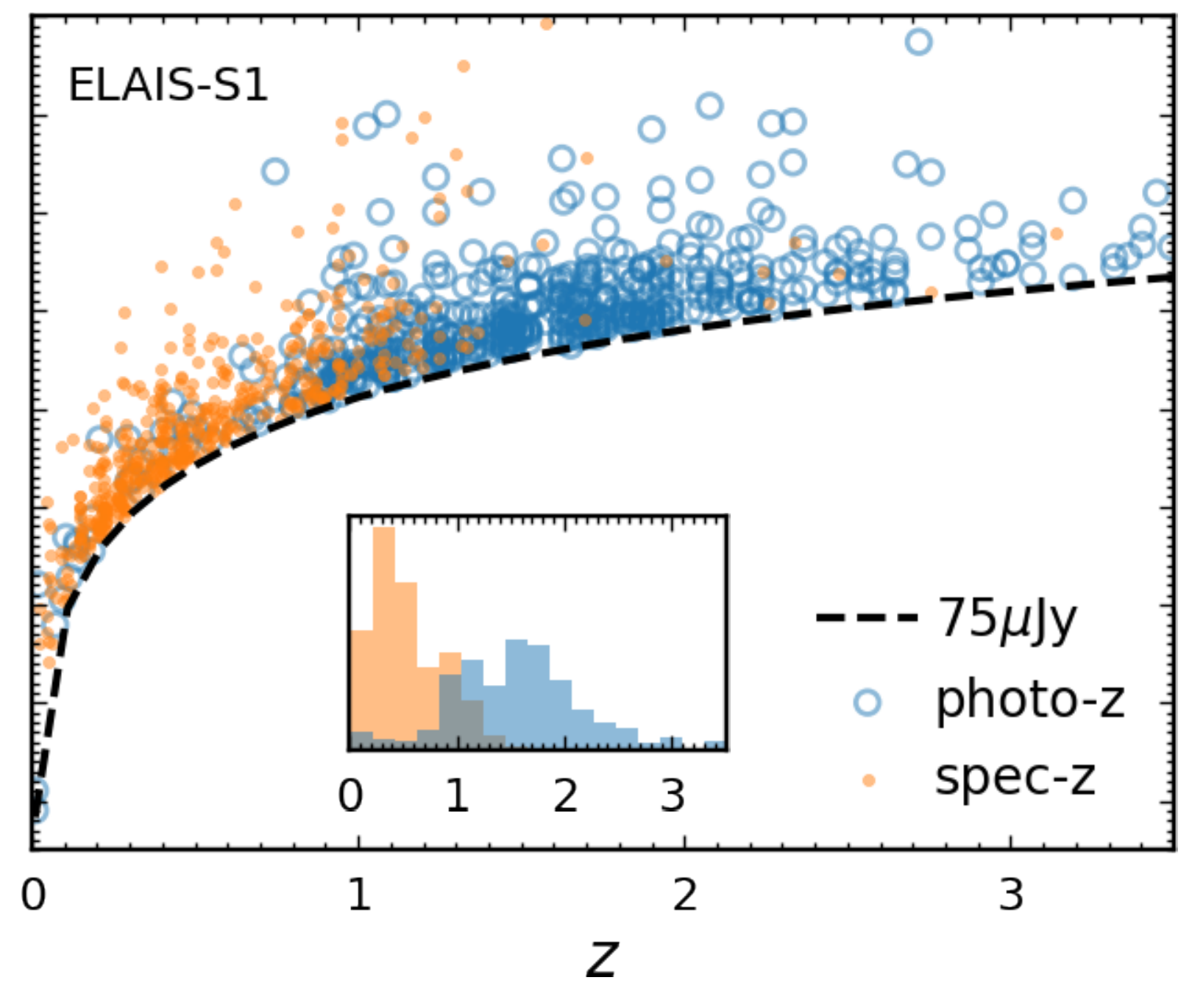}
\includegraphics[width=0.31\textwidth, clip]{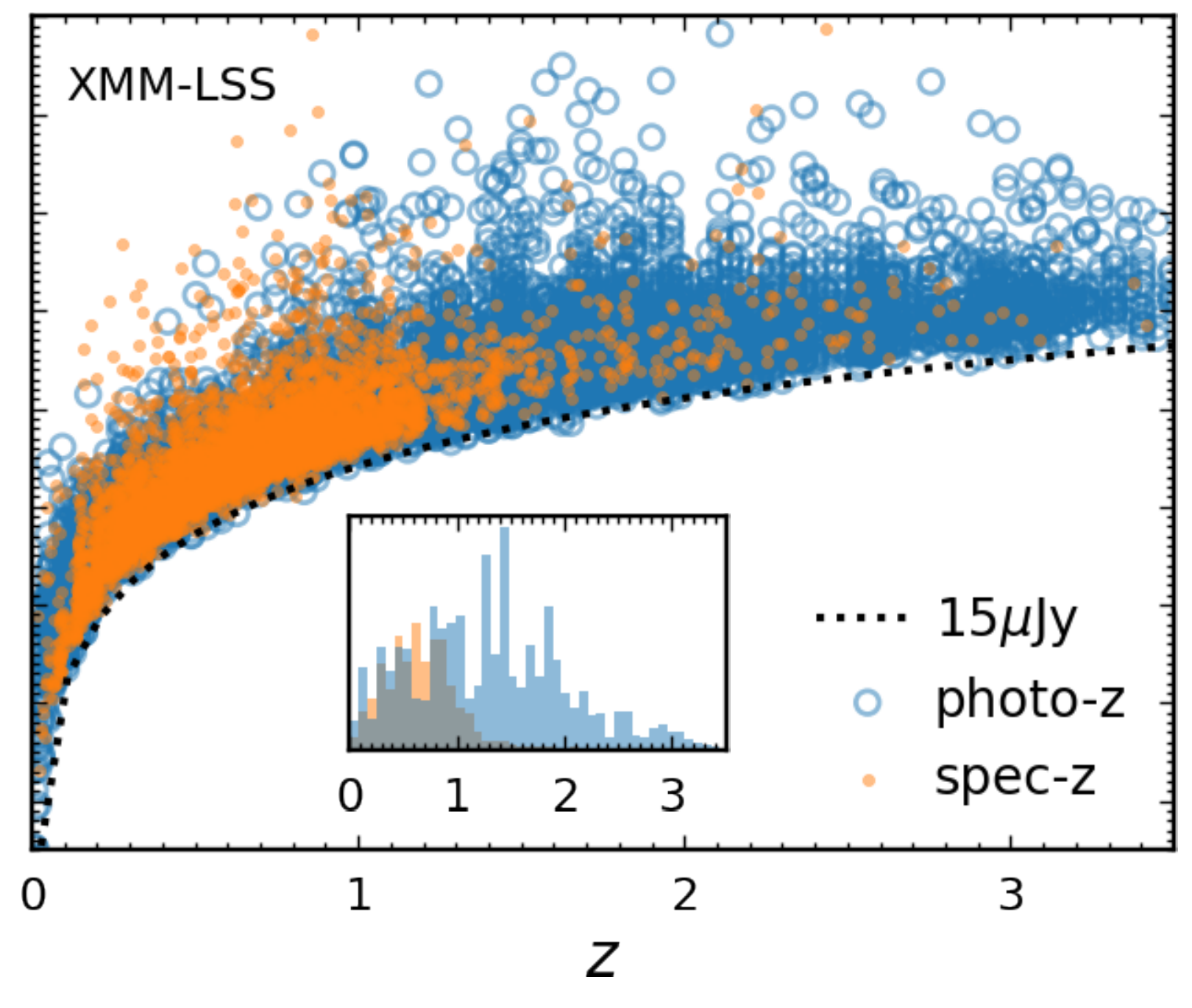}
\caption{The radio sources in the $L_\mathrm{5GHz}$-$z$ plane 
    for the W-CDF-S, ELAIS-S1, and XMM-LSS fields.
    The monochromatic radio luminosities at rest-frame 5 GHz are
    calculated assuming $\alpha_r=-0.7$.
    We show objects with photometric and spectroscopic redshifts
    using blue circles and orange dots, respectively.
    The inset of each panel shows the distributions of spectroscopic and photometric $z$ in orange and blue, respectively.
    The lower envelopes are indicated by dashed lines,
    which correspond to radio fluxes of 75~$\micro$Jy, 75~$\micro$Jy, and 15~$\micro$Jy (from left to right), respectively.
    }
\label{fig:zdist}
\end{figure*}

\subsection{Radio observations at higher and lower frequencies}
We also find the counterparts of our radio sources in catalogs at higher and lower radio frequencies in the three XMM-SERVS fields.
These radio surveys include 
the 2.3~GHz observations of the two ATLAS fields (\citealt{zinn2012}),
The Rapid ASKAP Continuum Survey (RACS; \citealt{mcconnell2020}) 
at 887.5 MHz that covers all three XMM-SERVS fields, 
and the LOFAR observations at 150 MHz of the XMM-LSS field (\citealt{hale2019}).

The 2.3~GHz sources in the W-CDF-S and ELAIS-S1 fields have been matched to 
1.4~GHz sources in the previous data release of ATLAS. 
To link these 2.3~GHz sources with 1.4~GHz sources from the ATLAS DR3,
we adopted a NN method with a distance cut of 5~arcsec, 
which results in 278 and 349 matches in the W-CDF-S and ELAIS-S1 fields, respectively.

We utilize the component-fitting results in the RACS and LOFAR catalogs to link them to the ATLAS/VLA/MIGHTEE surveys.
Specifically, we find ATLAS/VLA/MIGHTEE sources that are enclosed by 
the ellipse defined by major axes, minor axes, and the position angle of a component in the RACS/LOFAR catalog.
If only one ATLAS/VLA/MIGHTEE source is enclosed by the ellipse, we link it with the corresponding RACS/LOFAR component.
If more than one ATLAS/VLA/MIGHTEE source is enclosed by the ellipse and these ATLAS/VLA/MIGHTEE sources have one unique optical/IR counterpart, 
we also link these ATLAS/VLA/MIGHTEE sources with the corresponding RACS/LOFAR component.
Otherwise, if more than one ATLAS/VLA/MIGHTEE source is enclosed by the ellipse and 
these ATLAS/VLA/MIGHTEE sources have different optical/IR counterparts (including unknown optical/IR counterparts), 
we treat the RACS/LOFAR component as confused and remove it.
In total, we matched 382/190/323 RACS sources in the W-CDF-S/ELAIS-S1/XMM-LSS fields and 751 LOFAR sources in the XMM-LSS field.

\section{Multiwavelength properties of radio sources}
\label{sec:multi_wav}
\subsection{The main sample}
\label{sec:main_sample}
We define our main sample in the following way.
We require the optical/IR positions of the radio sources to be covered by the VIDEO survey
since the VIDEO fields have the richest multiwavelength coverage.
An object is considered covered if the value on the VIDEO confidence map is larger than the 5th percentile of all positive values.
Furthermore, we remove radio sources that are in the regions masked around bright stars.
Finally, we match the optical/IR counterparts with the stars compiled by \citet{zou2022}
and remove these radio stars since we are most interested in extragalactic radio sources.
In total, we removed 93 radio stars.
The results are shown in Table~\ref{tab:mainsample}.
The sky areas of the main sample are 3.22, 1.56, and 2.96 deg$^2$ in the W-CDF-S, ELAIS-S1, and XMM-LSS fields, 
leading to a total solid-angle coverage of 7.74~deg$^2$.
In comparison, the sky areas of the VIDEO data of the three fields are 4.5, 3, and 4.5 deg$^2$, respectively; 
therefore, about 65\% of the VIDEO fields are covered by the main radio catalog.

\subsection{Redshifts}
We use spectroscopic and high-quality UV-to-MIR photometric redshifts 
compiled by \citet{zou2022}.
There are 2324/2380, 1031/1041, and 16637/16985 objects that have redshift measurements in the \mbox{W-CDF-S}, ELAIS-S1, and XMM-LSS fields, respectively.
The fractions of spectroscopic redshifts are 42.3\%, 51.2\%, and 27.5\%, respectively.
About 75\% of the photometric redshifts have quality $Qz<1$ (good photometric redshifts).
In Fig.~\ref{fig:zdist}, we show radio sources with redshift measurements in the $L_\mathrm{5GHz}$-$z$ plane,
where $L_\mathrm{5GHz}$ is the rest-frame 5 GHz luminosity.
Since only a small portion of the radio sources have multiwavelength radio data,
a uniform radio spectral index of $\alpha_\mathrm{r}=-0.7$ (e.g. \citealt{smolcic2017a}) is adopted when we calculate $L_\mathrm{5GHz}$.

\subsection{Photometry from the X-rays to IR}
\label{sec:photometries}

We use X-ray data from the XMM-SERVS survey (\citealt{chen2018, ni2021}) 
to constrain the high-energy properties of the radio sources.
First, we define X-ray coverage to be if the hosts of radio sources have $>5$th percentile of full-band exposure time in each field.
Second, since the X-ray sources in the XMM-SERVS catalogs have been matched to their optical/IR counterparts, 
we link the X-ray and radio sources if their respective optical/IR positions are separated by $<1$~arcsec and 
the X-ray position is $<3$~arcsec away from the optical/IR position of the radio source.
Since the median X-ray positional error is 1.2~arcsec, the 3~arcsec cut ensures that the optical/IR position is 2.5~$\sigma$ from the X-ray position.
If the X-ray position alone is $<2$~arcsec from the optical/IR position of the radio source,
we also think the X-ray is associated with the radio sources.
This matching procedure results in 220/2285, 88/1018, and 965/16481 securely X-ray detected
radio-source host galaxies that are covered by the {\it XMM-Newton} observations.
Note that the X-ray detection fractions of the host galaxies of these faint radio sources are low, in the range of 5--10\%.
Aside from these securely X-ray detected sources, we also label sources that are securely \mbox{X-ray} undetected.
We masked out detected X-ray sources utilizing the XMM-SERVS catalog with circular regions with a radius of $10\log N_\mathrm{full}$~arcsec, 
where $N_\mathrm{full}$ is the full-band net counts. 
Generally, $N_\mathrm{full}=$ 10--10$^4$, rendering masks of the size of 10--40~arcsec.
These masks are large enough such that the brightness of the source is below that of background beyond these radii (\citealt{read2011}).
The host galaxies of radio sources that are covered by X-ray observations but in the source-free regions are securely undetected,
and there are 1753/2285, 814/1018, and 13193/16481 such objects in the W-CDF-S/ELAIS-S1/XMM-LSS fields.
Their X-ray properties are constrained later via stacking analyses in \S~\ref{sec:x_stack}.

For the multiwavelength data in the optical and NIR, 
we mainly use the {\it Tractor} catalogs (\citealt{zou2021tractor, nyland2023}), 
which cover $u$ through IRAC2.
In the main samples, 2326/2380, 1031/1041, 
and 16644/16985 objects have counterparts in the {\it Tractor} 
catalogs of the W-CDF-S, ELAIS-S1, and XMM-LSS fields.
The {\it Tractor} fluxes have an offset of $\approx0.1$ mag compared with the original VIDEO catalog (\citealt{nyland2017}) in the $J$ and $Ks$ bands, 
and we corrected these two bands with $+0.111$ mag and $-0.118$ mag in all three fields (cf. Table~3 of \citealt{nyland2017}).
Furthermore, we also find a notable offset for the CFHT $u$ band of the XMM-LSS field by comparing with the original catalog (\citealt{hudelot2012}),
and we correct this band by $-0.305$ mag, 
which is the median offset.
We find that the offsets are (sometimes) different for point sources and extended sources.  Therefore, improving the PSFs used by the {\it Tractor} might help to resolve the offsets.
The corrections we make are also backed by the {\sc cigale} fitting results (see \S~\ref{sec:cigale}).
If the uncorrected fluxes are used, 
the distributions of the SED fit residuals at these bands are not centered around zero.
Note that inconsistencies of similar magnitude between photometric 
catalogues derived from different techniques are also present in other fields (e.g. \citealt{davies2021}).
We supplement these data with the DES DR2 ($grizY$) photometry
if the {\it Tractor} photometry is missing.

For the two red IRAC channels (5.8 and 8.0 $\mu m$), we use the photometry from the Spitzer Data Fusion (\citealt{vaccari2015}).
Furthermore, to construct a uniform data set in the MIR, we additionally estimate 
the 5.8~$\mu$m and 8.0~$\mu$m band upper limits from the SWIRE survey in Appendix~\ref{sec:upp}.
Additionally, we use the WISE-3 (12$\mu m$) photometry from the AllWISE catalog (\citealt{wright2010}), and we 
use a zero point of 29.045~Jy to convert the WISE-3 magnitudes to energy flux densities (\citealt{wright2010}).
Finally, we use MIPS 24$\mu m$, PACS 100$\mu m$, PACS 160$\mu m$, SPIRE 250$\mu m$, SPIRE 350$\mu m$, and SPIRE 500$\mu m$
photometry from the HELP project (\citealt{shirley2021}), which uses the XID+ algorithm to deblend confused far-infrared (FIR)
sources.\footnote{We fixed issues with the XID+MIPS24 and XID+PACS catalogs in Appendix~\ref{sec:xid_m24}.}
Note we also utilize the measurements of the upper limits (from Appendix~\ref{sec:upp} or \citealt{zou2022}) 
in these FIR bands if they are not matched in the catalogs above.

\subsection{Radio-Selected AGNs}
\label{sec:radio_agns}
\begin{figure*}
\centering
\includegraphics[width=0.95\textwidth, clip]{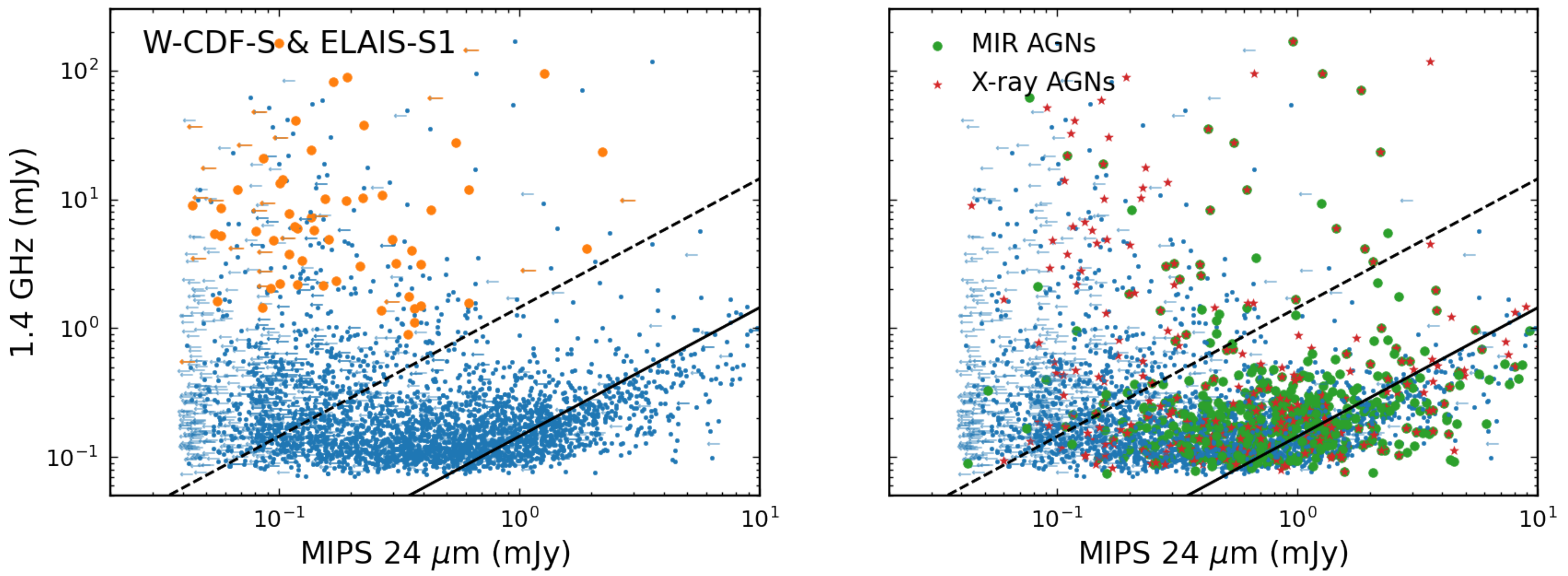}
\includegraphics[width=0.95\textwidth, clip]{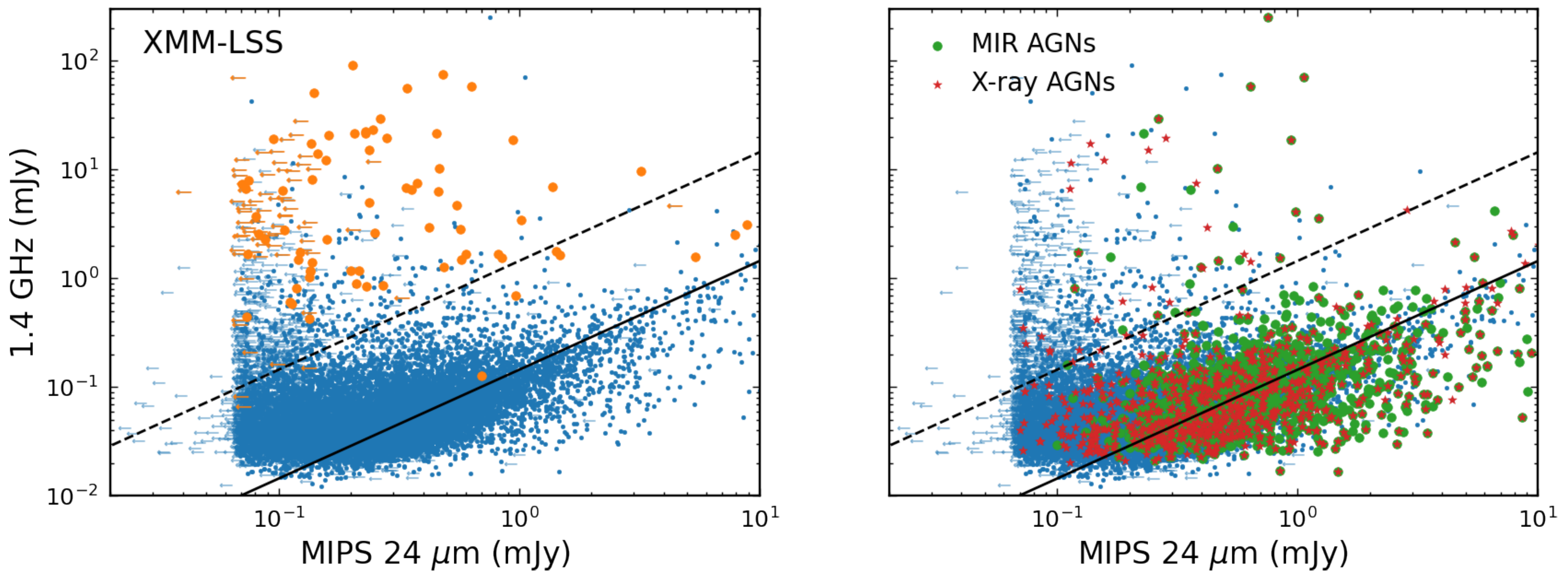}
\caption{Left: Radio (1.4 GHz) vs. FIR (24 $\mu$m) flux densities. Left arrows represent 24 $\mu$m upper limits. 
    Large orange dots are radio AGNs that are selected by radio morphology or spectral index, while the remaining radio sources are shown in blue.
    The black solid line shows $q_{24} = \log(f_\mathrm{24\mu m}/f_\mathrm{1.4GHz})= 0.84$ from \citet{appleton2004},
    which is expected for radio emission produced by star-forming processes.
    We shift radio flux densities predicted by $q_{24}$ 
    by a factor of 10 to select objects that show strong radio excess (i.e. RL AGNs), 
    which is shown as the dashed line.
    Right: Same as the left panels, 
    but MIR and X-ray selected AGNs are shown as large green dots 
    and red stars, respectively.
    Objects from the ATLAS fields (W-CDF-S and ELAIS-S1) are shown in the same plots.
    Most MIR and X-ray AGNs are radio quiet.
    }
\label{fig:fir_radio}
\end{figure*}

The radio emission from extragalactic sources could be produced by AGN jets and/or star-forming processes.
However, star-formation related radio emission is produced within the galaxy, 
has a typical radio spectral index of $\approx-0.7$ to $-0.8$, 
and is related the FIR emission (e.g. \citealt{condon1992, delhaize2017, tabatabaei2017, an2021}).
Therefore, we select radio(-loud) AGNs using radio morphology (lobe/jet structures), radio spectral index ($\alpha_r>-0.3$), 
or radio excess (sources with a radio excess have radio flux densities 10 or more times larger than those predicted by the 
$q_{24}$ parameter of \citealt{appleton2004}), where  $q_{24}=\log(S_{24\mathrm{\mu m}}/S_{1.4\mathrm{GHz}})$.
The results are given in Table~\ref{tab:ragn_table} and also shown in the left panels of Fig.~\ref{fig:fir_radio}.
There are 713, 275, and 827 radio AGNs in the W-CDF-S, ELAIS-S1, and XMM-LSS, respectively.
Note that the sky density of the MIGHTEE sources in the XMM-LSS field
is a factor of $\approx7$ higher than that of the ATLAS sources in the W-CDF-S and ELAIS-S1 fields,
but the sky density of radio AGNs for the former is only about 50\% higher than for the latter.
Therefore, as expected, radio sources fainter than the ATLAS sensitivity are dominated by galaxies and radio-quiet AGNs.
The faint radio AGNs below the ATLAS sensitivity are generally harder to identify and study in detail than the objects we select in this subsection.

\begin{figure}
\centering
\includegraphics[width=0.45\textwidth, clip]{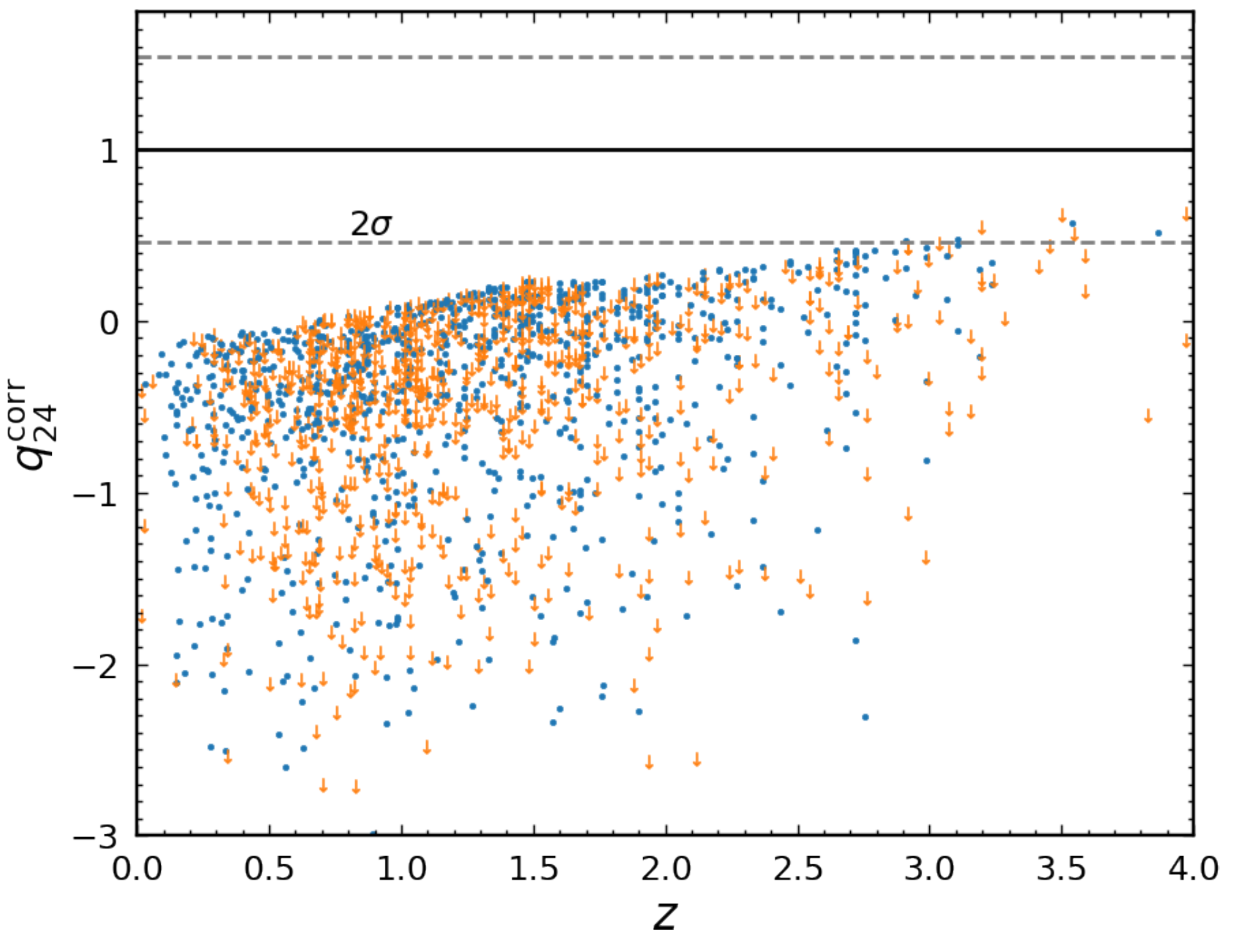}
\caption{The $K$-corrected $q_{24}$ parameter of the radio-excess AGNs, where we have assumed an M82-like SED in the MIR band.
    The solid and dashed horizontal lines represent the mean and $2\sigma$ 
    range of the $K$-corrected $q_{24}$ parameter of star-forming galaxies (\citealt{appleton2004}).
    Except for a few objects at $z>3$, the remaining objects show radio excesses at a $>2\sigma$ statistical-significance level after $K$ correction.
    }
\label{fig:q_corr}
\end{figure}

Note that the $q_{24}$ values we calculated are observed values (e.g. \citealt{middelberg2008}), and no $K$ correction was applied,
which depends upon redshift and the radio and MIR SEDs of the object.
The reliability of the AGNs selected by strong radio excess is not affected by the $K$ correction.
We assume a power-law spectral index of $-0.7$ in the radio band (e.g. \citealt{bonzini2013}), 
and an M82-like SED (e.g. \citealt{polletta2007, ibar2008}) in the MIR band.
We convolve the M82 SED with the MIPS 24~$\mu$m filter and normalize $q_{24}$ to 1 (\citealt{appleton2004}) at zero redshift.
Then, we calculate the $K$ correction of $q_{24}$ at different redshifts.
The $K$-corrected $q_{24}$ for radio-excess AGNs is shown in Fig.~\ref{fig:q_corr}, 
where the dashed lines represent the $2\sigma$ range of the $K$-corrected $q_{24}$ parameter (\citealt{appleton2004}).
Fig.~\ref{fig:q_corr} shows that these AGNs show radio excesses 
at a $2\sigma$ level after the $K$ correction of a star-forming galaxy, M82. 
The SEDs of radio AGNs might show a large diversity (see the left panel of Fig.~\ref{fig:stack_seds_agns}).
We therefore do not perform $K$ correction on the $q_{24}$ parameter here.
The $K$-corrected $q$ parameter, $q_\mathrm{FIR}$, is utilized in \S~\ref{sec:cigale}.
Furthermore, Fig.~\ref{fig:q_corr} indicates that we might miss some radio-excess AGNs at $z<2$.

\begin{figure*}
\centering
\includegraphics[width=0.29\textwidth, clip]{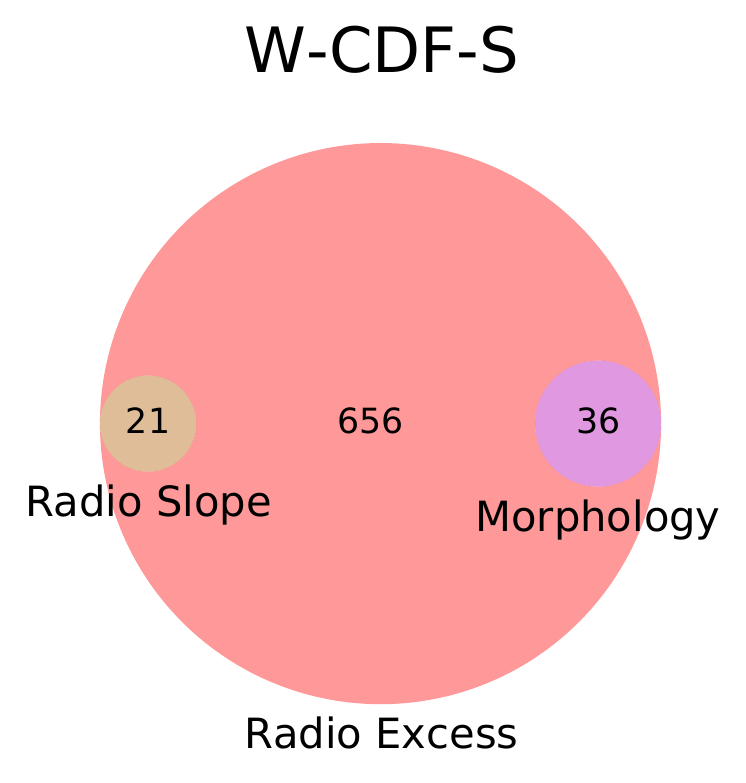}
\includegraphics[width=0.29\textwidth, clip]{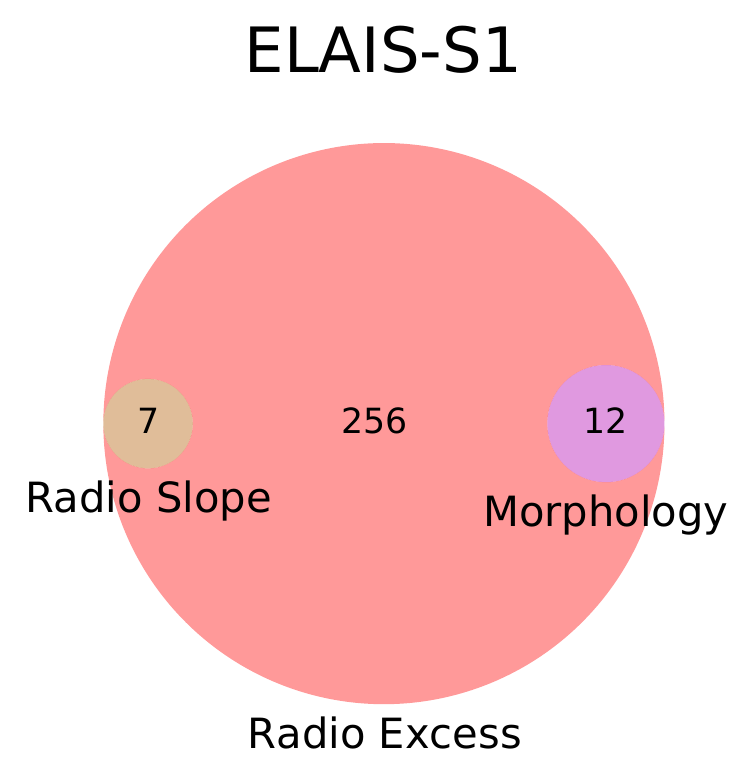}
\includegraphics[width=0.39\textwidth, clip]{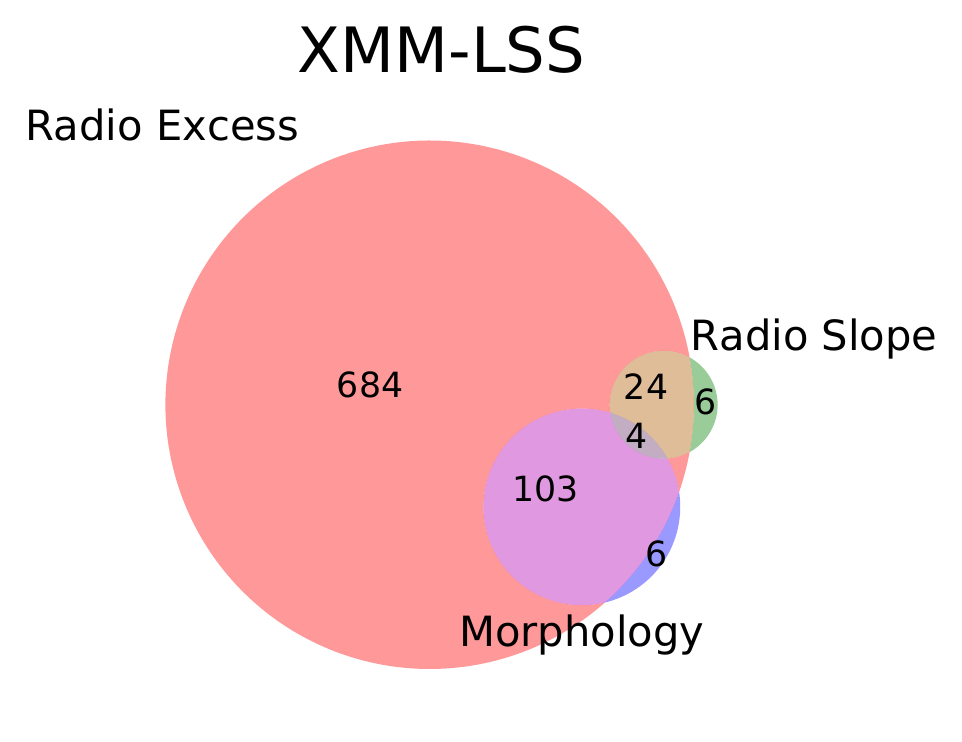}
\caption{Venn diagrams for the radio AGNs selected using various methods in the W-CDF-S, ELAIS-S1, and XMM-LSS fields.
    Note that the morphology selected radio AGNs are mainly lobe-dominated objects, 
    which usually have radio slope steeper than $\alpha_{\mathrm r}=-0.3$;
    thus, these two groups are largely disjoint.}
\label{fig:venn_r_agns}
\end{figure*}

In the right panels of Fig.~\ref{fig:fir_radio}, we show MIR and X-ray AGNs among the radio sources.
Most of these radio-detected MIR and X-ray AGNs are not radio-loud.
Note that large-scale lobes/jets are often luminous radio emitters,
and sources with measured spectral index in the radio band are also biased toward bright sources. 
Therefore, radio morphology/spectrum selected AGNs are almost always selected by the $q_{24}$ criterion as well.
We have created the Venn diagrams of radio AGNs selected using different methods in Fig.~\ref{fig:venn_r_agns}.

\begin{figure*}
\centering
\includegraphics[width=0.95\textwidth, clip]{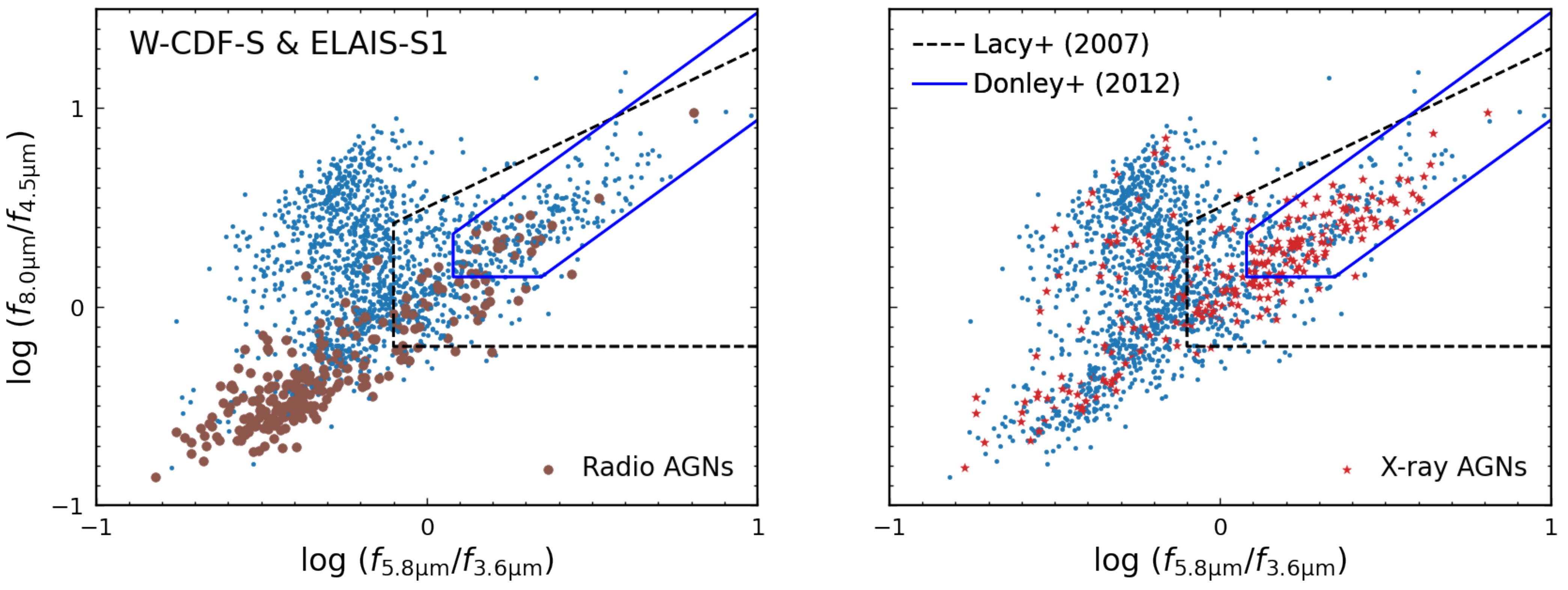}
\includegraphics[width=0.95\textwidth, clip]{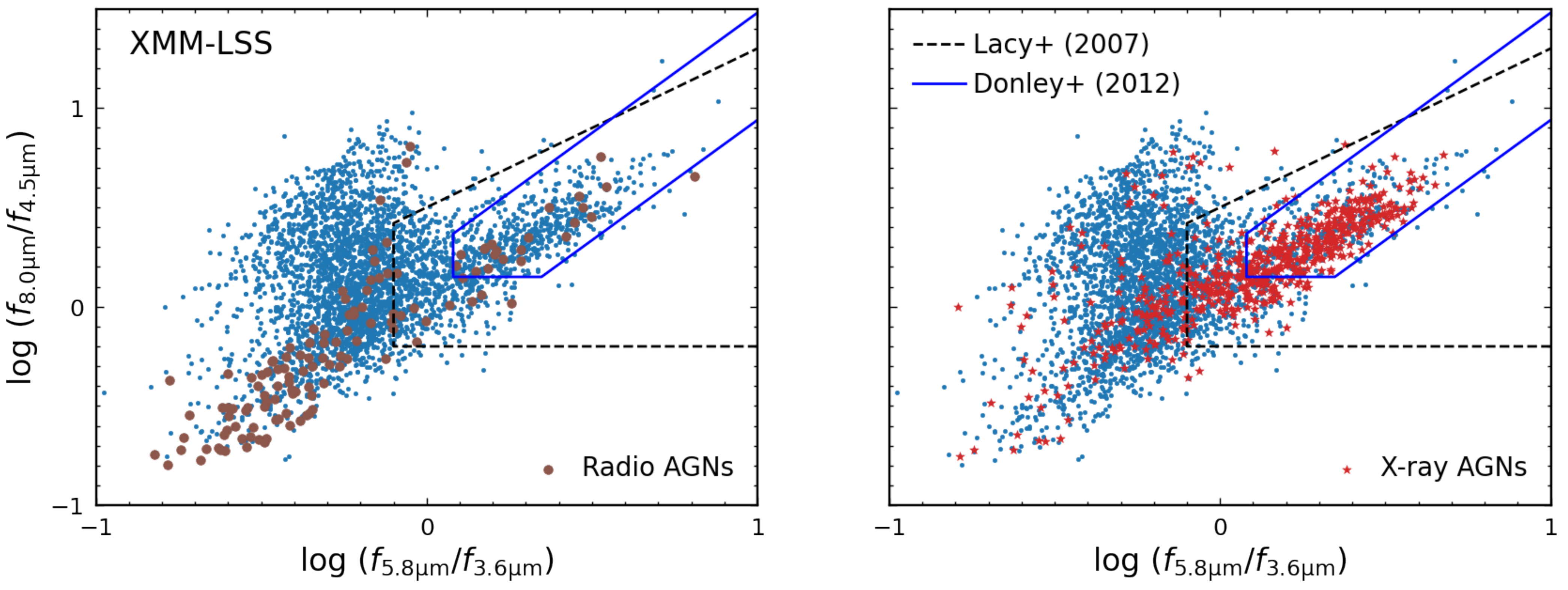}
\caption{Left: Radio sources in the color-color space defined by the four IRAC channels. 
    Radio AGNs are shown as large brown dots. 
    The MIR AGN selection criteria of \citet{lacy2007} and \citet{donley2012} are shown as black-dashed and blue-solid lines, respectively.
    Right: Same as the left panels, but X-ray AGNs are shown as red stars for comparison.
    Objects from the ATLAS fields (W-CDF-S and ELAIS-S1) are shown in the same plots.
    While X-ray AGNs are usually also MIR AGNs, most radio AGNs do not have MIR colors that are consistent with hot-dust emission.}
\label{fig:mir_wedge}
\end{figure*}

\begin{figure}
\centering
\includegraphics[width=0.45\textwidth, clip]{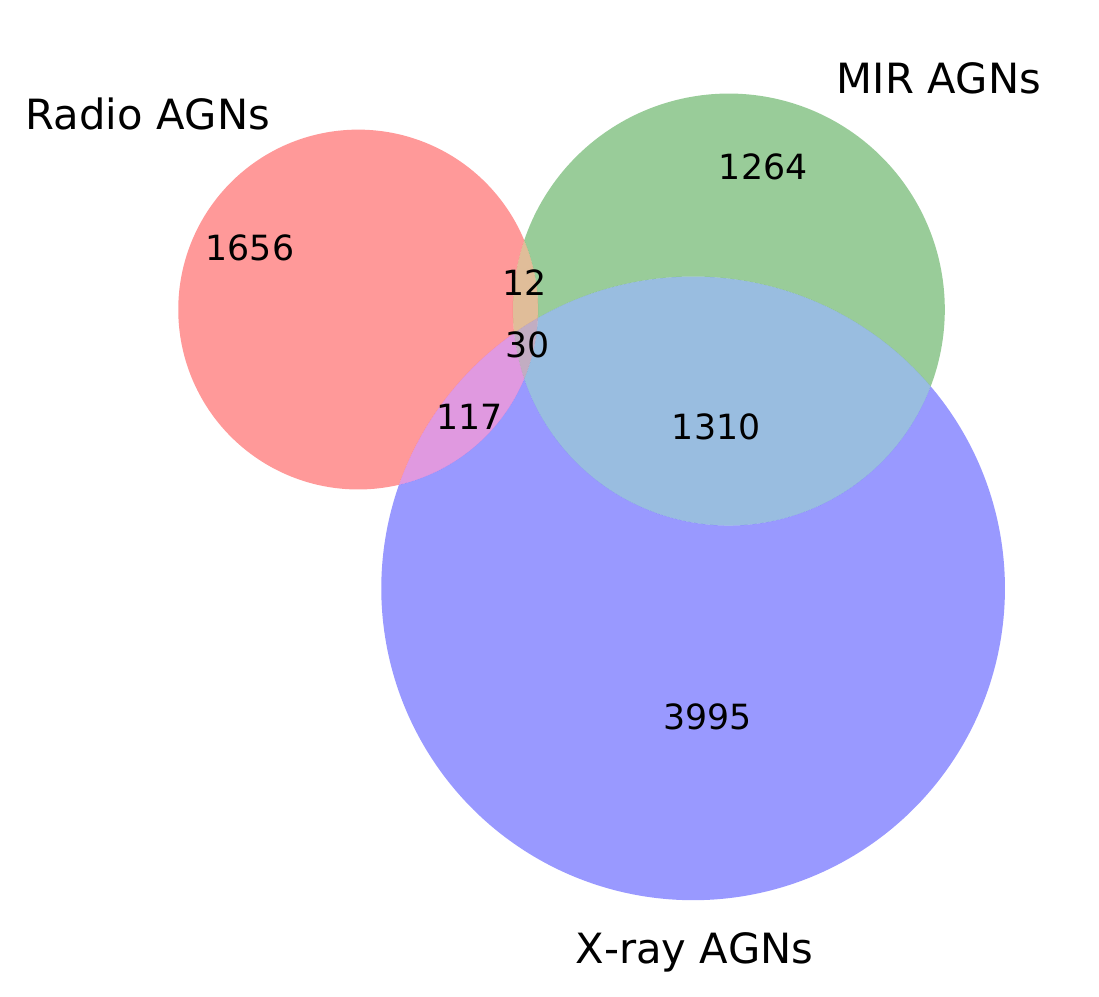}
\caption{Venn diagram for the radio, X-ray, and MIR (using the criteria of \citealt{donley2012}) selected AGNs in the 7.74 deg$^2$ region from which we select our main sample (see \S~\ref{sec:main_sample}).
Most radio-selected AGNs are not selected by X-ray or MIR methods.}
\label{fig:venn_full}
\end{figure}

\subsection{AGNs selected in other bands}
We compare radio AGNs with those selected in the X-ray and MIR in IRAC color-color space in Fig.~\ref{fig:mir_wedge}.
We list the number of X-ray and MIR AGNs among the radio sources in Table~\ref{tab:ragn}.
In Fig.~\ref{fig:mir_wedge}, radio AGNs do not overlap substantially with X-ray and MIR AGNs, which is consistent with the findings in \citet{whittam2022};
it is important to select AGNs using multiwavelength data.
Note that we have identified 1656 new radio AGNs in these fields not found by the X-ray or MIR selection approaches.
In Fig.~\ref{fig:venn_full}, we show a Venn diagram for AGNs selected using different methods in the 7.74 deg$^2$ region from which we selected our main sample.
The sky density of radio AGNs is $\approx249$ deg$^{-2}$, which is 0.74 of that of MIR AGNs and 0.35 that of X-ray AGNs.

\begin{figure*}
\centering
\includegraphics[width=0.45\textwidth, clip]{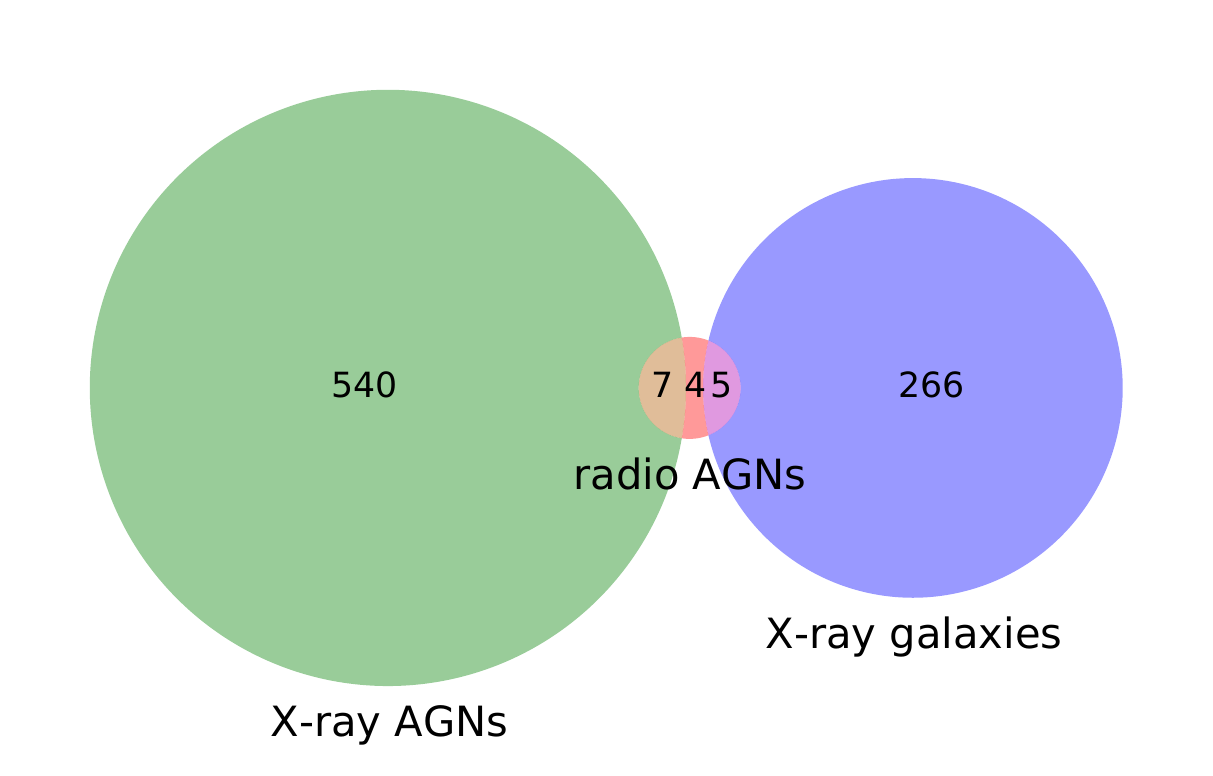}
\includegraphics[width=0.45\textwidth, clip]{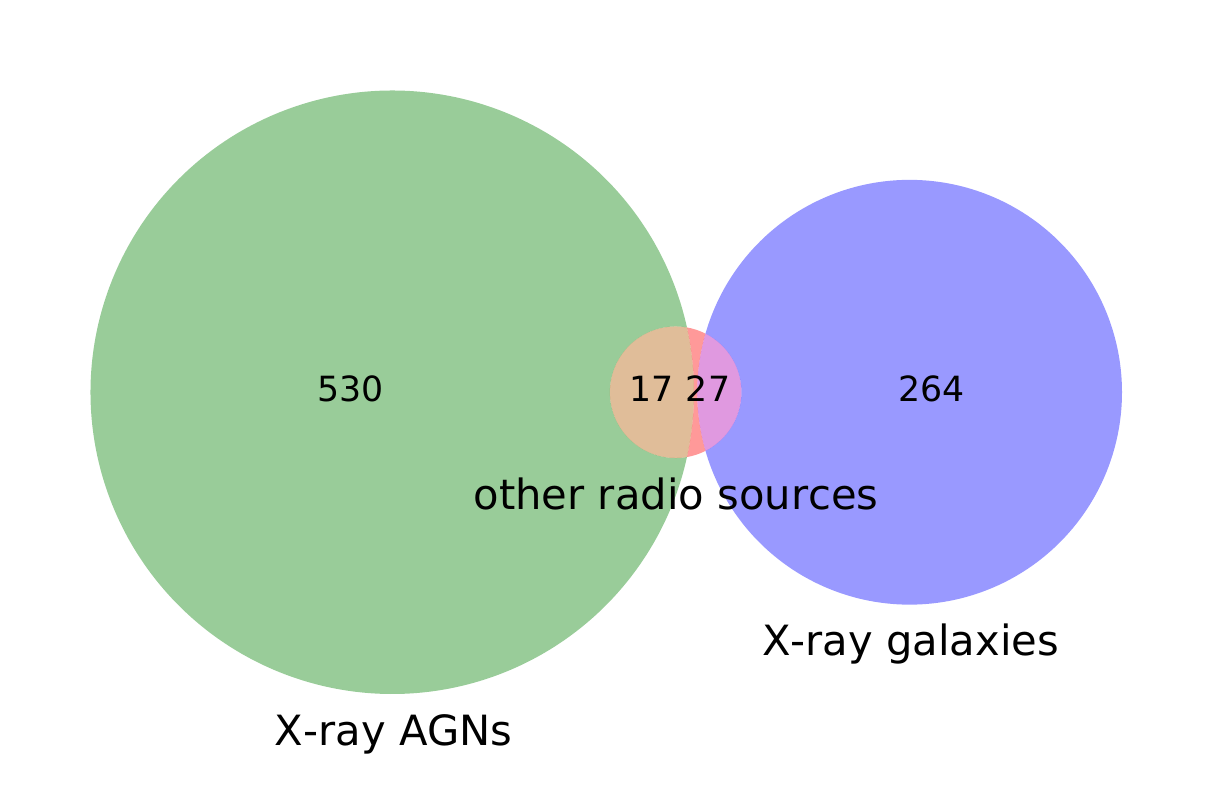}
\caption{Venn diagrams for the 42 ATLAS sources and 822 X-ray sources within the footprint of the deep {\it Chandra} coverage.
    There are 16 objects that are selected as radio AGNs (left), and the remaining 26 (right) might be radio-quiet AGNs and quiescent galaxies.}
\label{fig:venn}
\end{figure*}

We also compare our radio AGNs against the CDF-S \mbox{7-Ms} catalog (\citealt{luo2017}), 
which has achieved an unmatched sky density of reliable AGNs (23900 deg$^{-2}$).
The deep {\it Chandra} observations cover an area of about 484 arcmin$^2$, 
and we only focused on half of the area that has the top 50\% of exposure time.
There are 42 ATLAS radio sources within this sky region, 
36 of which are matched to the CDF-S 7-Ms catalog,
in which 24 and 12 objects are classified as AGNs and galaxies (see \S~4.5 of \citealt{luo2017}), respectively.
On the one hand, we selected 16/42 objects as radio AGNs, 
of which only 7 are classified as AGNs in the CDF-S 7-Ms catalog.
Inspecting the high-resolution VLA images (\citealt{miller2013}) of the remaining 9 radio AGNs, 
we found in one case the ATLAS radio source is affected by source confusion and is associated with four VLA components.
However, the brightest VLA source among the four still shows a radio excess and would be selected as an AGN.
Therefore, the number of radio AGNs not selected by the CDF-S 7-Ms catalog is not affected.
There are two radio AGNs that can be selected by their radio morphologies (note that they are also selected as RL AGNs using the $q_{24}$ criterion),
both of which are detected in the {\it Chandra} \mbox{7-Ms} observations but were classified as galaxies. 
These two radio AGNs are not MIR AGNs either.
Even the deepest X-ray survey missed a substantial fraction of radio-selected AGNs.
On the other hand, there are a large number of X-ray AGNs/galaxies that are not detected in the radio band.
We show in Fig.~\ref{fig:venn} the Venn diagrams that compare the AGN classification of the radio sources with those objects in the {\it Chandra} catalog.
Furthermore, we performed stacking analyses for
the 4 objects that are classified as radio AGNs but not detected in the CDF-S 7-Ms catalog,
and central point X-ray sources cannot be found in the stacked images for the non-detections.


\subsection{Stacked X-ray properties}
\label{sec:x_stack}

\begin{figure}
\centering
\includegraphics[width=0.45\textwidth, clip]{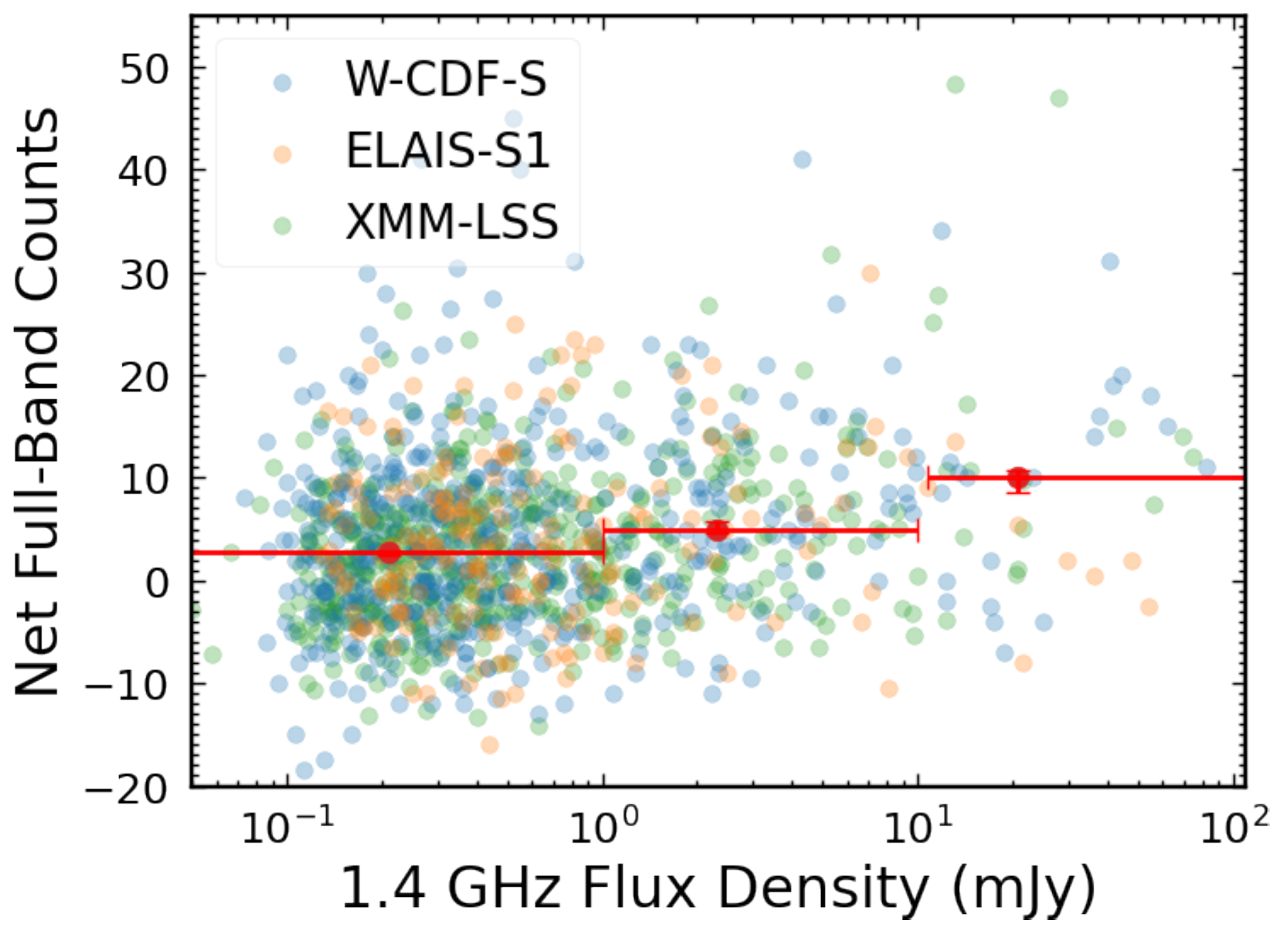}
\includegraphics[width=0.45\textwidth, clip]{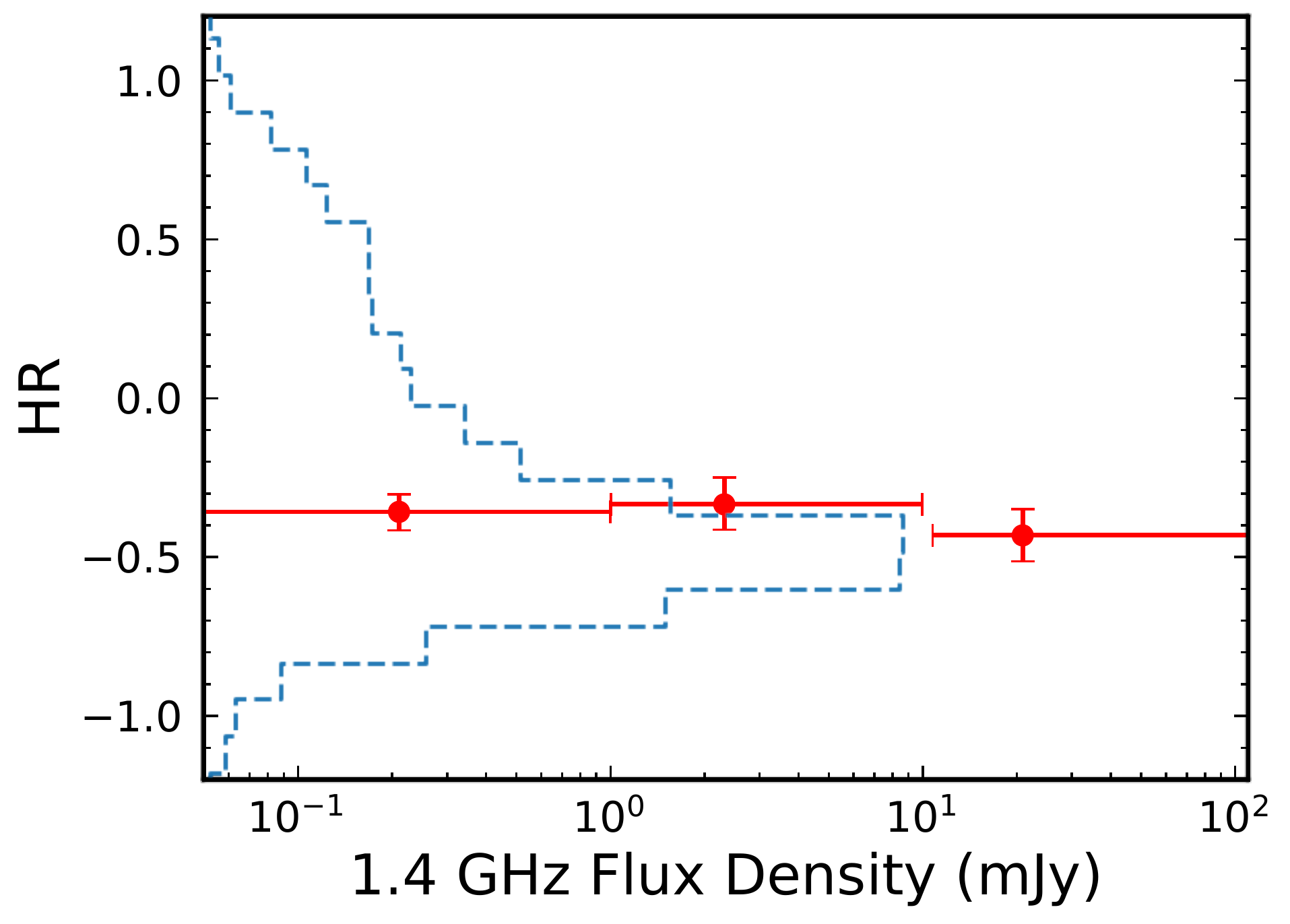}
\caption{The X-ray properties of the X-ray undetected RL AGNs resulting from stacking analyses.
    Top: The correlation between net full-band (0.5--10~keV) counts and the radio flux density of X-ray undetected RL AGNs.
The objects are grouped into three bins separated at 1 and 10 mJy. 
The red points with error bars represent the median values and bootstrapping-estimated $1\sigma$ interval.
Bottom: The red points with error bars are the average hardness ratio of the X-ray undetected RL AGNs in three $S_\mathrm{1.4GHz}$ bins.
The dashed curve shows the profile of the hardness-ratio distribution of the X-ray detected objects.
The broad-band X-ray spectra of the undetected objects are not substantially different from those of the detected objects.}
\label{fig:xmm_stack}
\end{figure}

For the (X-ray undetected) objects in the source-free 
regions (see \S~\ref{sec:photometries}), 
we performed stacking analyses to constrain their basic X-ray properties.
Note that the median full-band exposure times (combining the pn, MOS1, and MOS2 cameras) of these objects
are 89/87/100~ks in the W-CDF-S/ELAIS-S1/XMM-LSS fields.
Since the sensitivity improves with $\propto \sqrt{t_\mathrm{exp}}$ for {\it XMM-Newton} observations
with such exposures, we jointly analyze the X-ray data of the three fields since they have comparable depths.
Utilizing the science images in the soft (0.5--2~keV), hard (2--10~keV), and full (0.5--10~keV) bands,
we calculate the number of counts in the $3\times3$ pixels (i.e. $12\times12$ arcsec$^2$)
centered on the host positions. 
We estimated the background level using the median counts number for the
16 equal-size (i.e. $3\times3$ pixels) regions surrounding the $9\times9$ pixels centered on the host positions.
In the top panel of Fig~\ref{fig:xmm_stack}, 
we show the correlation between the full-band net counts of the X-ray undetected RL AGNs and their radio flux densities.
The X-ray and radio fluxes show a weak positive correlation.
The median net counts in the $S_\mathrm{1.4GHz}<1$~mJy (1624 objects), 1~mJy$<S_\mathrm{1.4GHz}<$10~mJy (272 objects), 
and $S_\mathrm{1.4GHz}>$10~mJy (47 objects) bins are 2.5, 5, and 10;
the X-ray flux increases by a factor of 4 typically
while the radio flux increases by a factor of $\approx100$.
The sub-linear correlation in Fig.~\ref{fig:xmm_stack} (top) is consistent with previous 
correlation analyses for RL-AGNs and radio-loud quasars (e.g. \citealt{brinkmann2000, miller2011, zhu2020}).
In comparison, the X-ray and radio emission are almost linearly correlated for RQ-AGNs and radio-quiet quasars (e.g. \citealt{brinkmann2000, laor2008}).
The 25th, 50th, and 75th percentiles of the corresponding net counts of the X-ray detected RL AGNs are 35.8, 65.8, and 141.7, respectively.
Therefore, the X-ray undetected objects are typically $>10$ times fainter than those of the detected objects.

We also calculate the hardness ratio to assess if the spectra of X-ray undetected RL AGNs are different from those of detected objects.
We define hardness ratio $\mathrm{HR}=\frac{H-S}{H+S}$, where $H$ and $S$ are the net counts in the hard and soft bands, respectively.
We stacked the hard-band and soft-band net counts in the three radio-flux bins and calculated corresponding hardness ratios.
The results are shown in the bottom panel of Fig.~\ref{fig:xmm_stack}, where the errors are estimated using bootstrapping.
For comparison, we also show the HR distribution of the X-ray detected RL AGNs.
The average HRs in the three flux bins are not significantly different from those of the detected objects, 
suggesting that their X-ray faintness is not mainly due to additional obscuration. 

\begin{table}
\centering
\caption{Comparison of radio AGNs with X-ray and MIR AGNs}
\label{tab:ragn}
\begin{threeparttable}[b]
\begin{tabularx}{\linewidth}{@{}Y@{}}
\begin{tabular}{lccc}
\hline
\hline
    Radio AGNs &  W-CDF-S& ELAIS-S1 & XMM-LSS \\
    \hline 
    Total & 713 & 275 & 827\\
    X-ray Detections & 66 & 28 & 62 \\
    X-ray AGNs & 65 & 20 & 62 \\
    Donley MIR AGNs & 19 & 4 & 19  \\
    Radio-Only\tnote{a} & 641 & 254 &761 \\
\hline 
\end{tabular}
\end{tabularx}
\begin{tablenotes}
\item[a] Radio-only refers to radio AGNs that are neither X-ray AGNs nor \citet{donley2012} MIR AGNs.
\end{tablenotes}
\end{threeparttable}
\end{table}

\begin{figure*}
\centering
\includegraphics[width=0.34\textwidth, clip]{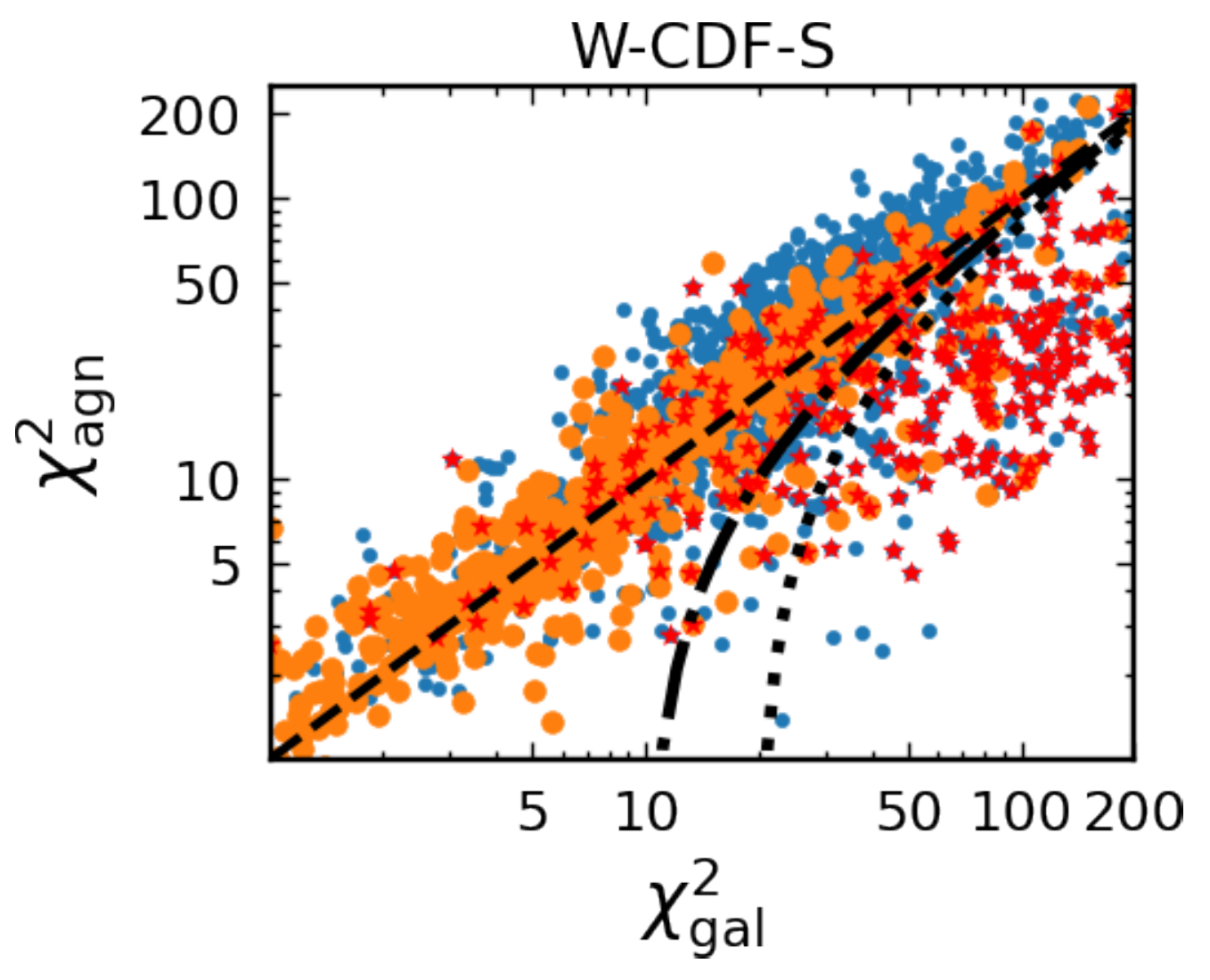}
\includegraphics[width=0.31\textwidth, clip]{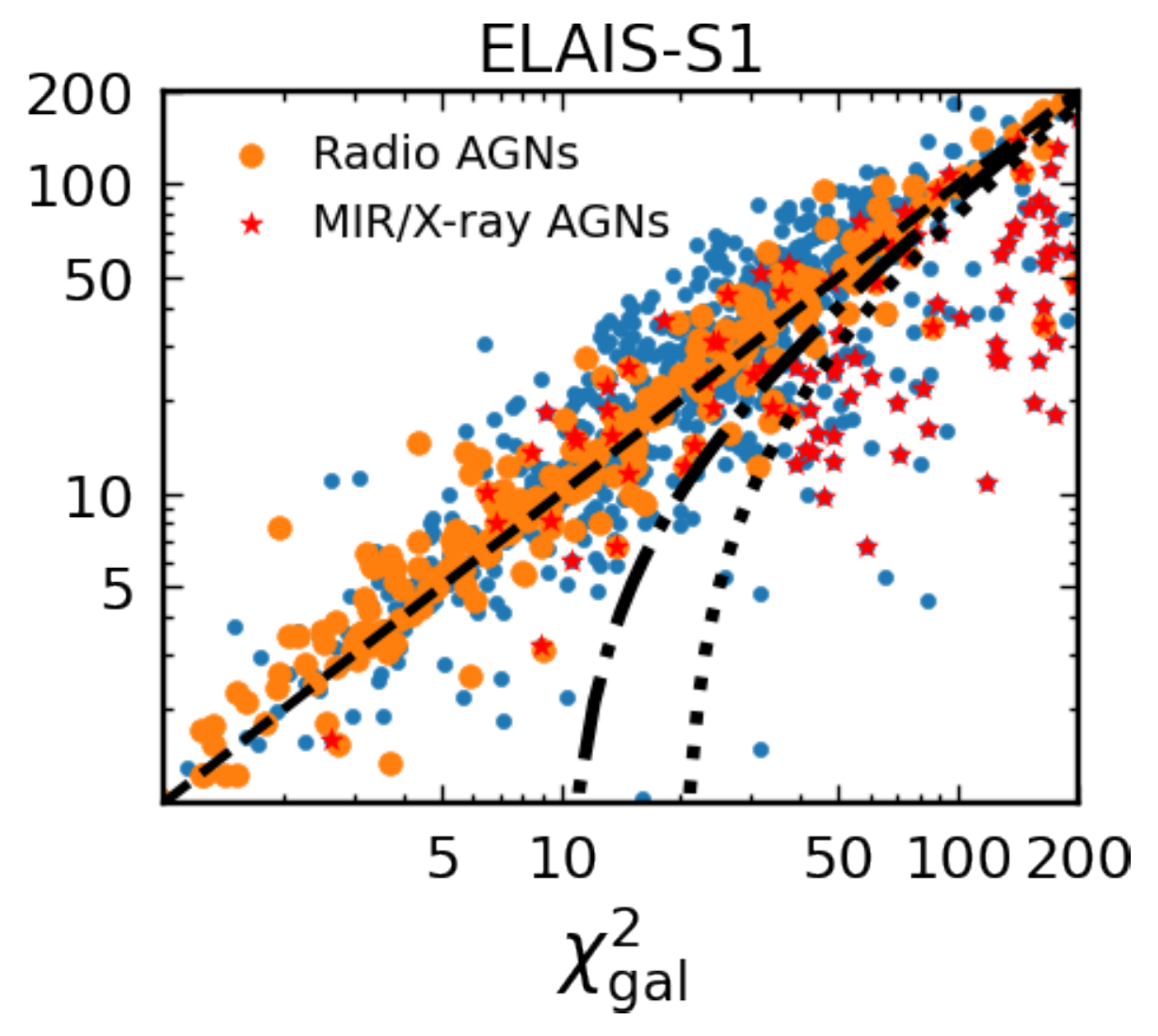}
\includegraphics[width=0.30\textwidth, clip]{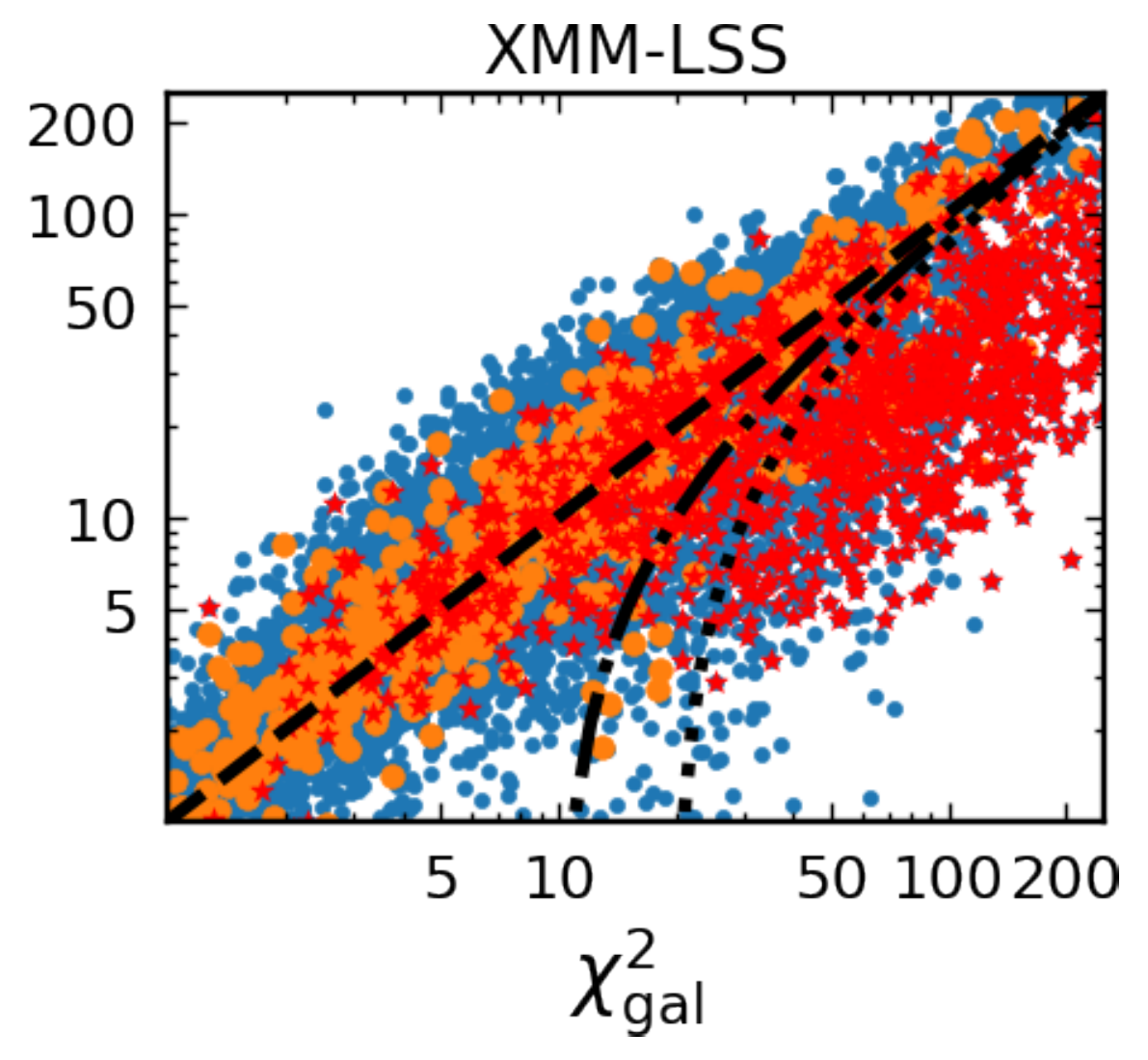}
    \caption{The $\chi^2$ of the UV-to-FIR SED fitting using galaxy-only ($x$-axis) and galaxy$+$AGN ($y$-axis) {\sc cigale} models for radio sources in the W-CDF-S/ELAIS-S1/XMM-LSS (left/middle/right) fields.
    Orange dots and red stars represent radio and MIR/X-ray AGNs, respectively, and the remaining objects are represented by blue dots.
    The dashed, dash-dotted, and dotted curves represent $\chi^2_\mathrm{agn}=\chi^2_\mathrm{gal}$, $\chi^2_\mathrm{agn}=\chi^2_\mathrm{gal}-10$, and 
    $\chi^2_\mathrm{agn}=\chi^2_\mathrm{gal}-20$, respectively.
    For radio AGNs, the galaxy-only models fit better their UV-to-FIR SEDs. 
    In contrast, models that have an AGN component are strongly preferred for most MIR/X-ray AGNs with $\chi^2_\mathrm{agn}<\chi^2_\mathrm{gal}-10$.}
\label{fig:cigale_chi_squared}
\end{figure*}

\begin{figure*}
\centering
\includegraphics[width=0.45\textwidth, clip]{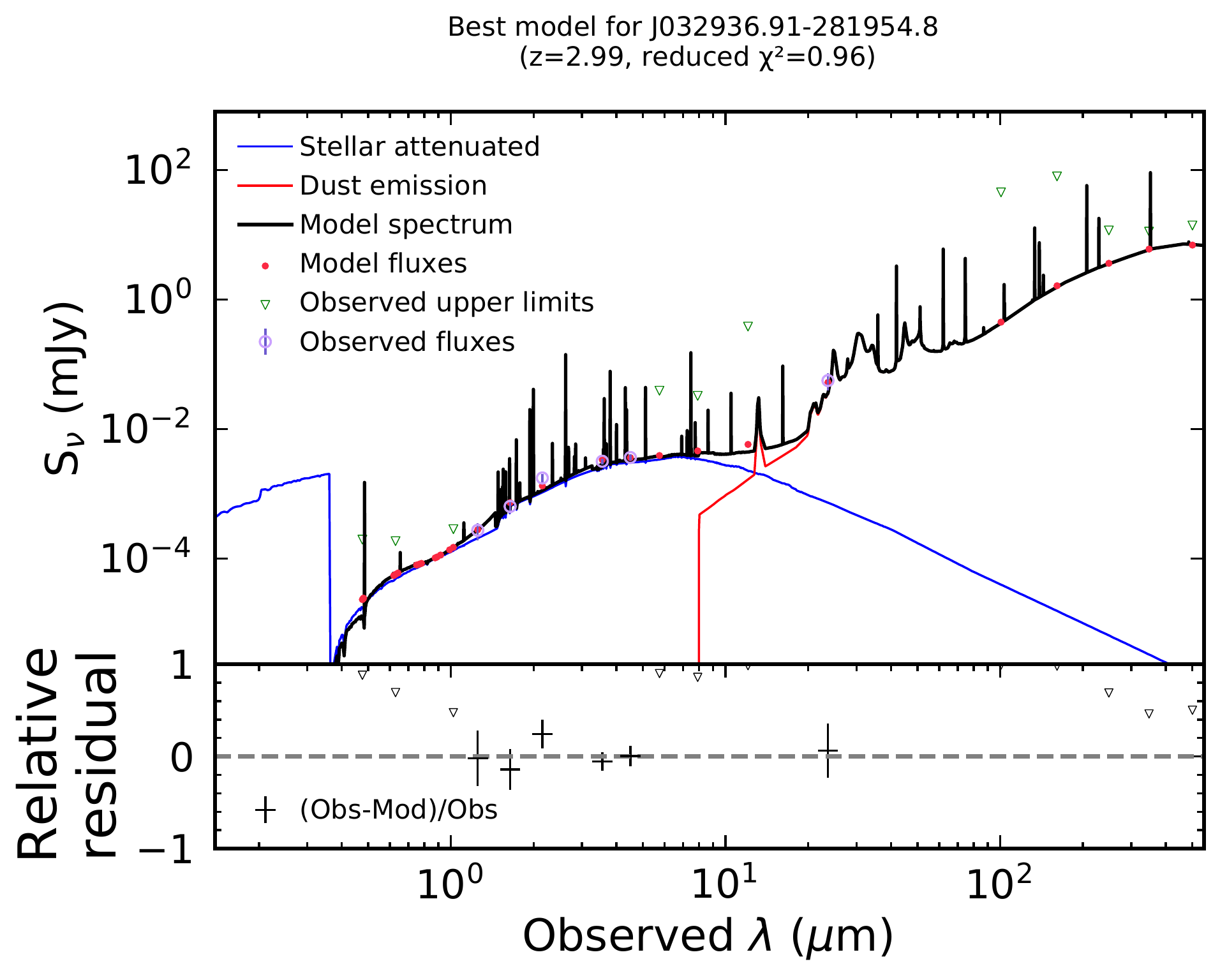}
\includegraphics[width=0.45\textwidth, clip]{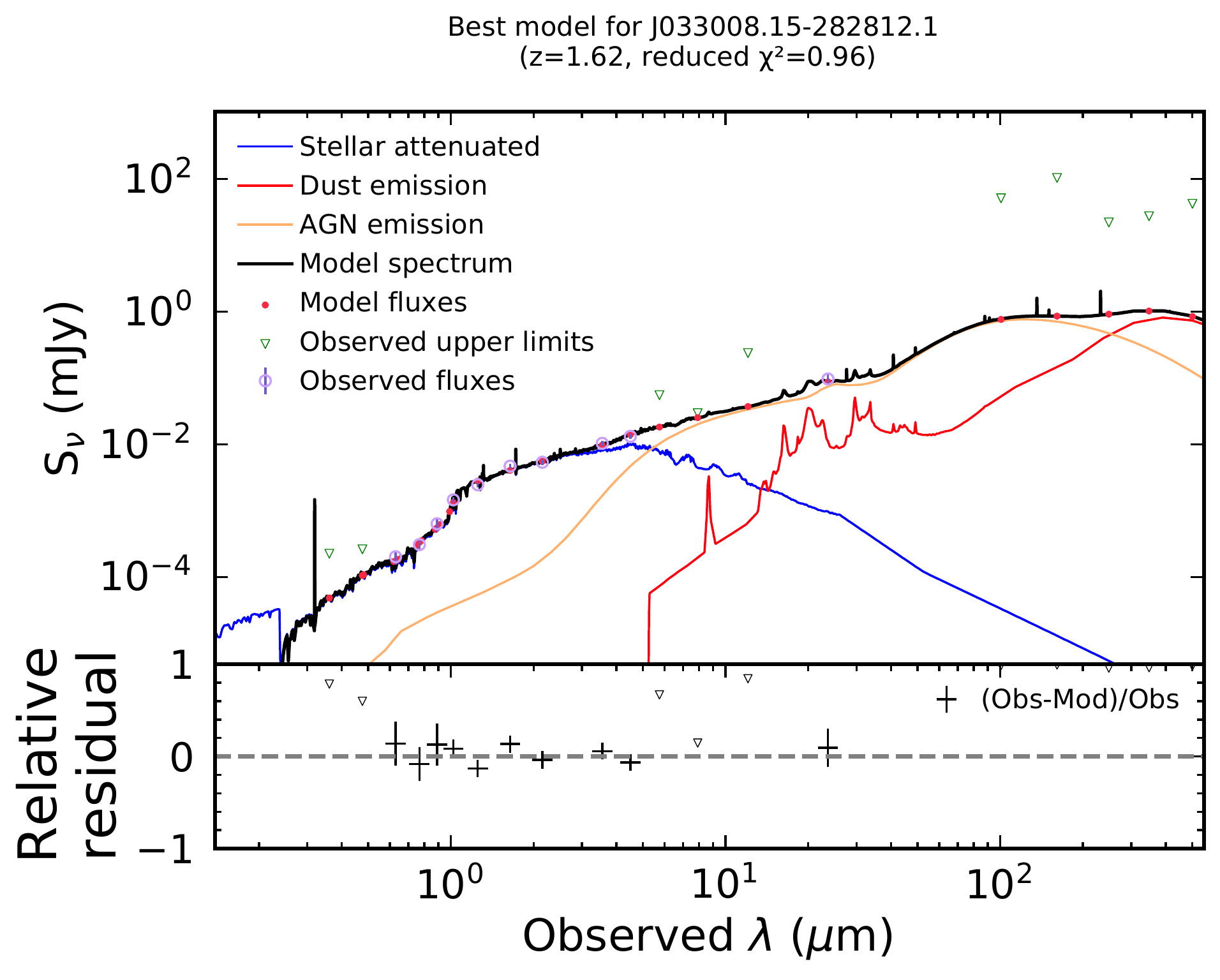} \\
\includegraphics[width=0.45\textwidth, clip]{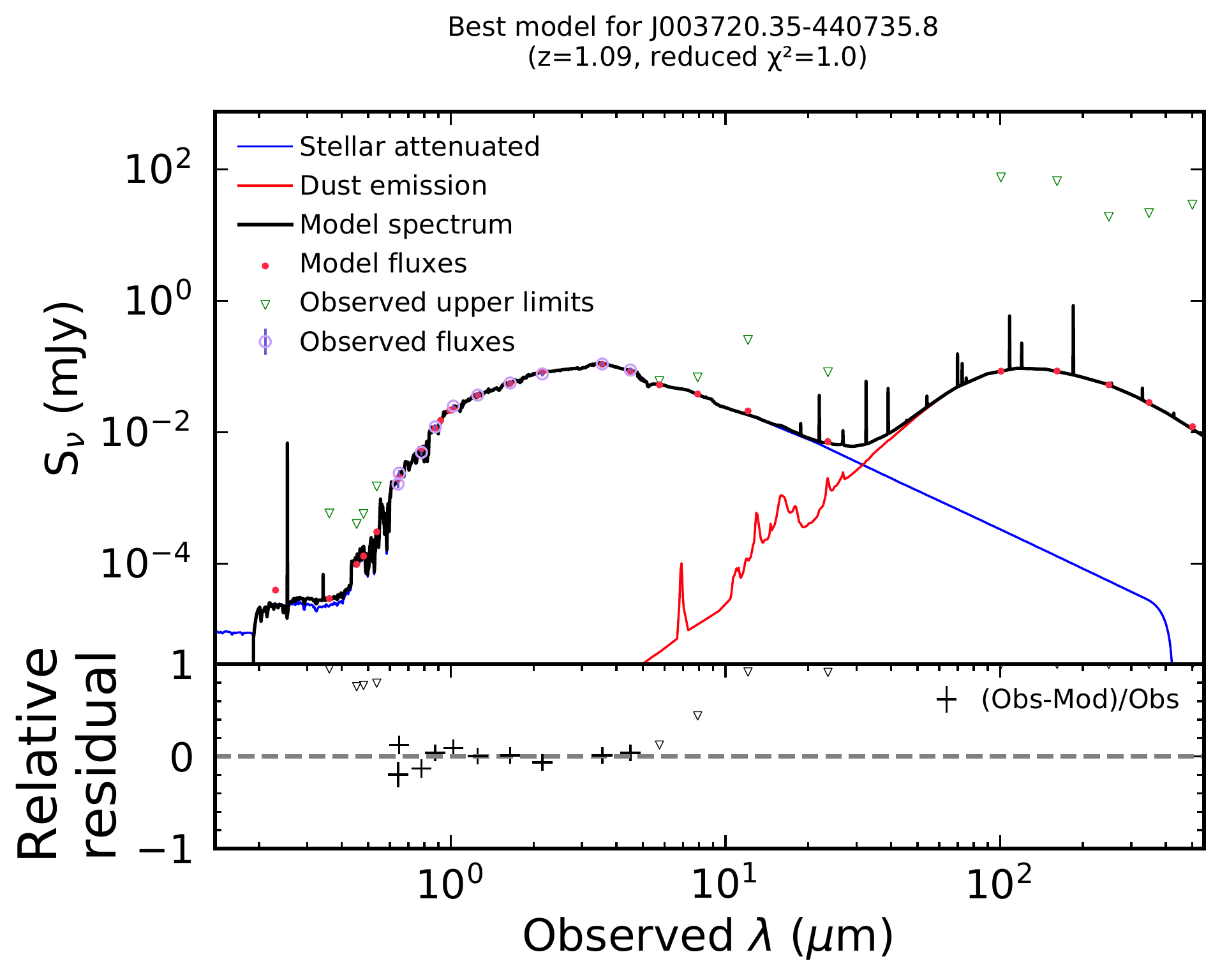}
\includegraphics[width=0.45\textwidth, clip]{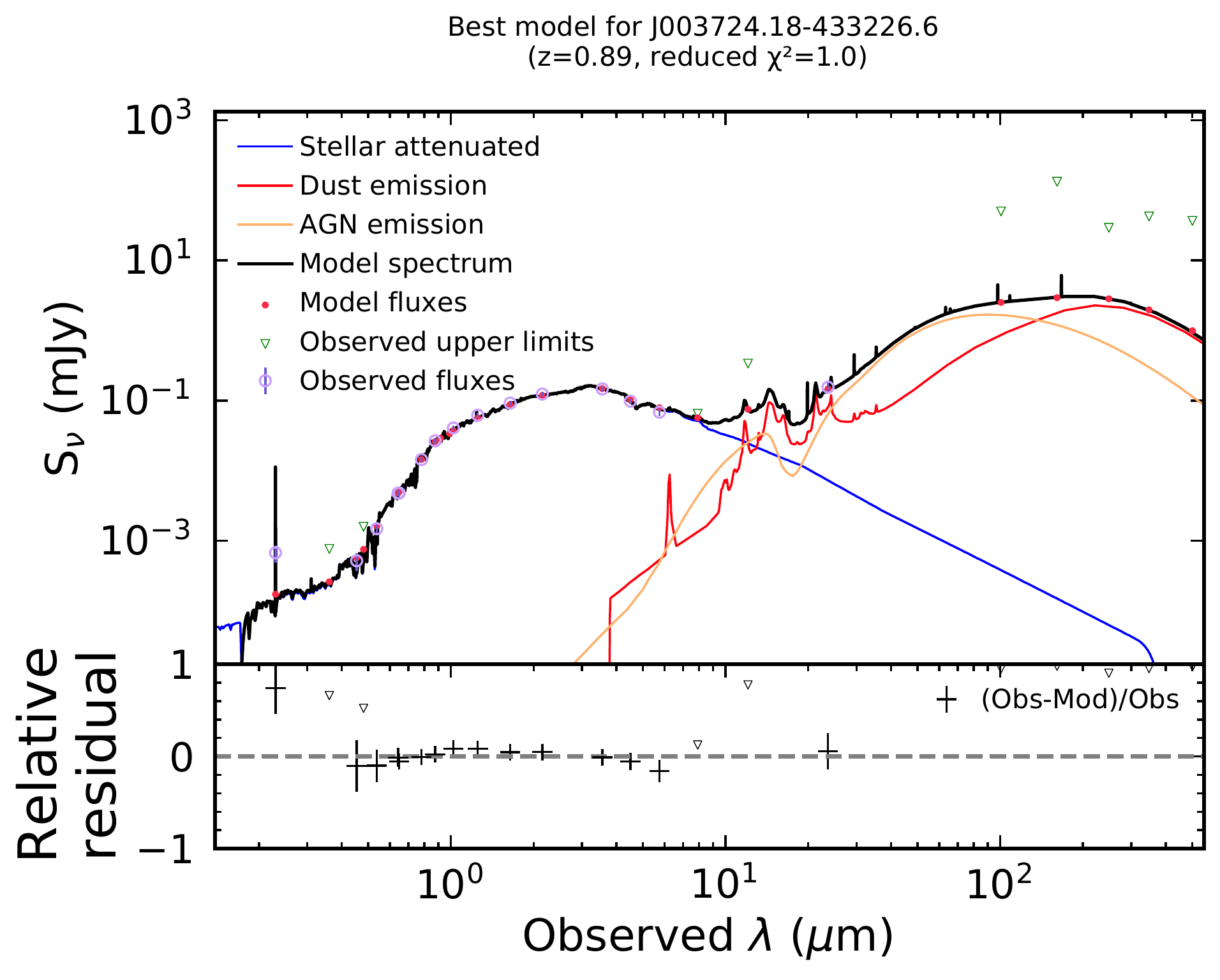} \\
\includegraphics[width=0.45\textwidth, clip]{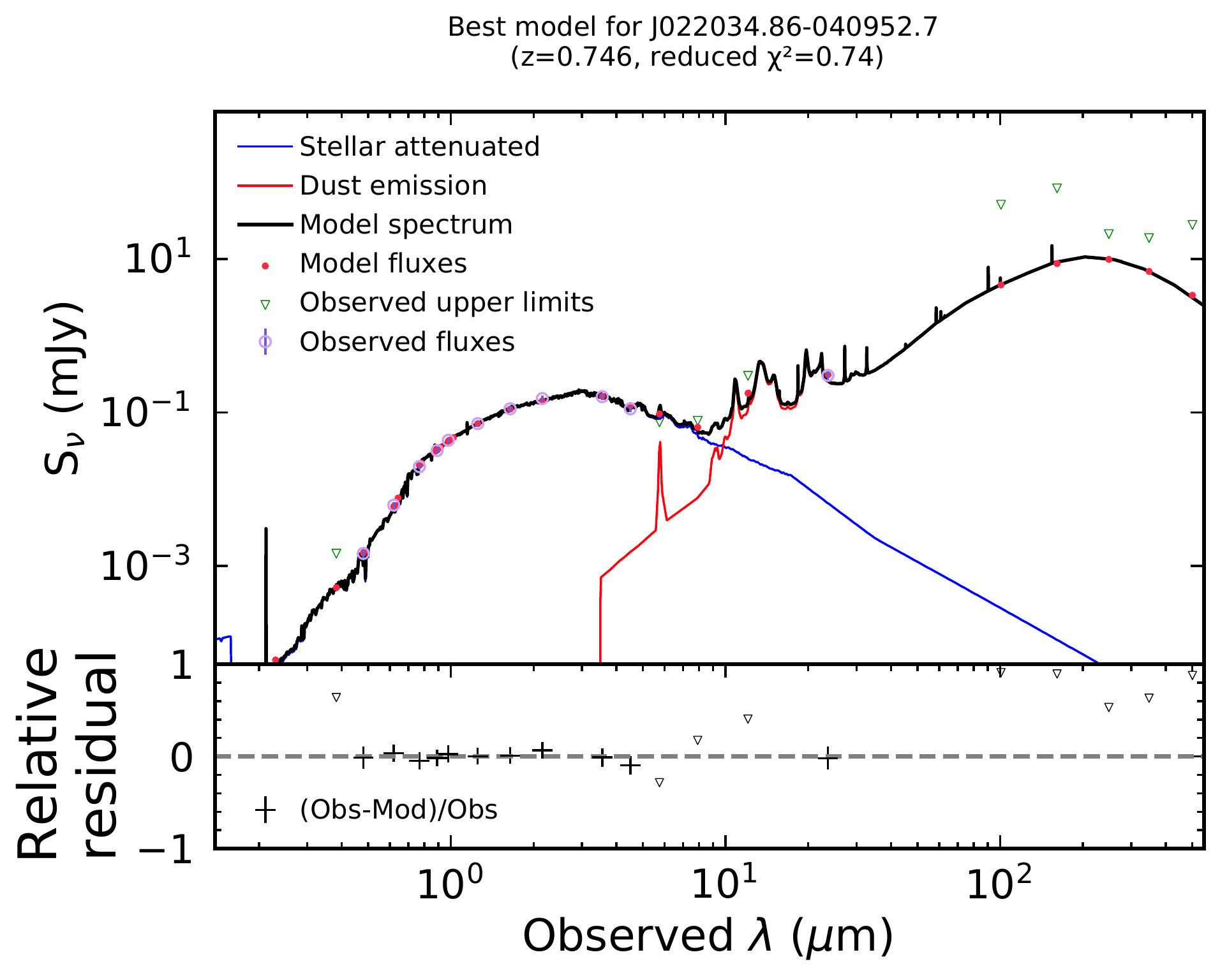}
\includegraphics[width=0.45\textwidth, clip]{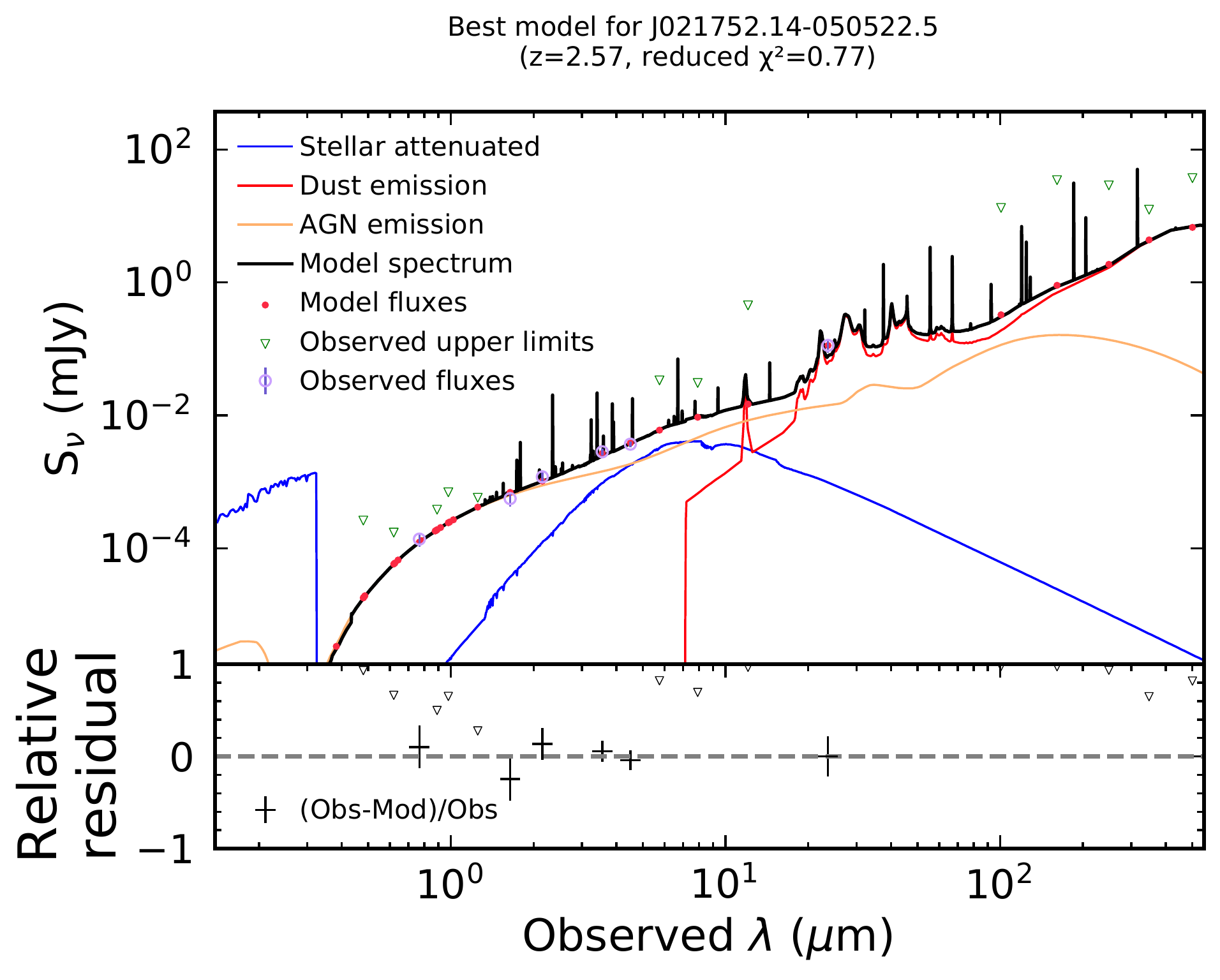}
    \caption{Examples of the SED fits for radio AGNs in the W-CDF-S/ELAIS-S1/XMM-LSS (top/middle/bottom) fields.
    The left and right columns are for fits that prefer
    galaxy-only ($\chi^2_\mathrm{gal}<\chi^2_\mathrm{agn}$)
    and galaxy+AGN ($\chi^2_\mathrm{gal}>\chi^2_\mathrm{agn}$) models, respectively.
    The stellar, dust, and AGN (only for the right panels) components are shown as blue, red, and orange curves, respectively.
    We omitted plotting the nebular emission which contributes little to the broad-band fluxes in these plots, though this is included in the fitting.
    These objects are selected at the median reduced $\chi^2$ in each field for the preferred model.
    J021752.14$-$050522.5 (bottom right) is also selected as an X-ray AGN.
    }
\label{fig:cigale_examples}
\end{figure*}

\begin{figure*}
\centering
\includegraphics[width=0.94\textwidth, clip]{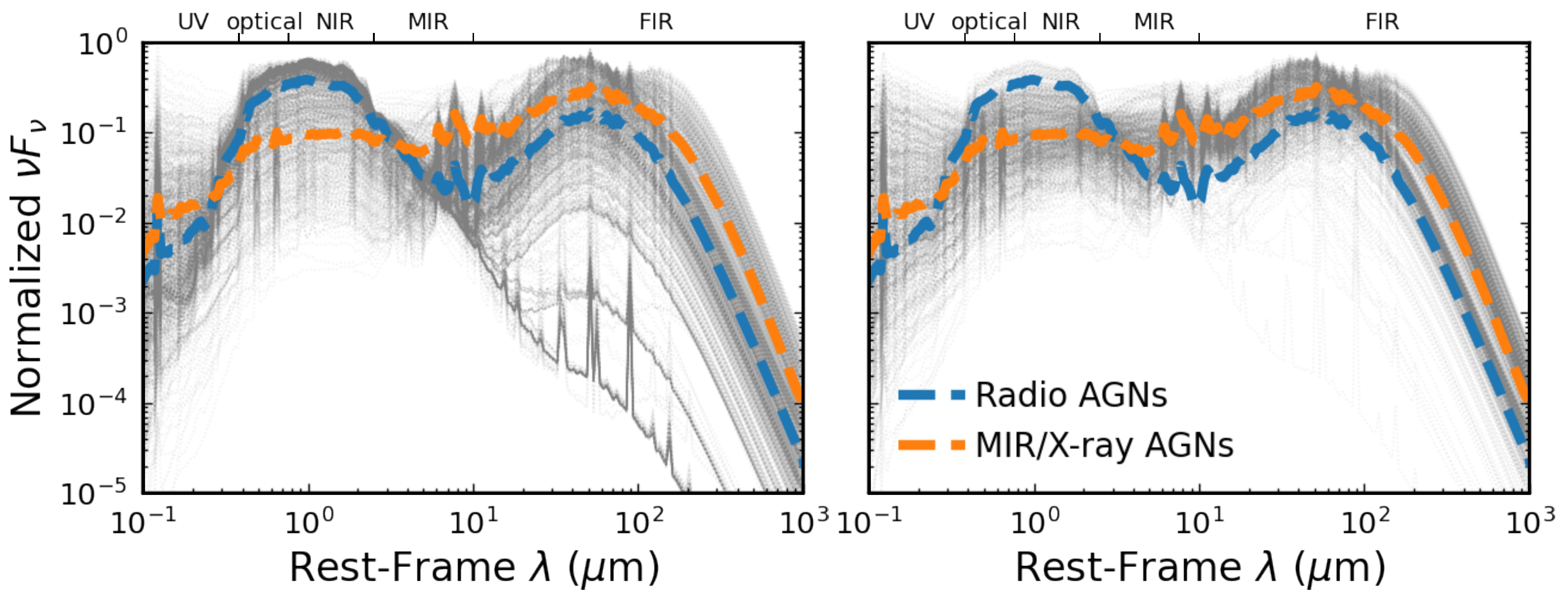}
    \caption{The composite median SEDs for MIR/X-ray AGNs (orange)
    and new AGNs that are only selected using radio methods (blue) are plotted in both panels.
Individual SEDs of radio and MIR/X-ray AGNs are shown in grey in the left and right panels, respectively.
    Only objects with $\chi^2_\nu<2$ are included.
    The host galaxies of radio AGNs are redder and bluer than MIR/X-ray AGNs in the UV and MIR bands, respectively.
    In the FIR band, even though the two composite median SEDs have similar shapes, 
    the host galaxies of a substantial fraction of the radio AGNs contain little dust (with weak FIR emission), 
    which is in constrast to the MIR/X-ray AGNs.}
\label{fig:stack_seds_agns}
\end{figure*}

\begin{figure*}
\centering
\includegraphics[width=0.33\textwidth, clip]{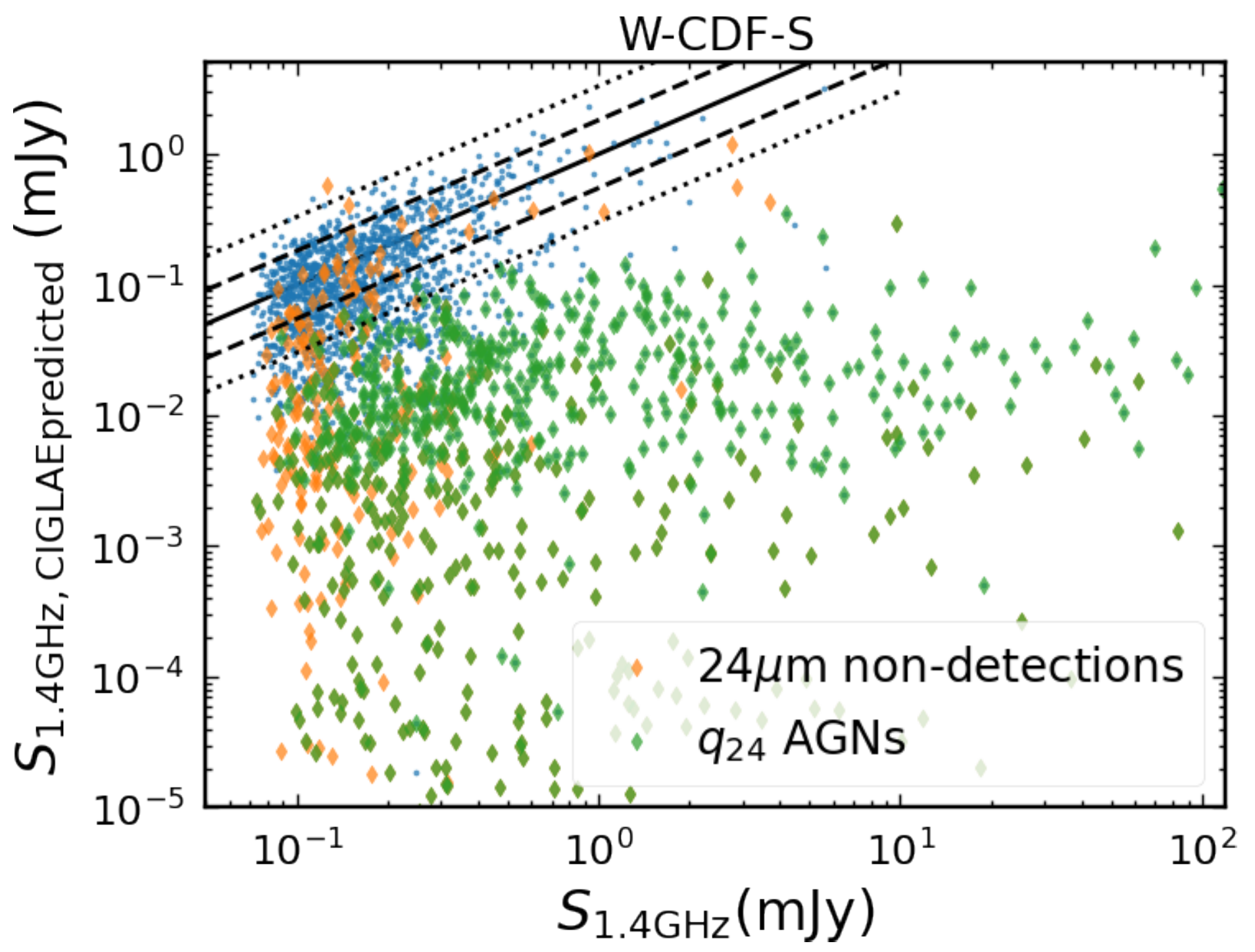}
\includegraphics[width=0.31\textwidth, clip]{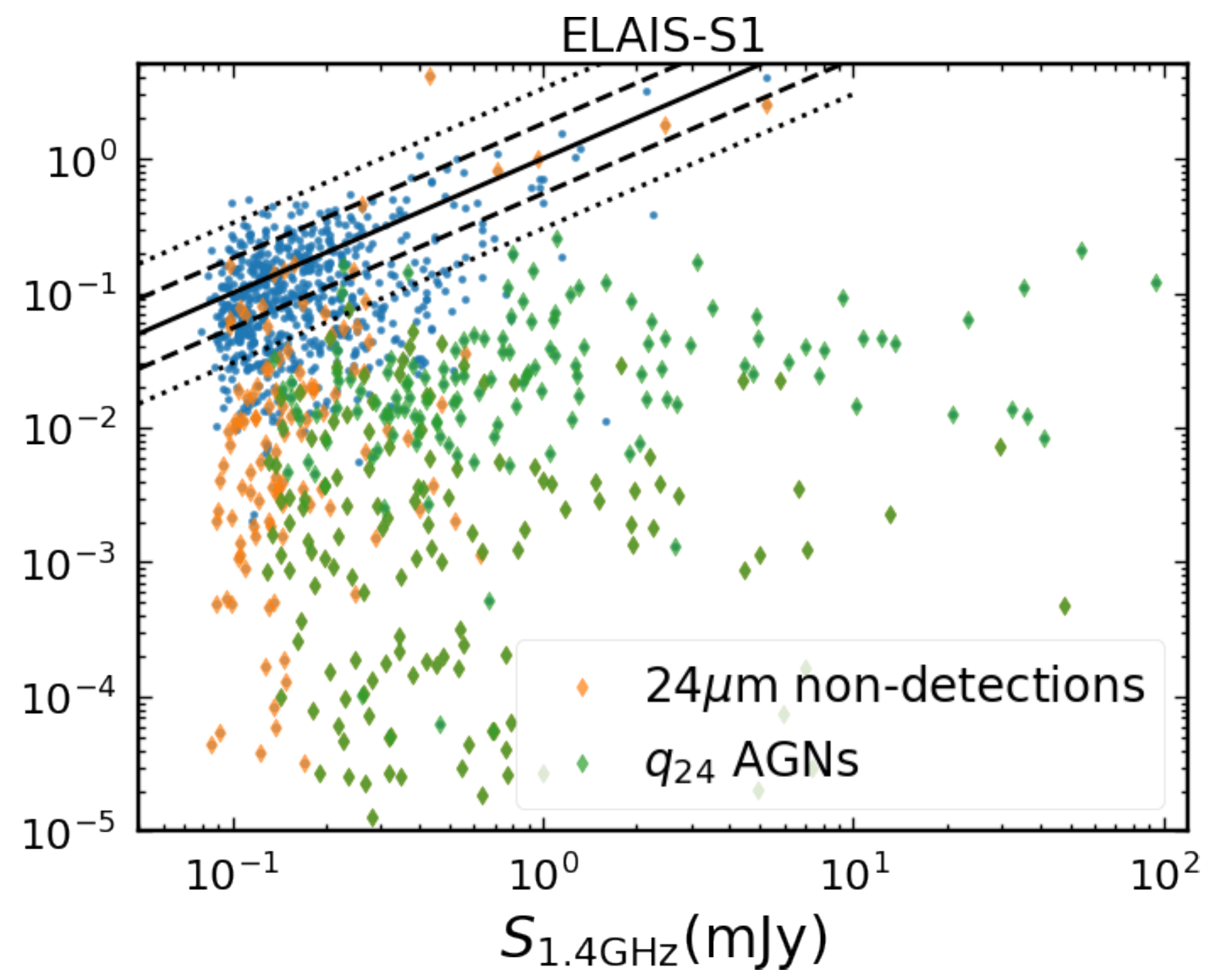}
\includegraphics[width=0.31\textwidth, clip]{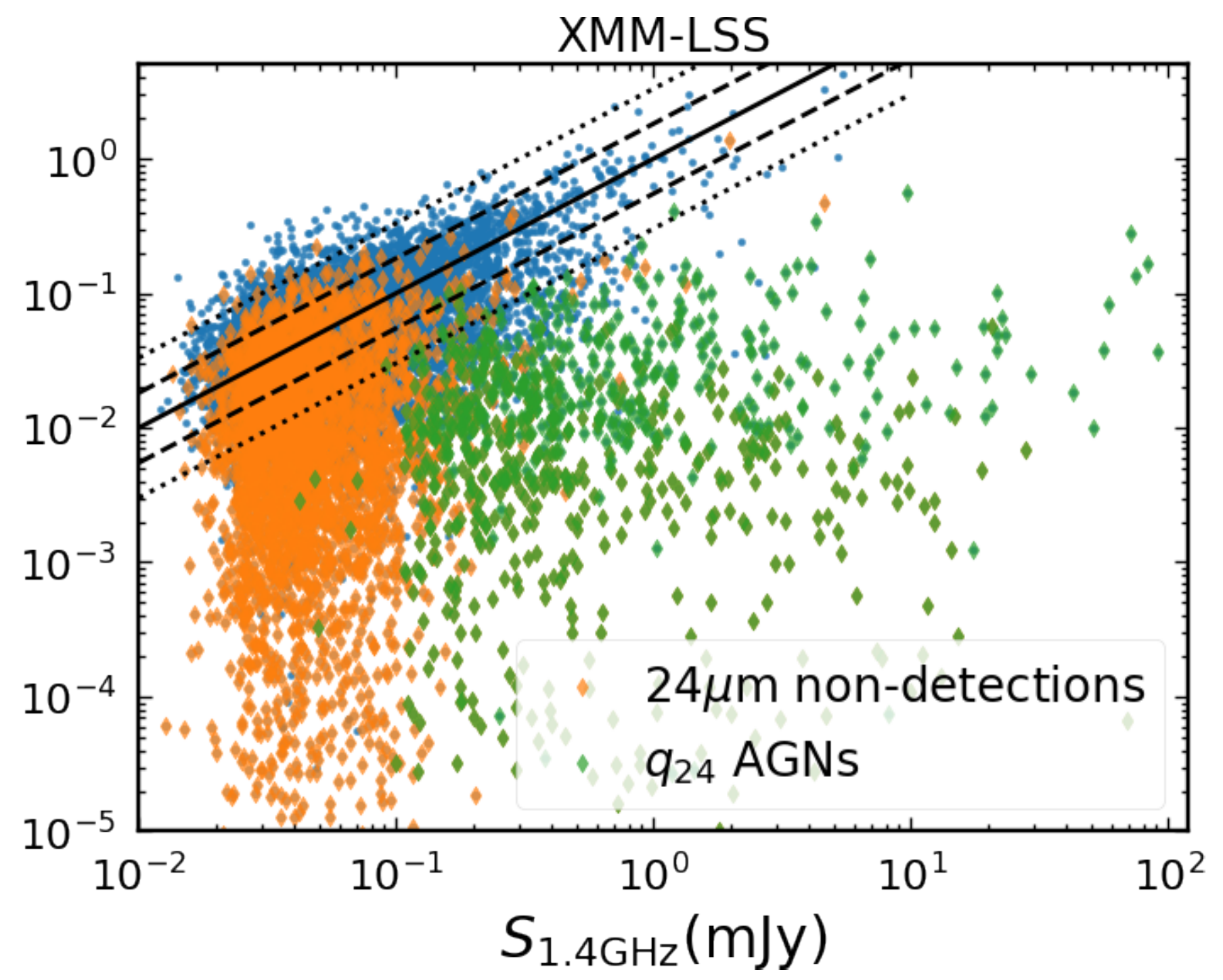}
\caption{The radio flux produced by star formation ($y$-axis) compared with the observed total radio flux ($x$-axis).
    The star-formation produced radio flux is estimated using the SED fitting results and the $q_\mathrm{FIR}$ parameter.
    Green diamonds represent the radio-excess AGNs selected using the $q_\mathrm{24}$ parameter (see \S~\ref{sec:radio_agns}).
    Orange diamonds represent objects that are not detected at MIPS 24$\mu$m, and, therefore, $q_\mathrm{24}$ parameters are upper limits.
    The remaining objects are blue dots. The $1:1$ line is shown as the black solid line in each panel.
    The dashed and dotted lines indicate the 1$\sigma$ and $2\sigma$ scatters of the $q_\mathrm{FIR}$ parameter (\citealt{yun2001}).
    The $q_\mathrm{24}$-selected radio-excess AGNs are generally $2\sigma$ away from the solid black line, 
    supporting the robustness of the $q_\mathrm{24}$-selection method.
    However, the $q_\mathrm{24}$-selection method would miss radio-excess AGNs below 0.1~mJy due to the flux limit of the 24~$\mu$m data.
    }
\label{fig:cigale_radio_excess}
\end{figure*}

\begin{table*}
\centering
\caption{The parameters of {\sc cigale} for the galaxy-only and galaxy+AGN models. 
    Unmentioned parameters are identical to those in Table~4 and Table~5 of \citet{zou2022}.}
\label{tab:cigal_params}
\begin{threeparttable}[b]
\begin{tabularx}{\linewidth}{@{}Y@{}}
\begin{tabular}{lcc}
\hline
\hline
    Module & Paramter & Value \\
\hline
    {\it galaxy-only model} \\
\hline
    \texttt{dale2014} & \texttt{alpha} & 1.0, 1.25, 1.5, 1.75, 2.0, 2.25, 2.5, 2.75, 3.0 \\
    \texttt{dustatt\_modified\_starburst} & \texttt{E\_BV\_lines} & 0.0001, 0.001, 0.01, 0.02, 0.04, 0.06, 0.08, 0.1, 0.15, 0.2, 0.3, 0.4, 0.5, 0.6, 0.8, 1, 1.2 \\
    \texttt{radio} & \texttt{qir\_sf} & 2.34 \\
            & \texttt{R\_agn} & 0.0 \\
\hline
    {\it galaxy+AGN model} \\
\hline
    \texttt{dale2014} & \texttt{alpha} & 1.0, 2.0, 3.0 \\
    \texttt{dustatt\_modified\_starburst} & \texttt{E\_BV\_lines} & 0.001, 0.01, 0.1, 0.2, 0.3, 0.4, 0.5, 0.6, 0.8, 1, 1.2 \\
    \texttt{skirtor2016} & \texttt{t} & 3, 7, 11 \\
    & \texttt{i} & 0, 10, 30, 50, 70, 90 \\
    & \texttt{fracAGN} & 0.1, 0.2, 0.4, 0.6, 0.8, 0.99 \\
\hline
\end{tabular}
\end{tabularx}
\end{threeparttable}
\end{table*}

\section{Discussion}
\label{sec:discussion}
\subsection{The SEDs of faint radio sources}
\label{sec:cigale}
We constrain the SEDs of the radio sources utilizing {\sc cigale} (v2022.1, \citealt{boquien2019, yang2020, yang2022}), 
from which we estimate the stellar masses, star-formation rates, and other physical parameters of the host galaxies.
\citet{zou2022} fit the SEDs of millions of objects in the three DDFs we have studied in this paper.
Our fitting methods generally follow those of \citet{zou2022} but with some modifications.
First, the photometric data are summarized in Table~1 of \citet{zou2022}.
We use a similar data set with the addition of the WISE-3 band ($12\mu$m; see \S~\ref{sec:photometries}), which fills the gap between the IRAC $8\mu$m and MIPS $24\mu$m bands.
Furthermore, we exclude the X-ray data from the fitting.
Only a small fraction of the radio sources have detected X-ray counterparts;
excluding the X-ray band reduces the required computation by a large factor.
Secondly, each object might have multiple optical fluxes in similar filters,
and \citet{zou2022} use all available photometric data in the fitting.
We select only the ``best'' fluxes in the order of HSC, VOICE, VIDEO, CFHTLS, and DES.
If a higher order flux is available for one of the $grizY$ bands, the remaining fluxes in the same band are ignored.
Thirdly, we utilize the \texttt{radio} module of {\sc cigale} to predict the radio emission that is produced by star formation.
In the \texttt{radio} module, \texttt{qir\_sf} represents the $q_\mathrm{FIR}$ parameter (\citealt{helou1985}),
\begin{equation}
    q_\mathrm{FIR} = \log \Big[\frac{F_\mathrm{FIR}/(3.75\times10^{12} \mathrm{Hz})}{S_\mathrm{1.4GHz}}\Big],
\end{equation}
where $F_\mathrm{FIR}$ is the rest-frame FIR flux that is estimated from the {\it IRAS} 60 and 100 $\mu$m fluxes, 
and $S_\mathrm{1.4GHz}$ the flux at rest-frame 1.4 GHz.
$q_\mathrm{FIR}$ is similar to the $q_\mathrm{24}$ parameter but uses longer-wavelength FIR fluxes that trace the star-forming processes better.
Furthermore, $q_\mathrm{FIR}$ is defined for rest-frame quantities, and the $K$ correction is avoided using the best-fit {\sc cigale} models.
The \texttt{radio} module has a parameter \texttt{R\_agn} 
that represents the radio-loudness parameter of AGNs, $R=L_\mathrm{5GHz}/L_\mathrm{2500\angstrom}$ (\citealt{yang2022}),
where $L_\mathrm{5GHz}$ and $L_\mathrm{2500\angstrom}$ are the monochromatic luminosities at rest-frame 5 GHz and 2500~\angstrom, respectively.
We fix \texttt{qir\_sf} $=2.34$ (e.g. \citealt{yun2001}) and \texttt{R\_agn} $=0$ 
in the parameter settings.\footnote{\label{ft:qir}The values of $q_\mathrm{FIR}$ from previous studies are 
generally in the range of 2--2.41 (cf. \citealt{sargent2010}). 
\citet{delvecchio2021} report a value of about 2.6 and find that $q_\mathrm{FIR}$ might
depend strongly on stellar mass and mildly on redshift, probably because the radio data (VLA 3GHz observations in the COSMOS field)
they use have a smaller beam size and are at a higher frequency than previous studies.}
Note that the radio band is not included in the fitting; the \texttt{radio} module simply calculates the radio fluxes at 1.4 GHz that are
expected to be produced by star formation, which are not compared with the observed fluxes in the fitting process.

Following \citet{zou2022}, the SED of each object is fit twice, 
once with a galaxy-only model and once with a galaxy+AGN model,
the parameter settings of which are similar to those in Table~4 and Table~5 of \citet{zou2022}, respectively.
In Table~\ref{tab:cigal_params}, we give the few parameters that are different from those in \citet{zou2022}.
It is a common practice to add systematic uncertainties to the flux-density 
errors in {\sc cigale}. However, we find this practice unnecessary (cf. Fig.~\ref{fig:cigale_examples}) for our 
datasets in part because an additional 3\%--9\% flux uncertainty has been added 
on top of the statistical uncertainties (in quadrature) in the {\it Tractor} photometry catalogues (K. Nyland 2023, private communication).
We turn this feature off by setting \texttt{additionalerror=0} (the default value is 10\%).
In Fig.~\ref{fig:cigale_chi_squared}, we compare the $\chi^2$ statistic of the best-fit galaxy-only model with that of the best-fit galaxy+AGN 
model.\footnote{In \citet{zou2022}, different best-fit models are compared using $\Delta$BIC, 
which is equal to $\chi^2_\mathrm{gal}-\chi^2_\mathrm{agn}-3\ln N_\mathrm{phot}$, where $N_\mathrm{phot}$ is the number of fluxes.}
Radio AGNs we selected in \S~\ref{sec:radio_agns} and AGNs that are selected using MIR/X-ray methods are represented by different symbols.
Since the parameter settings of the galaxy+AGN model are less dense than those of the galaxy-only model (\citealt{zou2022}),
the latter is not a special case of the former with zero AGN component.
Otherwise, $\chi^2_\mathrm{agn}\le\chi^2_\mathrm{gal}$ would always be true.
Notably, the majority of the MIR/X-ray AGNs prefer the galaxy+AGN model with $\chi^2_\mathrm{gal}>\chi^2_\mathrm{agn}$,
while radio AGNs generally prefer the galaxy-only model, 
indicating that the signatures of active nuclei are not strong for these AGNs at bands other than the radio.
The host-galaxy property (stellar mass and SFR) results from the {\sc cigale} fits are provided in Table~\ref{tab:ragn_table}.
For each object, we use the fitting results with the smaller $\chi^2$ between those of the galaxy-only and galaxy+AGN models.

We show typical SED fitting results for radio AGNs in Fig.~\ref{fig:cigale_examples}. 
For each field, we separate the objects that prefer the galaxy-only and galaxy+AGN models, 
and choose the object with the median reduced $\chi^2$ (i.e. $\chi^2_\nu$) 
in each group. Note that {\sc cigale} counts the number of data points including
upper limits when calculating the degrees of freedom.
We recalculate the reduced $\chi^2$ including only the detected fluxes.
Generally, the {\sc cigale} models reproduce well the observed fluxes from the UV-to-FIR bands.
The median reduced $\chi^2$ in Fig.~\ref{fig:cigale_examples} are close to unity.
We acknowledge that the {\it Spitzer}/PACS and {\it Herschel}/SPIRE data points are very often upper limits, and 
the FIR band is effectively constrained only by the MIPS~24~$\micro$Jy fluxes and 
UV/optical fluxes via the energy-balance principle of {\sc cigale} (\citealt{boquien2019, yang2020}).

We show the best-fit SEDs from 100~nm to 1000~$\mu$m of radio AGNs and MIR/X-ray AGNs in the left and right panels of Fig.~\ref{fig:stack_seds_agns}, respectively.
Even though the two populations do not substantially overlap (see Fig.~\ref{fig:venn_full}), we exclude objects that are both radio and MIR/\mbox{X-ray} AGNs in Fig.~\ref{fig:stack_seds_agns}.
Therefore, the radio AGNs here refer to the AGNs that are only selected in the radio band but not with other methods.
Furthermore, we exclude objects with $\chi^2_\nu\ge2$ to remove 
low-quality fits so that the model-dependent {\sc cigale} SEDs 
well depict the underlying true SEDs.
The SEDs of the remaining objects are first normalized to $\int_\mathrm{1000\mu m}^\mathrm{100nm}f_\nu d\nu=1$ following \citet{zou2022}.
Then, the SEDs are rebinned in rest-frame wavelength from 
1000~$\mu$m to 100~nm
with a bin size of 0.02~dex.
We calculate the composite median SEDs for the radio and MIR/X-ray AGNs and plot them in both panels of Fig.~\ref{fig:stack_seds_agns}.
The SEDs of typical radio AGNs show two prominent peaks in the optical/NIR and FIR bands, 
while those of the MIR/X-ray AGNs are flatter and show excess emission in the UV and MIR bands.
The contrast between the stacked SEDs of radio 
AGNs and MIR/X-ray AGNs is consistent with Fig.~\ref{fig:venn_full}
where radio AGNs overlap little with AGNs selected using MIR/X-ray methods,
probably because a substantial fraction of the 
radio AGNs we select are low-luminosity radio galaxies
without bright and blue accretion disks and hot, dusty tori.
Notably, the radio AGNs seem to contain a small but substantial population of objects that are dust-poor.

In Fig.~\ref{fig:cigale_radio_excess},
we compare the radio emission that is related to star formation in the host galaxies with the observed total radio fluxes.
The radio AGNs selected using the $q_\mathrm{24}$ parameter are generally consistent with the $q_\mathrm{FIR}$ parameter.
Furthermore, a larger value of $q_\mathrm{FIR}\approx2.6$ (see Footnote~\ref{ft:qir}) will shift the data points downward by $\approx0.3$ dex and move radio-excess AGNs further away from the black line. 
Most of the remaining, probably star formation-dominated, objects seem consistent with $q_\mathrm{FIR}=2.34$ and seem unlikely to tolerate a vertical shift of $\approx0.3$ dex.
At radio fluxes below 0.1~mJy, the $q_\mathrm{FIR}$ parameter shows a population of radio-excess AGNs that are missed 
by the $q_\mathrm{24}$ parameter due to the limited sensitivity of the 24$\mu$m data.

\begin{figure}
\centering
\includegraphics[width=0.48\textwidth, clip]{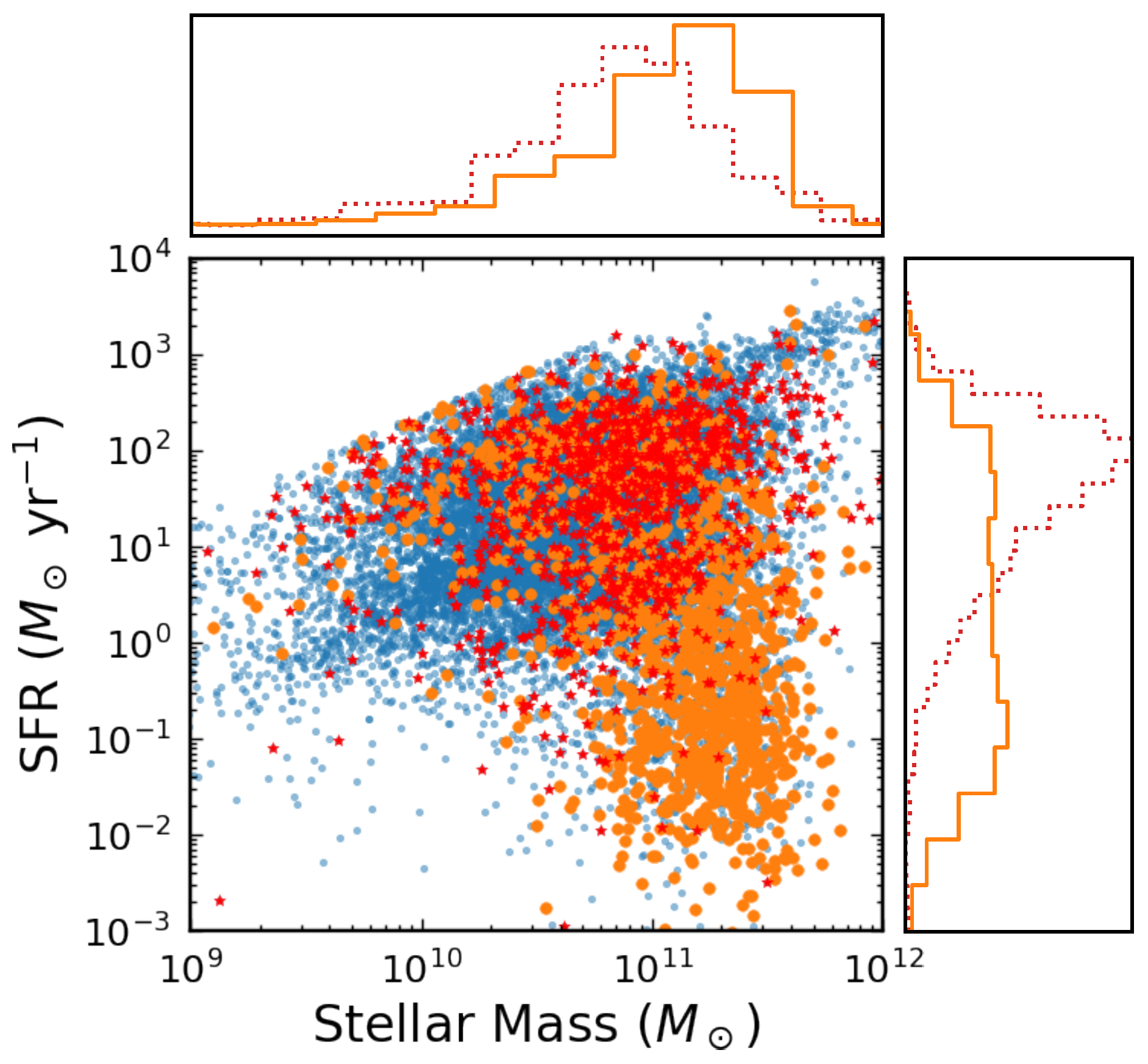}
    \caption{The SFRs and stellar masses resulting from SED fitting using {\sc cigale}.
    The orange dots and red stars are radio and MIR/X-ray AGNs, respectively. The remaining radio sources are in blue.
    We also show the SFR and stellar mass distributions for AGNs in the right and top panels, respectively.}
\label{fig:SFR_StellarMass}
\end{figure}

We show the ``Bayesian-like'' estimations of SFR and stellar mass from {\sc cigale} fits in Fig.~\ref{fig:SFR_StellarMass}.
The host galaxies of radio AGNs are generally more massive than those of MIR/X-ray AGNs. 
The SFRs of radio AGNs can reach as high as those of MIR/X-ray AGNs. However, a large fraction of radio AGNs have SFR $<1M_\odot$ yr$^{-1}$.

\begin{figure*}
\centering
\includegraphics[width=0.95\textwidth, clip]{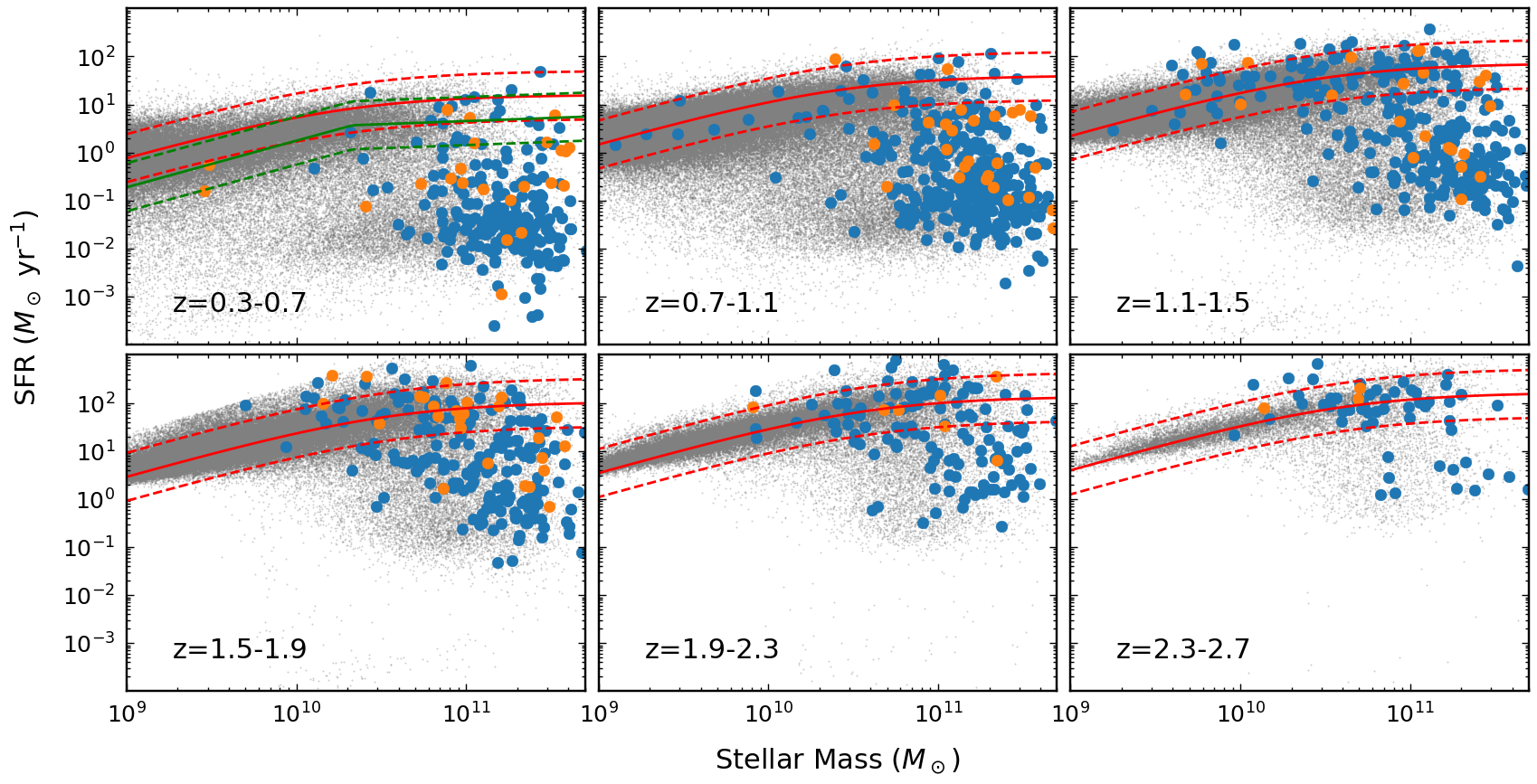}
    \caption{The SFRs vs. stellar masses for radio AGNs (blue) in six redshift bins 
    in comparison with the main sequence of star-forming galaxies.
    Large points are radio AGNs, and large points with orange color are also X-ray and/or MIR AGNs.
    The small grey dots are taken from the SED fitting results of \citet{zou2022}.
    The red solid curve in each panel represents the main sequence of \citet{popesso2022}.
    The green solid curve in the top-left panel represents the main sequence of \citet{leja2022}.
    We used dashed curves to represent a scatter of 0.3 dex for the main sequence in each panel.
    }
\label{fig:ms}
\end{figure*}

We further split the radio AGNs into six redshift bins from $z=0.3$ to 2.7 with a bin width of 0.4 and show the results 
in Fig.~\ref{fig:ms}.
We compare radio AGNs with the ``main sequence'' of \citet{popesso2022} in each redshift bin.
The SFR and $M_\odot$ for the VIDEO-selected galaxies in the three DDFs from \citet{zou2022} are shown as small grey points in Fig.~\ref{fig:ms},
which are consistent with the main sequence of \citet{popesso2022} except for the first bin (i.e. the top-left panel of Fig.~\ref{fig:ms}).
We also plot the main sequence of \citet{leja2022} in the first redshift bin in green, 
which is more consistent with the star-forming galaxies of \citet{zou2022}.
A progressively larger fraction of radio AGNs are hosted by main-sequence galaxies with increasing redshift;
while most radio AGNs live in massive, quiescent galaxies at $z<1.1$, 
star-forming galaxies dominate the host galaxies of radio AGNs at $z=1.9$--2.7, where the cosmic star formation peaks.
Note that our paper and \cite{zou2022} use similar data sets and methods to derive the SFR and stellar mass,
and our results are not affected by the definition of the main sequence from the literature. 
The main sequences of \citet{popesso2022} and \citet{leja2022} in Fig.~\ref{fig:ms} are only used to guide visualization. 
The large orange points in Fig.~\ref{fig:ms} show the radio AGNs that are also X-ray and/or MIR AGNs.
These objects are high-excitation radio galaxies (HERGs), and most of the large blue points are 
likely low-excitation radio galaxies (LERGs) because LERGs show little sign of active nuclei except in the radio band.
Using the MIGHTEE/COSMOS data, \citet{whittam2022} find that HERGs and LERGs show little difference in SFR and stellar mass,
which is in contrast with the results in the local universe. 
Our results seem consistent with those of \citet{whittam2022} in that there 
is no apparent difference between the host-galaxy SFRs and stellar masses of HERGs and LERGs at high redshift.

\subsection{The dependence of the incidence of radio AGNs on galaxy stellar mass and redshift}
\begin{figure}
\centering
\includegraphics[width=0.48\textwidth, clip]{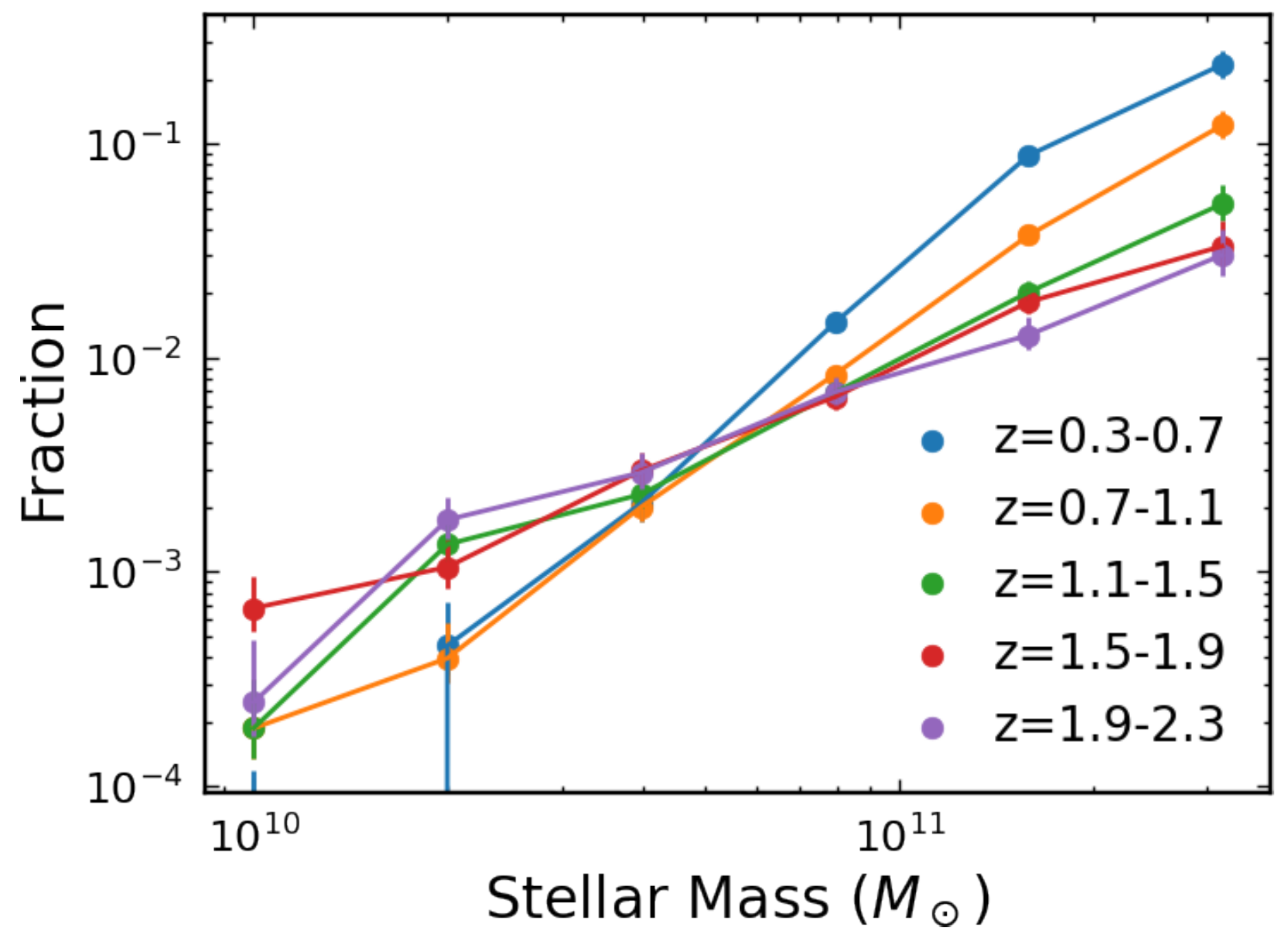}
\includegraphics[width=0.48\textwidth, clip]{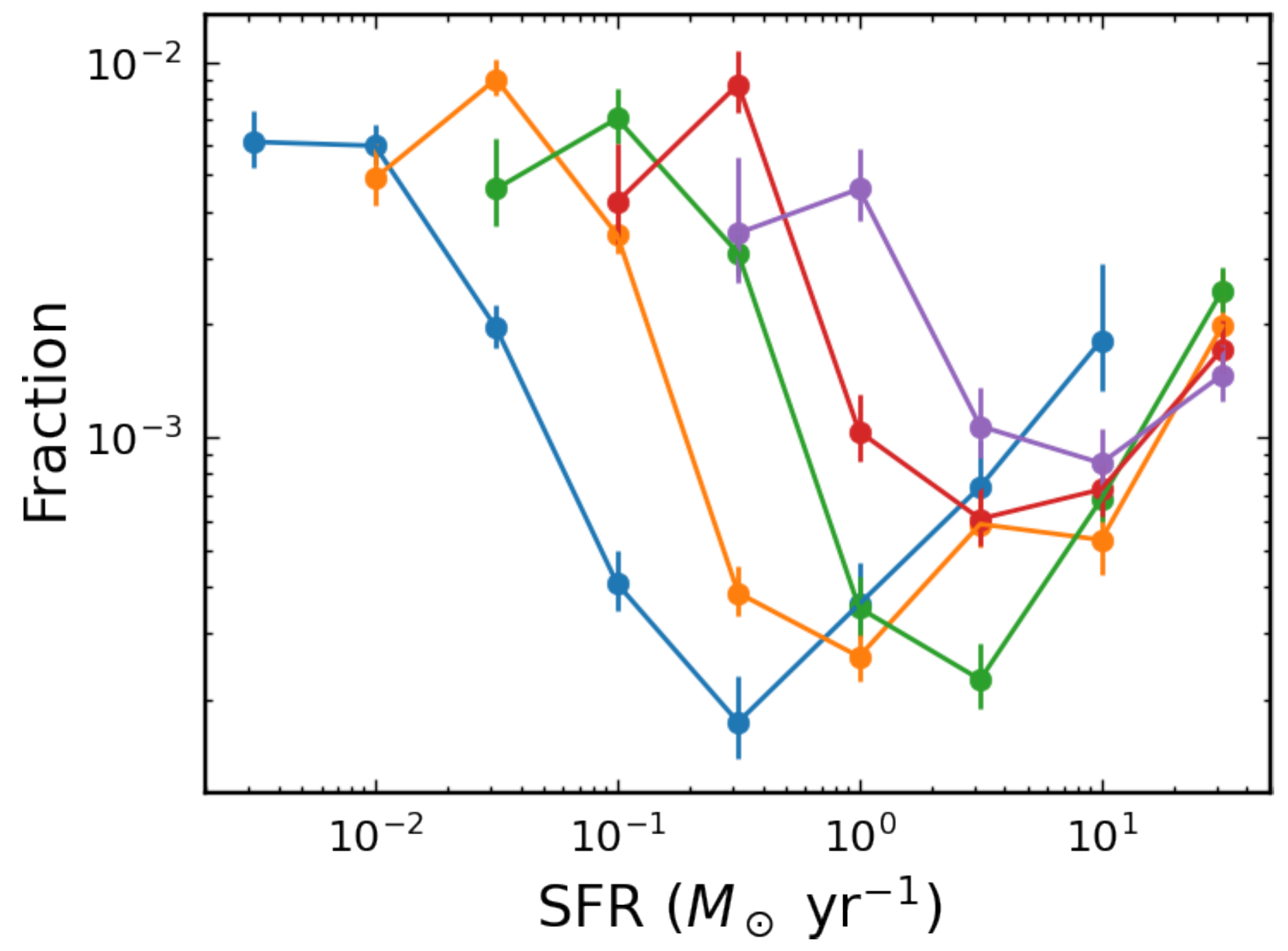}
\caption{Top: The dependence of radio AGN fraction upon stellar mass for galaxies with $0.3<z<2.3$ and $M_*>10^{10}M_\odot$ in the three DDFs we study in this paper.
In all redshift bins, the radio AGN fraction increases with stellar mass.
It also seems that the stellar-mass dependence evolves with redshift such that the dependence is shallower at high redshift.
Bottom: The radio AGN fraction generally first decreases and then increases with SFR in all redshift bins, 
    suggesting two types of host galaxies of radio AGNs, quiescent and star-forming galaxies.}
\label{fig:frac_StellarMass}
\end{figure}

We investigate how the properties of the galaxies (and redshift) could affect the incidence of radio AGNs using the SED fitting results of this paper and \citet{zou2022}.
We select the galaxy sample from \citet{zou2022} as in \S~\ref{sec:main_sample} and focus on galaxies with $0.3<z<2.3$ and $M_*>10^{10}M_\odot$.
Given the depth of the VIDEO survey, the sample is complete for $M_*>10^{10}M_\odot$ galaxies up to $z\gtrsim2$.
The galaxies are split into $z$-$M_*$ bins with a step of 0.4 in redshift and a step of 0.3 dex in stellar mass.
The fraction of radio AGNs in each bin is calculated. The errors of the radio AGN fraction are calculated
using \texttt{astropy.stats.binom\_conf\_interval} with the \texttt{flat} method (e.g. \citealt{cameron2011}).
The results are shown in the top panel of Fig.~\ref{fig:frac_StellarMass}.
The dependence of the radio AGN fraction on stellar mass is strong such that the fraction increases by 2--3 orders
of magnitude from $M_*=10^{10}M_\odot$ to $3\times 10^{11}M_\odot$.
However, the stellar-mass dependence in the $z>1.1$ bins is not as strong as in the $0.3<z<1.1$ bins.

In the bottom panel of Fig.~\ref{fig:frac_StellarMass}, we show the dependence of radio AGN fraction upon SFR.
The overall shape of the relation is consistent with a ``V'' shape throughout the redshift range we study, 
indicating that the fraction of radio AGNs first decreases and then increases with SFR.
The SFR and the radio-AGN fraction at the vertex of the ``V'' both increase with redshift.
The bottom panel of Fig.~\ref{fig:frac_StellarMass} is consistent with a scenario 
where the radio AGNs could be hosted by both quiescent and star-forming galaxies,
and the incidence of radio AGNs decreases with SFR in quiescent galaxies but increases with SFR in star-forming galaxies.
The triggering mechanism of radio AGNs might thus be different in quiescent galaxies and star-forming galaxies.
Furthermore, there are overall increases of SFR and radio AGN fraction for all types of galaxies 
from $z\sim0$ to $z=2$--3, shifting the vertex of the ``V''.

Thanks to the deep multiwavelength 
coverage in the three DDFs we study in this paper,
the incidence rate of radio AGNs 
has been constrained to a higher redshift 
than in previous work (e.g. \citealt{kondapally2022}).
The incidence of X-ray selected AGNs also increases with stellar mass. 
A strong positive evolution with redshift is found for X-ray AGNs with $M_*>10^{10}M_\odot$ (e.g. \citealt{yang2018, aird2019}), 
which is consistent with radio AGNs in the first two stellar-mass bins of the top panel of Fig.~\ref{fig:frac_StellarMass}.
However, for the most massive galaxies with $M_*>10^{11}M_\odot$, the radio AGN fraction decreases with redshift.
Furthermore, the X-ray AGN fraction increases almost linearly with the SFR (e.g. \citealt{aird2019}), 
which is not the case for radio AGNs, 
probably because most radio AGNs are hosted by quiescent galaxies and fueled by the cooling of hot gas.

\section{Summary and Future Prospects}
\label{sec:summary}
\subsection{Summary of main results}
We have performed deep radio AGN selection and characterization in three
well-studied multiwavelength survey fields (W-CDF-S, ELAIS-S1, and XMM-LSS)
spanning 11.3~deg$^2$ in total that will soon be extensively observed
by Rubin as DDFs. The deep radio data used are from the ATLAS/W-CDF-S,
ATLAS/ELAIS-S1, VLA/XMM-LSS, and MIGHTEE/XMM-LSS surveys. Our main results
are the following:

1. We have performed careful cross-matching of the 25,000 faint radio sources
  to optical, NIR, and MIR objects that has successfully identified reliable
  counterparts for most (93--96\%) of the radio sources; such counterparts
  are essential for AGN selection and characterization. See \S2 and \S3. 

2. We have constructed FIR-to-X-ray SEDs for the counterparts of the 22,000
  radio sources in our main (7.7~deg$^2$) sample. About 82\% of these radio
  sources have spectroscopic or high-quality photometric redshifts available,
  and these indicate that most of the radio sources lie at $z<2$. See \S4.2
  and \S4.3.

3. We have selected 1815 radio AGNs in our three fields based upon radio
  morphology (lobe/jet structures), radio spectral index ($\alpha_r$), and
  excess radio flux ($q_{24}$) criteria. As expected from other surveys,
  these radio AGNs usually do not overlap with AGNs selected using X-ray
  and MIR techniques. In fact, 1656 of the 1815 radio AGNs (91\%) had
  not been identified as AGNs previously based upon their X-ray and/or MIR
  emission. This represents a substantive increase of the total
  AGN population in these key sky fields. See \S4.4 and \S4.5.

4. Stacking of the sensitive XMM-SERVS X-ray data in these fields shows
  only a weak positive correlation between the X-ray and radio fluxes, and
  the average X-ray hardness ratios of the X-ray detected and X-ray
  undetected radio AGNs are statistically consistent.  See \S4.6.

5. We have fit the FIR-to-UV SEDs of all radio sources using {\sc cigale} from
  which we derive, e.g., host-galaxy stellar masses and star-formation
  rates. As expected, most radio AGNs prefer galaxy-only SED models
  statistically, indicating that AGN signatures for such sources are
  generally weak at bands other than the radio. See \S5.1.

6. We have measured the dependence of the radio AGN fraction upon
stellar mass and SFR out to $z\approx 2.3$. The fraction strongly
rises with stellar mass, although this rise is weaker at
higher redshifts. 
The dependence of radio AGN fraction appears to decrease with SFR in quiescent galaxies and increase with SFR in star-forming galaxies.
See \S~5.2.


\subsection{Future work}
\label{sec:futurework}
The 1815 radio AGNs selected in this work will be studied extensively over
the next decade by multiple facilities. Rubin, {\it Euclid}, the {\it Roman Space
Telescope}, and the Hawaii Two-0 Survey, for example, will obtain 
superb NIR-to-optical photometric and imaging data that will
further constrain their photometric redshifts, SEDs, and
morphologies. The resulting measurements of stellar masses,
star-formation rates, and host-galaxy structures will shed
light on the triggering mechanisms for radio AGNs to high
redshift.
Some of these photometric data
will also allow time-domain studies, although many of the radio AGNs have
host-dominated SEDs outside the radio band so are not expected to show
strong NIR-to-optical variability. Our radio AGNs will also be extensively
targeted by powerful spectroscopic surveys, including
the Deep Extragalactic Visible Legacy Survey (DEVILS),
the Multi-Object Optical and Near-Infrared Spectrograph (MOONS) survey,
the Subaru Prime Focus Spectrograph (PFS) survey, and
the Wide Area VISTA Extragalactic Survey (WAVES). 
These will provide precise redshifts as well as further characterize the
nature of our sources spectroscopically. The precise redshifts will allow
our radio AGNs to be placed in their large-scale cosmic environmental
context; such radio AGNs often trace groups and clusters. While rapid
black-hole growth does not appear to depend materially upon cosmic environment
once stellar mass is controlled (e.g., \citealt{yang2018, krishnan2020}),
large-sample studies are still needed to assess the environmental
dependence of jet-driven feedback by radio AGNs that generally represent
lower accretion-rate systems in a ``maintenance mode''.
Ultimately, deeper X-ray coverage of the LSST DDFs by missions including 
{\it Athena\/}, {\it STAR-X\/}, and {\it AXIS\/} should detect a larger
fraction of our radio AGNs, allowing their accretion properties to be
clarified. 

Many more radio sources in the W-CDF-S and ELAIS-S1 fields will be detected
as deeper radio observations are gathered; e.g., by the MIGHTEE survey, which should raise
the radio-source sky density by a factor of $\approx 7$. Based on our results
for the XMM-LSS field in \S4.5, however, the increment of radio AGNs will be
more modest---about a factor of 50\% (the remaining majority of radio sources will be
a combination of star-forming galaxies and radio-quiet AGNs). 
The MIGHTEE-H~{\sc i} project will provide a probe of the state and kinematics of neutral gas around RL AGNs via H~{\sc i} absorption, 
and the polarisation observations of MIGHTEE will provide another probe of intrinsic and environmental properties of RL AGNs.

\section*{Acknowledgements}
We thank the anonymous referee and the 
scientific editor, Tim Pearson, for their comments that improved the clarity of our manuscript.
SZ, WNB, WY, and FZ acknowledge support from NSF grant AST-2106990, NASA grant 80NSSC19K0961, Chandra X-ray Center grant AR1-22008X, and Penn State ACIS Instrument Team Contract SV4-74018 (issued by the Chandra X-ray Center, which is operated by the Smithsonian Astrophysical Observatory for and on behalf of NASA under contract NAS8-03060).
B.L. acknowledges financial support from the National Natural Science
Foundation of China grant 11991053.
Y.Q.X. acknowledges support from NSFC grants (12025303 and 11890693), the K.C. Wong Education Foundation, 
and the science research grants from the China Manned Space Project with NO. CMS-CSST-2021-A06.
The Chandra ACIS Team Guaranteed Time Observations (GTO) utilized were selected by the ACIS Instrument Principal Investigator, Gordon P. Garmire, currently of the Huntingdon Institute for X-ray Astronomy, LLC, which is under contract to the Smithsonian Astrophysical Observatory via Contract SV2-82024.

\section*{Data availability}
The catalogues released in this article are available as online supplementary material, 
and they can also be downloaded at {\url https://personal.psu.edu/wnb3/xmmservs/xmmservs.html}

\bibliographystyle{mnras}
\bibliography{mn} 

\begin{thebibliography}{}
\makeatletter
\relax
\def\mn@urlcharsother{\let\do\@makeother \do\$\do\&\do\#\do\^\do\_\do\%\do\~}
\def\mn@doi{\begingroup\mn@urlcharsother \@ifnextchar [ {\mn@doi@}
  {\mn@doi@[]}}
\def\mn@doi@[#1]#2{\def\@tempa{#1}\ifx\@tempa\@empty \href
  {http://dx.doi.org/#2} {doi:#2}\else \href {http://dx.doi.org/#2} {#1}\fi
  \endgroup}
\def\mn@eprint#1#2{\mn@eprint@#1:#2::\@nil}
\def\mn@eprint@arXiv#1{\href {http://arxiv.org/abs/#1} {{\tt arXiv:#1}}}
\def\mn@eprint@dblp#1{\href {http://dblp.uni-trier.de/rec/bibtex/#1.xml}
  {dblp:#1}}
\def\mn@eprint@#1:#2:#3:#4\@nil{\def\@tempa {#1}\def\@tempb {#2}\def\@tempc
  {#3}\ifx \@tempc \@empty \let \@tempc \@tempb \let \@tempb \@tempa \fi \ifx
  \@tempb \@empty \def\@tempb {arXiv}\fi \@ifundefined
  {mn@eprint@\@tempb}{\@tempb:\@tempc}{\expandafter \expandafter \csname
  mn@eprint@\@tempb\endcsname \expandafter{\@tempc}}}

\bibitem[\protect\citeauthoryear{{Abbott} et~al.,}{{Abbott}
  et~al.}{2021}]{abbott2021}
{Abbott} T.~M.~C.,  et~al., 2021, \mn@doi [\apjs] {10.3847/1538-4365/ac00b3},
  \href {https://ui.adsabs.harvard.edu/abs/2021ApJS..255...20A} {255, 20}

\bibitem[\protect\citeauthoryear{{Aird}, {Coil}  \& {Georgakakis}}{{Aird}
  et~al.}{2019}]{aird2019}
{Aird} J.,  {Coil} A.~L.,   {Georgakakis} A.,  2019, \mn@doi [\mnras]
  {10.1093/mnras/stz125}, \href
  {https://ui.adsabs.harvard.edu/abs/2019MNRAS.484.4360A} {484, 4360}

\bibitem[\protect\citeauthoryear{{An} et~al.,}{{An} et~al.}{2021}]{an2021}
{An} F.,  et~al., 2021, \mn@doi [\mnras] {10.1093/mnras/stab2290}, \href
  {https://ui.adsabs.harvard.edu/abs/2021MNRAS.507.2643A} {507, 2643}

\bibitem[\protect\citeauthoryear{{Appleton} et~al.,}{{Appleton}
  et~al.}{2004}]{appleton2004}
{Appleton} P.~N.,  et~al., 2004, \mn@doi [\apjs] {10.1086/422425}, \href
  {https://ui.adsabs.harvard.edu/abs/2004ApJS..154..147A} {154, 147}

\bibitem[\protect\citeauthoryear{{Barbieri} \& {Bertola}}{{Barbieri} \&
  {Bertola}}{1972}]{barbier1972}
{Barbieri} C.,  {Bertola} F.,  1972, \mn@doi [\mnras]
  {10.1093/mnras/156.4.399}, \href
  {https://ui.adsabs.harvard.edu/abs/1972MNRAS.156..399B} {156, 399}

\bibitem[\protect\citeauthoryear{{Berta} et~al.,}{{Berta}
  et~al.}{2006}]{berta2006}
{Berta} S.,  et~al., 2006, \mn@doi [\aap] {10.1051/0004-6361:20054548}, \href
  {https://ui.adsabs.harvard.edu/abs/2006A&A...451..881B} {451, 881}

\bibitem[\protect\citeauthoryear{{Bertin} \& {Arnouts}}{{Bertin} \&
  {Arnouts}}{1996}]{bertin1996}
{Bertin} E.,  {Arnouts} S.,  1996, \mn@doi [\aaps] {10.1051/aas:1996164}, \href
  {https://ui.adsabs.harvard.edu/abs/1996A&AS..117..393B} {117, 393}

\bibitem[\protect\citeauthoryear{{Bonzini}, {Padovani}, {Mainieri},
  {Kellermann}, {Miller}, {Rosati}, {Tozzi}  \& {Vattakunnel}}{{Bonzini}
  et~al.}{2013}]{bonzini2013}
{Bonzini} M.,  {Padovani} P.,  {Mainieri} V.,  {Kellermann} K.~I.,  {Miller}
  N.,  {Rosati} P.,  {Tozzi} P.,   {Vattakunnel} S.,  2013, \mn@doi [\mnras]
  {10.1093/mnras/stt1879}, \href
  {https://ui.adsabs.harvard.edu/abs/2013MNRAS.436.3759B} {436, 3759}

\bibitem[\protect\citeauthoryear{{Boquien}, {Burgarella}, {Roehlly}, {Buat},
  {Ciesla}, {Corre}, {Inoue}  \& {Salas}}{{Boquien} et~al.}{2019}]{boquien2019}
{Boquien} M.,  {Burgarella} D.,  {Roehlly} Y.,  {Buat} V.,  {Ciesla} L.,
  {Corre} D.,  {Inoue} A.~K.,   {Salas} H.,  2019, \mn@doi [\aap]
  {10.1051/0004-6361/201834156}, \href
  {https://ui.adsabs.harvard.edu/abs/2019A&A...622A.103B} {622, A103}

\bibitem[\protect\citeauthoryear{Bradley et~al.,}{Bradley
  et~al.}{2020}]{photutils}
Bradley L.,  et~al., 2020, astropy/photutils: 1.0.0,
  \mn@doi{10.5281/zenodo.4044744}, \url
  {https://doi.org/10.5281/zenodo.4044744}

\bibitem[\protect\citeauthoryear{{Brandt} et~al.,}{{Brandt}
  et~al.}{2018}]{brandt2018}
{Brandt} W.~N.,  et~al., 2018, arXiv e-prints, \href
  {https://ui.adsabs.harvard.edu/abs/2018arXiv181106542B} {p. arXiv:1811.06542}

\bibitem[\protect\citeauthoryear{{Brinkmann}, {Laurent-Muehleisen}, {Voges},
  {Siebert}, {Becker}, {Brotherton}, {White}  \& {Gregg}}{{Brinkmann}
  et~al.}{2000}]{brinkmann2000}
{Brinkmann} W.,  {Laurent-Muehleisen} S.~A.,  {Voges} W.,  {Siebert} J.,
  {Becker} R.~H.,  {Brotherton} M.~S.,  {White} R.~L.,   {Gregg} M.~D.,  2000,
  \aap, \href {https://ui.adsabs.harvard.edu/abs/2000A&A...356..445B} {356,
  445}

\bibitem[\protect\citeauthoryear{{Brusa} et~al.,}{{Brusa}
  et~al.}{2007}]{brusa2007}
{Brusa} M.,  et~al., 2007, \mn@doi [\apjs] {10.1086/516575}, \href
  {https://ui.adsabs.harvard.edu/abs/2007ApJS..172..353B} {172, 353}

\bibitem[\protect\citeauthoryear{{Cameron}}{{Cameron}}{2011}]{cameron2011}
{Cameron} E.,  2011, \mn@doi [\pasa] {10.1071/AS10046}, \href
  {https://ui.adsabs.harvard.edu/abs/2011PASA...28..128C} {28, 128}

\bibitem[\protect\citeauthoryear{{Chang} et~al.,}{{Chang}
  et~al.}{2017}]{chang2017}
{Chang} Y.-Y.,  et~al., 2017, \mn@doi [\apjs] {10.3847/1538-4365/aa97da}, \href
  {https://ui.adsabs.harvard.edu/abs/2017ApJS..233...19C} {233, 19}

\bibitem[\protect\citeauthoryear{{Chen} et~al.,}{{Chen}
  et~al.}{2018}]{chen2018}
{Chen} C. T.~J.,  et~al., 2018, \mn@doi [\mnras] {10.1093/mnras/sty1036}, \href
  {https://ui.adsabs.harvard.edu/abs/2018MNRAS.478.2132C} {478, 2132}

\bibitem[\protect\citeauthoryear{{Ciliegi}, {Zamorani}, {Hasinger}, {Lehmann},
  {Szokoly}  \& {Wilson}}{{Ciliegi} et~al.}{2003}]{ciliegi2003}
{Ciliegi} P.,  {Zamorani} G.,  {Hasinger} G.,  {Lehmann} I.,  {Szokoly} G.,
  {Wilson} G.,  2003, \mn@doi [\aap] {10.1051/0004-6361:20021721}, \href
  {https://ui.adsabs.harvard.edu/abs/2003A&A...398..901C} {398, 901}

\bibitem[\protect\citeauthoryear{{Condon}}{{Condon}}{1992}]{condon1992}
{Condon} J.~J.,  1992, \mn@doi [\araa] {10.1146/annurev.aa.30.090192.003043},
  \href {https://ui.adsabs.harvard.edu/abs/1992ARA&A..30..575C} {30, 575}

\bibitem[\protect\citeauthoryear{{Davies} et~al.,}{{Davies}
  et~al.}{2021}]{davies2021}
{Davies} L.~J.~M.,  et~al., 2021, \mn@doi [\mnras] {10.1093/mnras/stab1601},
  \href {https://ui.adsabs.harvard.edu/abs/2021MNRAS.506..256D} {506, 256}

\bibitem[\protect\citeauthoryear{{Delhaize} et~al.,}{{Delhaize}
  et~al.}{2017}]{delhaize2017}
{Delhaize} J.,  et~al., 2017, \mn@doi [\aap] {10.1051/0004-6361/201629430},
  \href {https://ui.adsabs.harvard.edu/abs/2017A&A...602A...4D} {602, A4}

\bibitem[\protect\citeauthoryear{{Delvecchio} et~al.,}{{Delvecchio}
  et~al.}{2021}]{delvecchio2021}
{Delvecchio} I.,  et~al., 2021, \mn@doi [\aap] {10.1051/0004-6361/202039647},
  \href {https://ui.adsabs.harvard.edu/abs/2021A&A...647A.123D} {647, A123}

\bibitem[\protect\citeauthoryear{{Donley} et~al.,}{{Donley}
  et~al.}{2012}]{donley2012}
{Donley} J.~L.,  et~al., 2012, \mn@doi [\apj] {10.1088/0004-637X/748/2/142},
  \href {https://ui.adsabs.harvard.edu/abs/2012ApJ...748..142D} {748, 142}

\bibitem[\protect\citeauthoryear{{Fan}, {Budav{\'a}ri}, {Norris}  \&
  {Hopkins}}{{Fan} et~al.}{2015}]{fan2015}
{Fan} D.,  {Budav{\'a}ri} T.,  {Norris} R.~P.,   {Hopkins} A.~M.,  2015,
  \mn@doi [\mnras] {10.1093/mnras/stv994}, \href
  {https://ui.adsabs.harvard.edu/abs/2015MNRAS.451.1299F} {451, 1299}

\bibitem[\protect\citeauthoryear{{Fan}, {Budav{\'a}ri}, {Norris}  \&
  {Basu}}{{Fan} et~al.}{2020}]{fan2020}
{Fan} D.,  {Budav{\'a}ri} T.,  {Norris} R.~P.,   {Basu} A.,  2020, \mn@doi
  [\mnras] {10.1093/mnras/staa2447}, \href
  {https://ui.adsabs.harvard.edu/abs/2020MNRAS.498..565F} {498, 565}

\bibitem[\protect\citeauthoryear{{Franzen} et~al.,}{{Franzen}
  et~al.}{2015}]{franzen2015}
{Franzen} T.~M.~O.,  et~al., 2015, \mn@doi [\mnras] {10.1093/mnras/stv1866},
  \href {https://ui.adsabs.harvard.edu/abs/2015MNRAS.453.4020F} {453, 4020}

\bibitem[\protect\citeauthoryear{{Hale} et~al.,}{{Hale}
  et~al.}{2019}]{hale2019}
{Hale} C.~L.,  et~al., 2019, \mn@doi [\aap] {10.1051/0004-6361/201833906},
  \href {https://ui.adsabs.harvard.edu/abs/2019A&A...622A...4H} {622, A4}

\bibitem[\protect\citeauthoryear{{Hales} et~al.,}{{Hales}
  et~al.}{2014}]{hales2014}
{Hales} C.~A.,  et~al., 2014, \mn@doi [\mnras] {10.1093/mnras/stu576}, \href
  {https://ui.adsabs.harvard.edu/abs/2014MNRAS.441.2555H} {441, 2555}

\bibitem[\protect\citeauthoryear{{Hancock}, {Trott}  \&
  {Hurley-Walker}}{{Hancock} et~al.}{2018}]{hancock2018}
{Hancock} P.~J.,  {Trott} C.~M.,   {Hurley-Walker} N.,  2018, \mn@doi
  [Publications of the Astronomical Society of Australia]
  {10.1017/pasa.2018.3}, \href
  {https://ui.adsabs.harvard.edu/#abs/2018PASA...35...11H} {35, e011}

\bibitem[\protect\citeauthoryear{{Helou}, {Soifer}  \&
  {Rowan-Robinson}}{{Helou} et~al.}{1985}]{helou1985}
{Helou} G.,  {Soifer} B.~T.,   {Rowan-Robinson} M.,  1985, \mn@doi [\apjl]
  {10.1086/184556}, \href
  {https://ui.adsabs.harvard.edu/abs/1985ApJ...298L...7H} {298, L7}

\bibitem[\protect\citeauthoryear{{Herschel Point Source Catalogue Working
  Group} et~al.,}{{Herschel Point Source Catalogue Working Group}
  et~al.}{2020}]{psc2020}
{Herschel Point Source Catalogue Working Group} et~al., 2020, VizieR Online
  Data Catalog, \href {https://ui.adsabs.harvard.edu/abs/2020yCat.8106....0H}
  {p. VIII/106}

\bibitem[\protect\citeauthoryear{{Heywood}, {Hale}, {Jarvis}, {Makhathini},
  {Peters}, {Sebokolodi}  \& {Smirnov}}{{Heywood} et~al.}{2020}]{heywood2020}
{Heywood} I.,  {Hale} C.~L.,  {Jarvis} M.~J.,  {Makhathini} S.,  {Peters}
  J.~A.,  {Sebokolodi} M.~L.~L.,   {Smirnov} O.~M.,  2020, \mn@doi [\mnras]
  {10.1093/mnras/staa1770}, \href
  {https://ui.adsabs.harvard.edu/abs/2020MNRAS.496.3469H} {496, 3469}

\bibitem[\protect\citeauthoryear{{Heywood} et~al.,}{{Heywood}
  et~al.}{2022}]{heywood2022}
{Heywood} I.,  et~al., 2022, \mn@doi [\mnras] {10.1093/mnras/stab3021}, \href
  {https://ui.adsabs.harvard.edu/abs/2022MNRAS.509.2150H} {509, 2150}

\bibitem[\protect\citeauthoryear{{Hickox} et~al.,}{{Hickox}
  et~al.}{2009}]{hickox2009}
{Hickox} R.~C.,  et~al., 2009, \mn@doi [\apj] {10.1088/0004-637X/696/1/891},
  \href {https://ui.adsabs.harvard.edu/abs/2009ApJ...696..891H} {696, 891}

\bibitem[\protect\citeauthoryear{{Hudelot} et~al.,}{{Hudelot}
  et~al.}{2012}]{hudelot2012}
{Hudelot} P.,  et~al., 2012, VizieR Online Data Catalog, \href
  {https://ui.adsabs.harvard.edu/abs/2012yCat.2317....0H} {p. II/317}

\bibitem[\protect\citeauthoryear{{Ibar} et~al.,}{{Ibar}
  et~al.}{2008}]{ibar2008}
{Ibar} E.,  et~al., 2008, \mn@doi [\mnras] {10.1111/j.1365-2966.2008.13077.x},
  \href {https://ui.adsabs.harvard.edu/abs/2008MNRAS.386..953I} {386, 953}

\bibitem[\protect\citeauthoryear{{Ivezi{\'c}} et~al.,}{{Ivezi{\'c}}
  et~al.}{2019}]{ivezic2019}
{Ivezi{\'c}} {\v{Z}}.,  et~al., 2019, \mn@doi [\apj]
  {10.3847/1538-4357/ab042c}, \href
  {https://ui.adsabs.harvard.edu/abs/2019ApJ...873..111I} {873, 111}

\bibitem[\protect\citeauthoryear{{Jarvis} et~al.,}{{Jarvis}
  et~al.}{2013}]{jarvis2013}
{Jarvis} M.~J.,  et~al., 2013, \mn@doi [\mnras] {10.1093/mnras/sts118}, \href
  {https://ui.adsabs.harvard.edu/abs/2013MNRAS.428.1281J} {428, 1281}

\bibitem[\protect\citeauthoryear{{Jarvis} et~al.,}{{Jarvis}
  et~al.}{2016}]{jarvis2016}
{Jarvis} M.,  et~al., 2016, in MeerKAT Science: On the Pathway to the SKA. p.~6
  (\mn@eprint {arXiv} {1709.01901})

\bibitem[\protect\citeauthoryear{{Kondapally} et~al.,}{{Kondapally}
  et~al.}{2022}]{kondapally2022}
{Kondapally} R.,  et~al., 2022, \mn@doi [\mnras] {10.1093/mnras/stac1128},
  \href {https://ui.adsabs.harvard.edu/abs/2022MNRAS.513.3742K} {513, 3742}

\bibitem[\protect\citeauthoryear{{Krishnan} et~al.,}{{Krishnan}
  et~al.}{2020}]{krishnan2020}
{Krishnan} C.,  et~al., 2020, \mn@doi [\mnras] {10.1093/mnras/staa815}, \href
  {https://ui.adsabs.harvard.edu/abs/2020MNRAS.494.1693K} {494, 1693}

\bibitem[\protect\citeauthoryear{{Lacy}, {Petric}, {Sajina}, {Canalizo},
  {Storrie-Lombardi}, {Armus}, {Fadda}  \& {Marleau}}{{Lacy}
  et~al.}{2007}]{lacy2007}
{Lacy} M.,  {Petric} A.~O.,  {Sajina} A.,  {Canalizo} G.,  {Storrie-Lombardi}
  L.~J.,  {Armus} L.,  {Fadda} D.,   {Marleau} F.~R.,  2007, \mn@doi [\aj]
  {10.1086/509617}, \href
  {https://ui.adsabs.harvard.edu/abs/2007AJ....133..186L} {133, 186}

\bibitem[\protect\citeauthoryear{{Lacy} et~al.,}{{Lacy}
  et~al.}{2021}]{lacy2021}
{Lacy} M.,  et~al., 2021, \mn@doi [\mnras] {10.1093/mnras/staa3714}, \href
  {https://ui.adsabs.harvard.edu/abs/2021MNRAS.501..892L} {501, 892}

\bibitem[\protect\citeauthoryear{{Laor} \& {Behar}}{{Laor} \&
  {Behar}}{2008}]{laor2008}
{Laor} A.,  {Behar} E.,  2008, \mn@doi [\mnras]
  {10.1111/j.1365-2966.2008.13806.x}, \href
  {https://ui.adsabs.harvard.edu/\#abs/2008MNRAS.390..847L} {390, 847}

\bibitem[\protect\citeauthoryear{{Leja} et~al.,}{{Leja}
  et~al.}{2022}]{leja2022}
{Leja} J.,  et~al., 2022, \mn@doi [\apj] {10.3847/1538-4357/ac887d}, \href
  {https://ui.adsabs.harvard.edu/abs/2022ApJ...936..165L} {936, 165}

\bibitem[\protect\citeauthoryear{{Lonsdale} et~al.,}{{Lonsdale}
  et~al.}{2003}]{lonsdale2003}
{Lonsdale} C.~J.,  et~al., 2003, \mn@doi [\pasp] {10.1086/376850}, \href
  {https://ui.adsabs.harvard.edu/abs/2003PASP..115..897L} {115, 897}

\bibitem[\protect\citeauthoryear{{Luo} et~al.,}{{Luo} et~al.}{2010}]{luo2010}
{Luo} B.,  et~al., 2010, \mn@doi [\apjs] {10.1088/0067-0049/187/2/560}, \href
  {https://ui.adsabs.harvard.edu/abs/2010ApJS..187..560L} {187, 560}

\bibitem[\protect\citeauthoryear{{Luo} et~al.,}{{Luo} et~al.}{2017}]{luo2017}
{Luo} B.,  et~al., 2017, \mn@doi [\apjs] {10.3847/1538-4365/228/1/2}, \href
  {https://ui.adsabs.harvard.edu/abs/2017ApJS..228....2L} {228, 2}

\bibitem[\protect\citeauthoryear{{MIPS Instrument and MIPS Instrument Support
  Teams}}{{MIPS Instrument and MIPS Instrument Support
  Teams}}{2021}]{mipsHandBook}
{MIPS Instrument and MIPS Instrument Support Teams} 2021, MIPS Instrument
  Handbook, \mn@doi{10.26131/IRSA488}, \url
  {https://catcopy.ipac.caltech.edu/dois/doi.php?id=10.26131/IRSA488}

\bibitem[\protect\citeauthoryear{{Mao}, {Sharp}, {Saikia}, {Norris},
  {Johnston-Hollitt}, {Middelberg}  \& {Lovell}}{{Mao} et~al.}{2010}]{mao2010}
{Mao} M.~Y.,  {Sharp} R.,  {Saikia} D.~J.,  {Norris} R.~P.,  {Johnston-Hollitt}
  M.,  {Middelberg} E.,   {Lovell} J. E.~J.,  2010, \mn@doi [\mnras]
  {10.1111/j.1365-2966.2010.16853.x}, \href
  {https://ui.adsabs.harvard.edu/abs/2010MNRAS.406.2578M} {406, 2578}

\bibitem[\protect\citeauthoryear{{Mao} et~al.,}{{Mao} et~al.}{2012}]{mao2012}
{Mao} M.~Y.,  et~al., 2012, \mn@doi [\mnras]
  {10.1111/j.1365-2966.2012.21913.x}, \href
  {https://ui.adsabs.harvard.edu/abs/2012MNRAS.426.3334M} {426, 3334}

\bibitem[\protect\citeauthoryear{{Marchesi} et~al.,}{{Marchesi}
  et~al.}{2016}]{marchesi2016}
{Marchesi} S.,  et~al., 2016, \mn@doi [\apj] {10.3847/0004-637X/817/1/34},
  \href {https://ui.adsabs.harvard.edu/abs/2016ApJ...817...34M} {817, 34}

\bibitem[\protect\citeauthoryear{{Mauduit} et~al.,}{{Mauduit}
  et~al.}{2012}]{mauduit2012}
{Mauduit} J.~C.,  et~al., 2012, \mn@doi [\pasp] {10.1086/666945}, \href
  {https://ui.adsabs.harvard.edu/abs/2012PASP..124..714M} {124, 714}

\bibitem[\protect\citeauthoryear{{McConnell} et~al.,}{{McConnell}
  et~al.}{2020}]{mcconnell2020}
{McConnell} D.,  et~al., 2020, \mn@doi [\pasa] {10.1017/pasa.2020.41}, \href
  {https://ui.adsabs.harvard.edu/abs/2020PASA...37...48M} {37, e048}

\bibitem[\protect\citeauthoryear{{Middelberg} et~al.,}{{Middelberg}
  et~al.}{2008}]{middelberg2008}
{Middelberg} E.,  et~al., 2008, \mn@doi [\aj] {10.1088/0004-6256/135/4/1276},
  \href {https://ui.adsabs.harvard.edu/abs/2008AJ....135.1276M} {135, 1276}

\bibitem[\protect\citeauthoryear{{Miller}, {Brandt}, {Schneider}, {Gibson},
  {Steffen}  \& {Wu}}{{Miller} et~al.}{2011}]{miller2011}
{Miller} B.~P.,  {Brandt} W.~N.,  {Schneider} D.~P.,  {Gibson} R.~R.,
  {Steffen} A.~T.,   {Wu} J.,  2011, \mn@doi [\apj]
  {10.1088/0004-637X/726/1/20}, \href
  {https://ui.adsabs.harvard.edu/#abs/2011ApJ...726...20M} {726, 20}

\bibitem[\protect\citeauthoryear{{Miller} et~al.,}{{Miller}
  et~al.}{2013}]{miller2013}
{Miller} N.~A.,  et~al., 2013, \mn@doi [\apjs] {10.1088/0067-0049/205/2/13},
  \href {https://ui.adsabs.harvard.edu/abs/2013ApJS..205...13M} {205, 13}

\bibitem[\protect\citeauthoryear{{Nakos} et~al.,}{{Nakos}
  et~al.}{2009}]{nakos2009}
{Nakos} T.,  et~al., 2009, \mn@doi [\aap] {10.1051/0004-6361:200809584}, \href
  {https://ui.adsabs.harvard.edu/abs/2009A&A...494..579N} {494, 579}

\bibitem[\protect\citeauthoryear{{Naylor}, {Broos}  \& {Feigelson}}{{Naylor}
  et~al.}{2013}]{naylor2013}
{Naylor} T.,  {Broos} P.~S.,   {Feigelson} E.~D.,  2013, \mn@doi [\apjs]
  {10.1088/0067-0049/209/2/30}, \href
  {https://ui.adsabs.harvard.edu/abs/2013ApJS..209...30N} {209, 30}

\bibitem[\protect\citeauthoryear{{Ni} et~al.,}{{Ni} et~al.}{2021}]{ni2021}
{Ni} Q.,  et~al., 2021, \mn@doi [\apjs] {10.3847/1538-4365/ac0dc6}, \href
  {https://ui.adsabs.harvard.edu/abs/2021ApJS..256...21N} {256, 21}

\bibitem[\protect\citeauthoryear{{Norris} et~al.,}{{Norris}
  et~al.}{2006}]{norris2006}
{Norris} R.~P.,  et~al., 2006, \mn@doi [\aj] {10.1086/508275}, \href
  {https://ui.adsabs.harvard.edu/abs/2006AJ....132.2409N} {132, 2409}

\bibitem[\protect\citeauthoryear{{Nyland} et~al.,}{{Nyland}
  et~al.}{2017}]{nyland2017}
{Nyland} K.,  et~al., 2017, \mn@doi [\apjs] {10.3847/1538-4365/aa6fed}, \href
  {https://ui.adsabs.harvard.edu/abs/2017ApJS..230....9N} {230, 9}

\bibitem[\protect\citeauthoryear{{Nyland}, {Lacy}, {Brandt}, {Yang}, {Ni},
  {Sajina}, {Zou}  \& {Vaccari}}{{Nyland} et~al.}{2023}]{nyland2023}
{Nyland} K.,  {Lacy} M.,  {Brandt} W.~N.,  {Yang} G.,  {Ni} Q.,  {Sajina} A.,
  {Zou} F.,   {Vaccari} M.,  2023, \mn@doi [Research Notes of the American
  Astronomical Society] {10.3847/2515-5172/acbc72}, \href
  {https://ui.adsabs.harvard.edu/abs/2023RNAAS...7...33N} {7, 33}

\bibitem[\protect\citeauthoryear{{Padovani}}{{Padovani}}{2016}]{padovani2016}
{Padovani} P.,  2016, \mn@doi [\aapr] {10.1007/s00159-016-0098-6}, \href
  {https://ui.adsabs.harvard.edu/abs/2016A&ARv..24...13P} {24, 13}

\bibitem[\protect\citeauthoryear{{Padovani} et~al.,}{{Padovani}
  et~al.}{2017}]{padovani2017}
{Padovani} P.,  et~al., 2017, \mn@doi [\aapr] {10.1007/s00159-017-0102-9},
  \href {https://ui.adsabs.harvard.edu/abs/2017A&ARv..25....2P} {25, 2}

\bibitem[\protect\citeauthoryear{{Pineau}, {Motch}, {Carrera}, {Della Ceca},
  {Derri{\`e}re}, {Michel}, {Schwope}  \& {Watson}}{{Pineau}
  et~al.}{2011}]{pineau2011}
{Pineau} F.~X.,  {Motch} C.,  {Carrera} F.,  {Della Ceca} R.,  {Derri{\`e}re}
  S.,  {Michel} L.,  {Schwope} A.,   {Watson} M.~G.,  2011, \mn@doi [\aap]
  {10.1051/0004-6361/201015141}, \href
  {https://ui.adsabs.harvard.edu/abs/2011A&A...527A.126P} {527, A126}

\bibitem[\protect\citeauthoryear{{Pineau} et~al.,}{{Pineau}
  et~al.}{2017}]{pineau2017}
{Pineau} F.~X.,  et~al., 2017, \mn@doi [\aap] {10.1051/0004-6361/201629219},
  \href {https://ui.adsabs.harvard.edu/abs/2017A&A...597A..89P} {597, A89}

\bibitem[\protect\citeauthoryear{{Polletta} et~al.,}{{Polletta}
  et~al.}{2007}]{polletta2007}
{Polletta} M.,  et~al., 2007, \mn@doi [\apj] {10.1086/518113}, \href
  {https://ui.adsabs.harvard.edu/abs/2007ApJ...663...81P} {663, 81}

\bibitem[\protect\citeauthoryear{{Popesso} et~al.,}{{Popesso}
  et~al.}{2022}]{popesso2022}
{Popesso} P.,  et~al., 2022, arXiv e-prints, \href
  {https://ui.adsabs.harvard.edu/abs/2022arXiv220310487P} {p. arXiv:2203.10487}

\bibitem[\protect\citeauthoryear{{Poulain} et~al.,}{{Poulain}
  et~al.}{2020}]{poulain2020}
{Poulain} M.,  et~al., 2020, \mn@doi [\aap] {10.1051/0004-6361/201937108},
  \href {https://ui.adsabs.harvard.edu/abs/2020A&A...634A..50P} {634, A50}

\bibitem[\protect\citeauthoryear{{Read}, {Rosen}, {Saxton}  \&
  {Ramirez}}{{Read} et~al.}{2011}]{read2011}
{Read} A.~M.,  {Rosen} S.~R.,  {Saxton} R.~D.,   {Ramirez} J.,  2011, \mn@doi
  [\aap] {10.1051/0004-6361/201117525}, \href
  {https://ui.adsabs.harvard.edu/abs/2011A&A...534A..34R} {534, A34}

\bibitem[\protect\citeauthoryear{{SWIRE Project}}{{SWIRE
  Project}}{2020}]{swireProj}
{SWIRE Project} 2020, The Spitzer Wide-area InfraRed Extragalactic Survey,
  \mn@doi{10.26131/IRSA406}, \url
  {https://catcopy.ipac.caltech.edu/dois/doi.php?id=10.26131/IRSA406}

\bibitem[\protect\citeauthoryear{{Salvato} et~al.,}{{Salvato}
  et~al.}{2018}]{salvato2018}
{Salvato} M.,  et~al., 2018, \mn@doi [\mnras] {10.1093/mnras/stx2651}, \href
  {https://ui.adsabs.harvard.edu/abs/2018MNRAS.473.4937S} {473, 4937}

\bibitem[\protect\citeauthoryear{{Sargent} et~al.,}{{Sargent}
  et~al.}{2010}]{sargent2010}
{Sargent} M.~T.,  et~al., 2010, \mn@doi [\apjs] {10.1088/0067-0049/186/2/341},
  \href {https://ui.adsabs.harvard.edu/abs/2010ApJS..186..341S} {186, 341}

\bibitem[\protect\citeauthoryear{{Scolnic} et~al.,}{{Scolnic}
  et~al.}{2018}]{scolnic2018}
{Scolnic} D.~M.,  et~al., 2018, arXiv e-prints, \href
  {https://ui.adsabs.harvard.edu/abs/2018arXiv181200516S} {p. arXiv:1812.00516}

\bibitem[\protect\citeauthoryear{{Shirley} et~al.,}{{Shirley}
  et~al.}{2019}]{shirley2019}
{Shirley} R.,  et~al., 2019, \mn@doi [\mnras] {10.1093/mnras/stz2509}, \href
  {https://ui.adsabs.harvard.edu/abs/2019MNRAS.490..634S} {490, 634}

\bibitem[\protect\citeauthoryear{{Shirley} et~al.,}{{Shirley}
  et~al.}{2021}]{shirley2021}
{Shirley} R.,  et~al., 2021, \mn@doi [\mnras] {10.1093/mnras/stab1526}, \href
  {https://ui.adsabs.harvard.edu/abs/2021MNRAS.507..129S} {507, 129}

\bibitem[\protect\citeauthoryear{{Smol{\v{c}}i{\'c}}
  et~al.,}{{Smol{\v{c}}i{\'c}} et~al.}{2017a}]{smolcic2017a}
{Smol{\v{c}}i{\'c}} V.,  et~al., 2017a, \mn@doi [\aap]
  {10.1051/0004-6361/201628704}, \href
  {https://ui.adsabs.harvard.edu/abs/2017A&A...602A...1S} {602, A1}

\bibitem[\protect\citeauthoryear{{Smol{\v{c}}i{\'c}}
  et~al.,}{{Smol{\v{c}}i{\'c}} et~al.}{2017b}]{smolcic2017}
{Smol{\v{c}}i{\'c}} V.,  et~al., 2017b, \mn@doi [\aap]
  {10.1051/0004-6361/201630223}, \href
  {https://ui.adsabs.harvard.edu/abs/2017A&A...602A...2S} {602, A2}

\bibitem[\protect\citeauthoryear{{Spitzer Science Center and Infrared Science
  Archive}}{{Spitzer Science Center and Infrared Science
  Archive}}{2020}]{seipProj}
{Spitzer Science Center and Infrared Science Archive} 2020, Spitzer Enhanced
  Imaging Products, \mn@doi{10.26131/IRSA433}, \url
  {https://catcopy.ipac.caltech.edu/dois/doi.php?id=10.26131/IRSA433}

\bibitem[\protect\citeauthoryear{{Surace}, {Shupe}, {Fang}, {Evans}, {Alexov},
  {Frayer}, {Lonsdale}  \& {SWIRE Team}}{{Surace} et~al.}{2005}]{surace2005}
{Surace} J.~A.,  {Shupe} D.~L.,  {Fang} F.,  {Evans} T.,  {Alexov} A.,
  {Frayer} D.,  {Lonsdale} C.~J.,   {SWIRE Team} 2005, in American Astronomical
  Society Meeting Abstracts. p. 63.01

\bibitem[\protect\citeauthoryear{{Sutherland} \& {Saunders}}{{Sutherland} \&
  {Saunders}}{1992}]{sutherland1992}
{Sutherland} W.,  {Saunders} W.,  1992, \mn@doi [\mnras]
  {10.1093/mnras/259.3.413}, \href
  {https://ui.adsabs.harvard.edu/abs/1992MNRAS.259..413S} {259, 413}

\bibitem[\protect\citeauthoryear{{Tabatabaei} et~al.,}{{Tabatabaei}
  et~al.}{2017}]{tabatabaei2017}
{Tabatabaei} F.~S.,  et~al., 2017, \mn@doi [\apj]
  {10.3847/1538-4357/836/2/185}, \href
  {https://ui.adsabs.harvard.edu/abs/2017ApJ...836..185T} {836, 185}

\bibitem[\protect\citeauthoryear{{Tasse}, {Le Borgne}, {R{\"o}ttgering},
  {Best}, {Pierre}  \& {Rocca-Volmerange}}{{Tasse} et~al.}{2008}]{tasse2008}
{Tasse} C.,  {Le Borgne} D.,  {R{\"o}ttgering} H.,  {Best} P.~N.,  {Pierre} M.,
    {Rocca-Volmerange} B.,  2008, \mn@doi [\aap] {10.1051/0004-6361:20078453},
  \href {https://ui.adsabs.harvard.edu/abs/2008A&A...490..879T} {490, 879}

\bibitem[\protect\citeauthoryear{{Timlin} et~al.,}{{Timlin}
  et~al.}{2016}]{timlin2016}
{Timlin} J.~D.,  et~al., 2016, \mn@doi [\apjs] {10.3847/0067-0049/225/1/1},
  \href {https://ui.adsabs.harvard.edu/abs/2016ApJS..225....1T} {225, 1}

\bibitem[\protect\citeauthoryear{{Vaccari}}{{Vaccari}}{2015}]{vaccari2015}
{Vaccari} M.,  2015, in The Many Facets of Extragalactic Radio Surveys: Towards
  New Scientific Challenges. p.~27 (\mn@eprint {arXiv} {1604.02353})

\bibitem[\protect\citeauthoryear{{Weston}, {Seymour}, {Gulyaev}, {Norris},
  {Banfield}, {Vaccari}, {Hopkins}  \& {Franzen}}{{Weston}
  et~al.}{2018}]{weston2018}
{Weston} S.~D.,  {Seymour} N.,  {Gulyaev} S.,  {Norris} R.~P.,  {Banfield} J.,
  {Vaccari} M.,  {Hopkins} A.~M.,   {Franzen} T.~M.~O.,  2018, \mn@doi [\mnras]
  {10.1093/mnras/stx2562}, \href
  {https://ui.adsabs.harvard.edu/abs/2018MNRAS.473.4523W} {473, 4523}

\bibitem[\protect\citeauthoryear{{Whittam} et~al.,}{{Whittam}
  et~al.}{2022}]{whittam2022}
{Whittam} I.~H.,  et~al., 2022, \mn@doi [\mnras] {10.1093/mnras/stac2140},
  \href {https://ui.adsabs.harvard.edu/abs/2022MNRAS.516..245W} {516, 245}

\bibitem[\protect\citeauthoryear{{Wolf} et~al.,}{{Wolf}
  et~al.}{2004}]{wolf2004}
{Wolf} C.,  et~al., 2004, \mn@doi [\aap] {10.1051/0004-6361:20040525}, \href
  {https://ui.adsabs.harvard.edu/abs/2004A&A...421..913W} {421, 913}

\bibitem[\protect\citeauthoryear{{Wright} et~al.,}{{Wright}
  et~al.}{2010}]{wright2010}
{Wright} E.~L.,  et~al., 2010, \mn@doi [\aj] {10.1088/0004-6256/140/6/1868},
  \href {https://ui.adsabs.harvard.edu/#abs/2010AJ....140.1868W} {140, 1868}

\bibitem[\protect\citeauthoryear{{Yang}, {Brandt}, {Darvish}, {Chen}, {Vito},
  {Alexander}, {Bauer}  \& {Trump}}{{Yang} et~al.}{2018}]{yang2018}
{Yang} G.,  {Brandt} W.~N.,  {Darvish} B.,  {Chen} C. T.~J.,  {Vito} F.,
  {Alexander} D.~M.,  {Bauer} F.~E.,   {Trump} J.~R.,  2018, \mn@doi [\mnras]
  {10.1093/mnras/sty1910}, \href
  {https://ui.adsabs.harvard.edu/abs/2018MNRAS.480.1022Y} {480, 1022}

\bibitem[\protect\citeauthoryear{{Yang} et~al.,}{{Yang}
  et~al.}{2020}]{yang2020}
{Yang} G.,  et~al., 2020, \mn@doi [\mnras] {10.1093/mnras/stz3001}, \href
  {https://ui.adsabs.harvard.edu/abs/2020MNRAS.491..740Y} {491, 740}

\bibitem[\protect\citeauthoryear{{Yang} et~al.,}{{Yang}
  et~al.}{2022}]{yang2022}
{Yang} G.,  et~al., 2022, \mn@doi [\apj] {10.3847/1538-4357/ac4971}, \href
  {https://ui.adsabs.harvard.edu/abs/2022ApJ...927..192Y} {927, 192}

\bibitem[\protect\citeauthoryear{{Yun}, {Reddy}  \& {Condon}}{{Yun}
  et~al.}{2001}]{yun2001}
{Yun} M.~S.,  {Reddy} N.~A.,   {Condon} J.~J.,  2001, \mn@doi [\apj]
  {10.1086/323145}, \href
  {https://ui.adsabs.harvard.edu/abs/2001ApJ...554..803Y} {554, 803}

\bibitem[\protect\citeauthoryear{{Zhu}, {Brandt}, {Luo}, {Wu}, {Xue}  \&
  {Yang}}{{Zhu} et~al.}{2020}]{zhu2020}
{Zhu} S.~F.,  {Brandt} W.~N.,  {Luo} B.,  {Wu} J.,  {Xue} Y.~Q.,   {Yang} G.,
  2020, \mn@doi [\mnras] {10.1093/mnras/staa1411}, \href
  {https://ui.adsabs.harvard.edu/abs/2020MNRAS.496..245Z} {496, 245}

\bibitem[\protect\citeauthoryear{{Zinn}, {Middelberg}, {Norris}, {Hales}, {Mao}
   \& {Randall}}{{Zinn} et~al.}{2012}]{zinn2012}
{Zinn} P.~C.,  {Middelberg} E.,  {Norris} R.~P.,  {Hales} C.~A.,  {Mao} M.~Y.,
   {Randall} K.~E.,  2012, \mn@doi [\aap] {10.1051/0004-6361/201219349}, \href
  {https://ui.adsabs.harvard.edu/abs/2012A&A...544A..38Z} {544, A38}

\bibitem[\protect\citeauthoryear{{Zou} et~al.,}{{Zou}
  et~al.}{2021}]{zou2021tractor}
{Zou} F.,  et~al., 2021, \mn@doi [Research Notes of the American Astronomical
  Society] {10.3847/2515-5172/abe769}, \href
  {https://ui.adsabs.harvard.edu/abs/2021RNAAS...5...31Z} {5, 31}

\bibitem[\protect\citeauthoryear{{Zou} et~al.,}{{Zou} et~al.}{2022}]{zou2022}
{Zou} F.,  et~al., 2022, \mn@doi [\apjs] {10.3847/1538-4365/ac7bdf}, \href
  {https://ui.adsabs.harvard.edu/abs/2022ApJS..262...15Z} {262, 15}

\makeatother
\end{thebibliography}

\appendix

\section{The optimal likelihood-ratio cut}
\label{sec:appendix1}
We calculate the ratio of the likelihood that 
the MLRC is a genuine counterpart over the likelihood that
the MLRC is a confusing object,
\begin{align}
    \label{eq:mlrc_r}
    \mathcal{L}=\frac{Q\frac{f(r)}{g(r)}\frac{q(m)}{n(m)}}{Q\int_{L(m',r')<L_\mathrm{max}}q(m')f(r')dm'dr'+(1-Q)}.
\end{align}
    In a special case where the 
photometric properties are ignored, the likelihood ratio is
\begin{equation}
    \label{eq:optcut}
    \mathcal L = \frac{Qe^{-r^2/2}}{2\lambda [Qe^{-r^2/2} + (1-Q)]},
\end{equation}
which is consistent with Section 2.2 of \citet{sutherland1992}.
The optimal $L_{\max}$ that balances the number of matched genuine counterparts and the number of false positives 
can be found by solving $\mathcal{L}(L_{\max})=1$.\footnote{\label{fn:optcut} It is straightforward to solve Eq.~\ref{eq:optcut} and find that the optimal distance cut for the classic NN method is $r=\sqrt{2\ln\big(\frac{Q}{1-Q}\frac{1-2\lambda}{2\lambda}\big)}$.
The optimal distance cut does not exist if $\lambda\ge0.5$, which is rarely the case in practice.}
We denote this cut $L_{\mathcal{L}=1}$,
which depends upon $\lambda$ and, therefore, upon the positional uncertainties of radio sources.
Equivalently, we can use $\mathcal{L}>1$ directly to screen MLRCs without finding the value of $L_{\mathcal L=1}$.
The reliability of the MLRC is
\begin{align}
\mathcal{R} = {\mathcal{L}}/{(\mathcal{L}+1)},
\end{align}
and $\mathcal L>1$ is equivalent to $\mathcal{R}>0.5$.
Therefore, our optimal cut maximizes $N_{\rm T}-N_{\rm F}$, where $N_{\rm T}$ is the number of matched genuine counterparts 
and $N_{\rm F}$ the number of false positives.

\begin{figure}
\centering
\includegraphics[width=0.48\textwidth, clip]{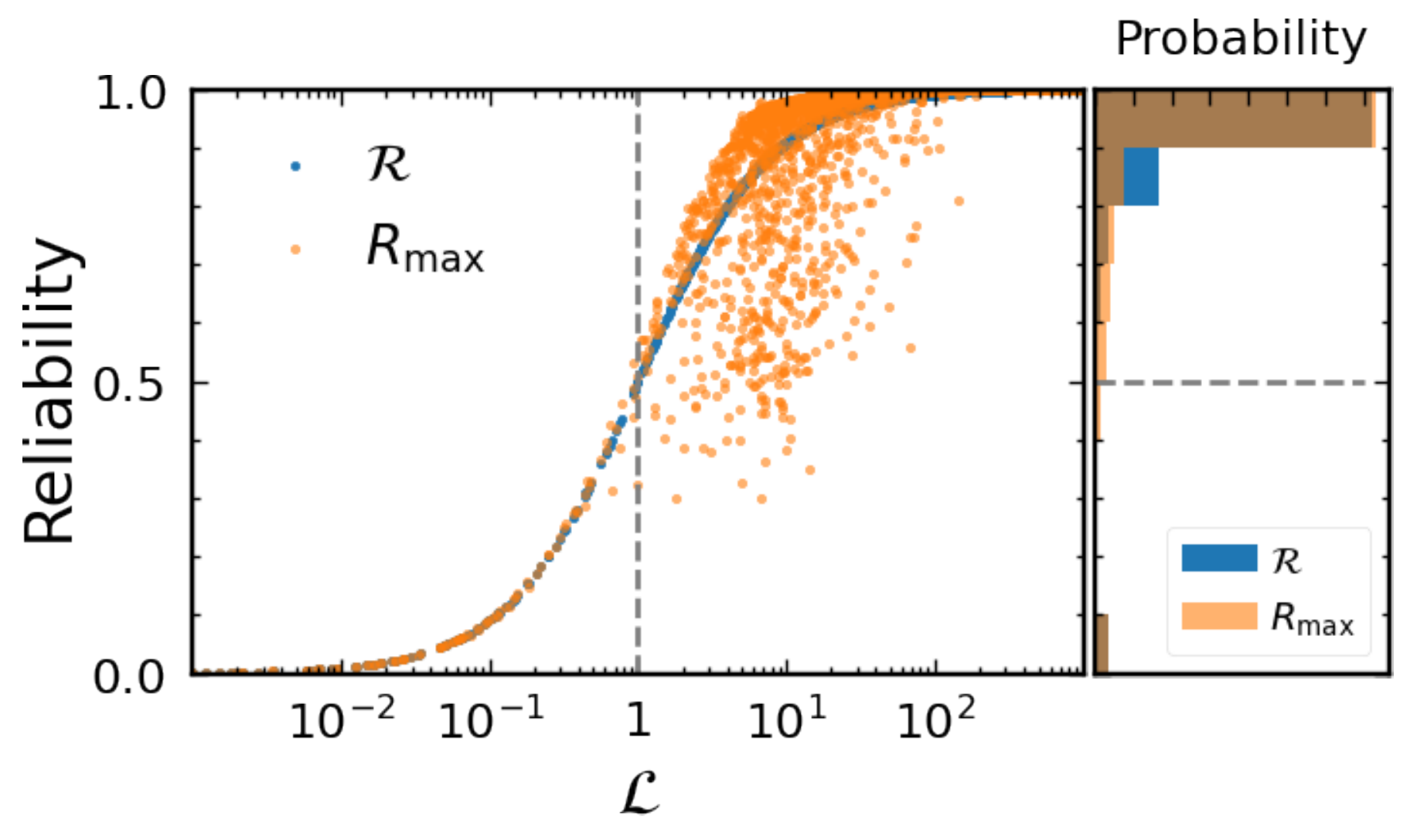}
    \caption{Left: The resulting $\mathcal{L}$, $\mathcal{R}$, and $R_{\max}$ of the ATLAS-VIDEO matching in the W-CDF-S field.
    Right: The distributions of $\mathcal{R}$ and $R_{\max}$.
    The cuts of $\mathcal{R}>0.5$ and $R_{\max}>0.5$ select similar sets of counterparts.}
\label{fig:reliability}
\end{figure}

$\mathcal{R}$ and $R_{\max}$ are expected to be consistent since they describe the same quantity (\citealt{sutherland1992});
the former utilizes the ``nearest neighbour'' and relies more on prior knowledge (i.e. $Q$ and positional error and photometric distributions)
while the latter utilizes all candidates and relies less on prior knowledge.
As an example, we show the resulting $\mathcal{L}$, $\mathcal{R}$, and $R_{\max}$ of the ATLAS-VIDEO/W-CDF-S matching in Fig.~\ref{fig:reliability}.
Therefore, $\mathcal{L}>1$, $\mathcal{R}>0.5$, and $R_{\max}>0.5$ 
corresponds (approximately) to the same optimal cut of $L_{\max}$.
In practice, $R_{\max}$ is not only easier to calculate but also less likely to be affected by source confusion than $\mathcal{R}$.
For confused radio sources associated with multiple close counterparts,
$R_\mathrm{\max}$ is easily diluted to $<0.5$ while $\mathcal{R}$ could still be large.

Using our optimal likelihood-ratio cut, it is straightforward to predict the completeness, 
\begin{multline}
     \mathcal C=1-\Big[Q\int_{L(m,r)\le L_{\mathcal{L}=1}}q(m)f(r)dmdr  + (1-Q)\Big]  \\
    \cdot e^{-\int_{L(m,r)>L_{\mathcal{L}=1} }n(m)g(r)dmdr}.
    \label{eq:completeness}
\end{multline}
Then, the purity, $\mathcal P$, is predicted by calculating
\begin{multline}
    \mathcal P\cdot \mathcal C=Q\int_{L(m,r)>L_{\mathcal{L}=1}}q(m)f(r)\\ \cdot e^{-\int_{L(m',r')>L(m, r)}n(m')g(r')dm'dr'}dmdr.
    \label{eq:purity}
\end{multline}
Note that simulations are often used to assess the completeness and purity of
cross-matching results in the literature.
These simulations are Monte-Carlo equivalents of the integrations in Eqs.~\ref{eq:completeness} and \ref{eq:purity}.
However, since simulations are rarely sufficiently realistic to correct for the issues we mention 
in \S~\ref{sec:source_stack} and, especially, Footnote~\ref{ft:lambda},
the estimated completeness and purity are likely not as accurate as those from Eqs.~\ref{eq:completeness} and \ref{eq:purity}.
The two equations above are general, and $L_{\mathcal{L}=1}$ can be replaced by other $L_{\max}$ cuts.

\section{Results of the probabilistic cross matching}
\label{sec:full_lrx}
In Table~\ref{tab:full_lrx}, we provide the full results of the probabilistic cross 
matching of \S~\ref{sec:xm_results}, 
which includes the reliability, $R_i$, of all candidates.
\begin{table*}
\centering
\caption{The full results of the probability cross matching. Only the top 5 rows are shown.
    The full table is available as online supplementary material.
    Column (1): Radio catalog. 
    Column (2): Optical/IR catalog. 
    Column (3): ID of the radio source.
    Column (4)(5): The RA and Dec of the radio source.
    Column (6)(7): The positional errors of the radio source in units of arcsec.
    Column (8)(9): The RA and Dec of the optical/IR candidate.
    Column (10): A flag indicating if the radio source is in the region masked by bright stars.
    Column (11)(12): The likelihood ratio and corresponding reliability of the optical/IR candidate. See Eqs.~\ref{eq:lr} and \ref{eq:r}.
    Column (13): A flag indicating if the optical/IR is the MLRC among all candidates.
    Column (14): The likelihood ratio of the MLRC. See Eq.~\ref{eq:mlrc_r}.
    Column (15): A flag indicating if the radio source is found to be confused by visual inspection. See \S~\ref{sec:vishost}.}
\label{tab:full_lrx}
\begin{threeparttable}[b]
\begin{tabularx}{\linewidth}{@{}Y@{}}
\begin{tabular}{lccccccccc}
\hline
\hline
    Radio & OIR & ID & RA & Dec & $\sigma_{\alpha}$ & $\sigma_{\delta}$ & RA & Dec & StarMask  \\
    &  & &(Radio) &(Radio)&arcsec &arcsec&(OIR)&(OIR)&\\
    (1)& (2) & (3)&(4) &(5)&(6) &(7)&(8)&(9)&(10) \\
\hline
    ATLAS/W-CDF-S &  DES & CI0002  & 51.47524& $-$28.69829& 0.49  &  1.04 &   51.47522 &  $-$28.69864 & 0 \\
    ATLAS/W-CDF-S &  DES & CI0004  & 51.48881& $-$28.05203& 0.35  &  0.65 &   51.48885 &  $-$28.05247 & 0 \\
    ATLAS/W-CDF-S &  DES & CI0005  & 51.49254& $-$28.19756& 0.56  &  1.24 &   51.49310 &  $-$28.19769 & 0 \\
    ATLAS/W-CDF-S &  DES & CI0006  & 51.49392& $-$27.42286& 0.93  &  2.16 &   51.49381 &  $-$27.42157 & 0 \\
    ATLAS/W-CDF-S &  DES & CI0006  & 51.49392& $-$27.42286& 0.93  &  2.16 &   51.49492 &  $-$27.42537 & 0  \\
\\
    $L_i$ & $R_i$ & MLRC & $\mathcal{L}$ & Confused \\
    &&&& &\\
    (11)&(12)&(13)&(14) & (15)\\
    6.76&   0.95& 1 &  10.58& 0\\
    9.13&   0.96& 1 &  16.64 & 0 \\
    0.03&   0.089& 1 &   0.083 & 0 \\
    0.23&   0.088& 0 &  - & 0 \\
    0.00&   0.000& 0 &  - & 0 \\
\hline
\end{tabular}
\end{tabularx}
\end{threeparttable}
\end{table*}




\section{Upper limits in the IRAC 5.8 micron and 8.0 micron bands}
\label{sec:upp}
The IR bands are important for assessing the levels of AGN and SF activity.
For objects that are not detected in the two redder {\it Spitzer}/IRAC channels, we calculate their flux upper limits.
We downloaded the IRAC 5.8/8.0 $\micro$m science mosaics of the most recent public release of SWIRE (\citealt{swireProj}).
Note that the native units of the downloaded mosaics are 1~MJy~steradian$^{-1}$, which can be converted to more convenient units using
1~MJy~steradian$^{-1}=8.46$~$\micro$Jy~pixel$^{-2}$ given the pixel scale of 0.6 arcsec (\citealt{timlin2016}).
Masks of bright stars created using the method of \citet{mauduit2012} are always applied in the following steps.
After masking out sources detected in the corresponding catalog, we estimated the background of each image using the {\sc SExtractorBackground} method implemented in {\sc astropy/photutils} (\citealt{photutils}),
which is then subtracted from the image.
Then, to minimize the effects of the emission from detected sources on undetected sources, we remove the fluxes of detected sources from the images.
We used the PSF-photometry method of {\sc photutils} and set initial source positions at those of the detected sources.
The IRAC point response functions (PRFs) are utilized in this procedure, after which we obtain residual images.
The residual images of different tiles are combined into single images for each band (each field consists of 12--16 tiles).
Then, we performed forced photometry at the positions of VIDEO sources using circular apertures of radius of 1.9 arcsec.
The `subpixel' method of {\sc photutils} that splits each pixel into $5\times5$ subpixels is adopted in the aperture photometry, 
which is consistent with that of {\sc SExtractor} (\citealt{bertin1996}).

\begin{figure}
\centering
\includegraphics[width=0.45\textwidth, clip]{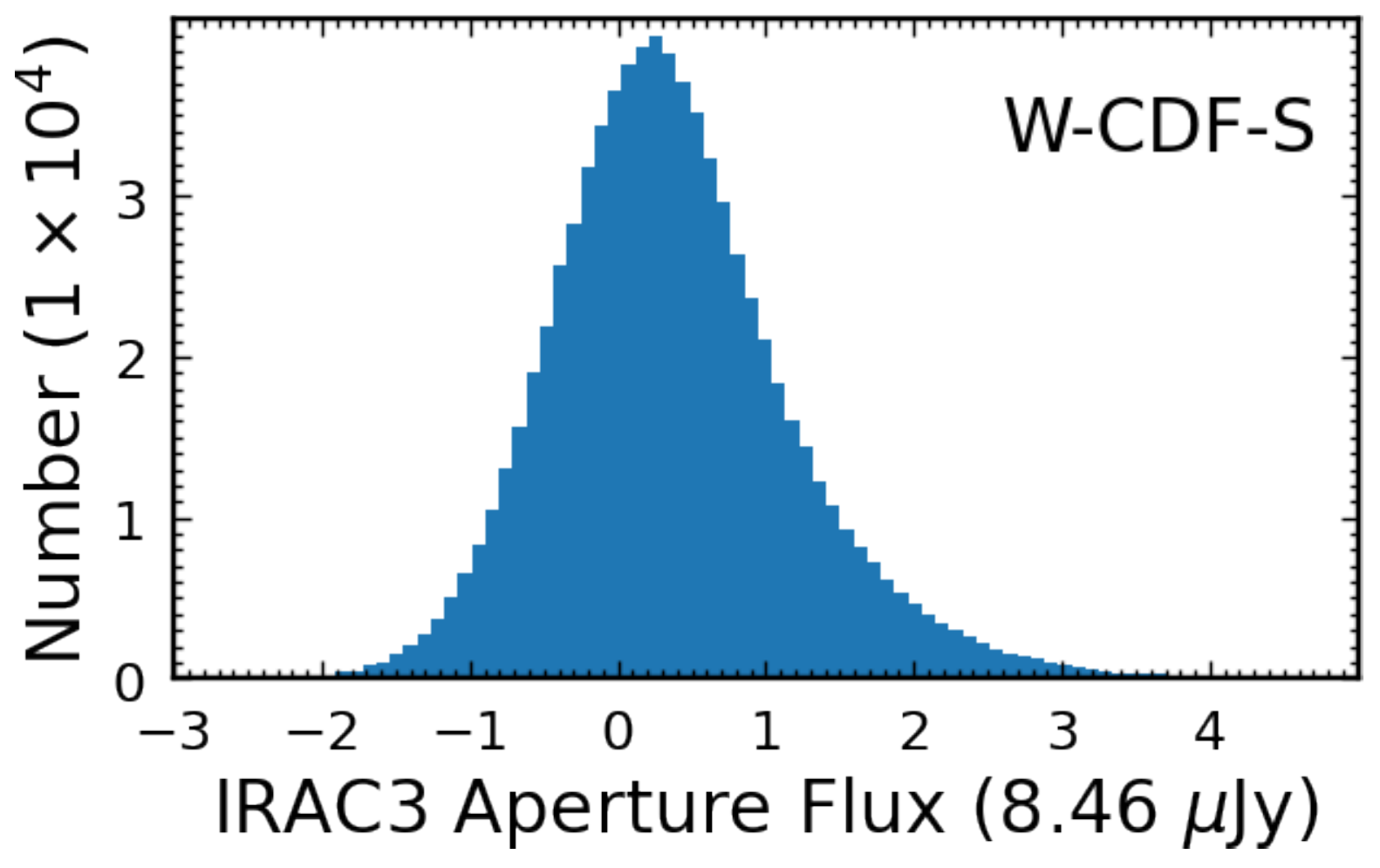}
\caption{An example of the SWIRE aperture flux for VIDEO objects 
that are not detected in the IRAC 5.8 $\mu$m (channel 3) and IRAC 8.0 $\mu$m (channel 4) bands.
This plot shows the distribution for the W-CDF-S field in the IRAC 5.8 $\mu$m band.  }
\label{fig:irac3_flux}
\end{figure}

The resulting flux density distribution is positively skewed and
peaked at a value slightly above zero (see Fig.~\ref{fig:irac3_flux}),
indicating weak emission from real celestial objects above background.
We estimated a global $1\sigma$ level of background fluctuations for each band for
each field from the negative part of the flux distribution by calculating the square root of the 2nd moment.
Then, utilizing the uncertainty maps as relative weight, we calculated the local background fluctuation level to 
take the spatial variation into account.
The flux upper limit at each VIDEO position is then estimated adopting Eq. (13) of \citet{zou2022}.
Note that a correction reflecting the finite size of apertures is also applied (\citealt{surace2005}).

\section{MIPS 24 micron PACS 100 micron, and PACS 160 micron fluxes}
\label{sec:xid_m24}
\begin{figure}
\centering
\includegraphics[width=0.45\textwidth, clip]{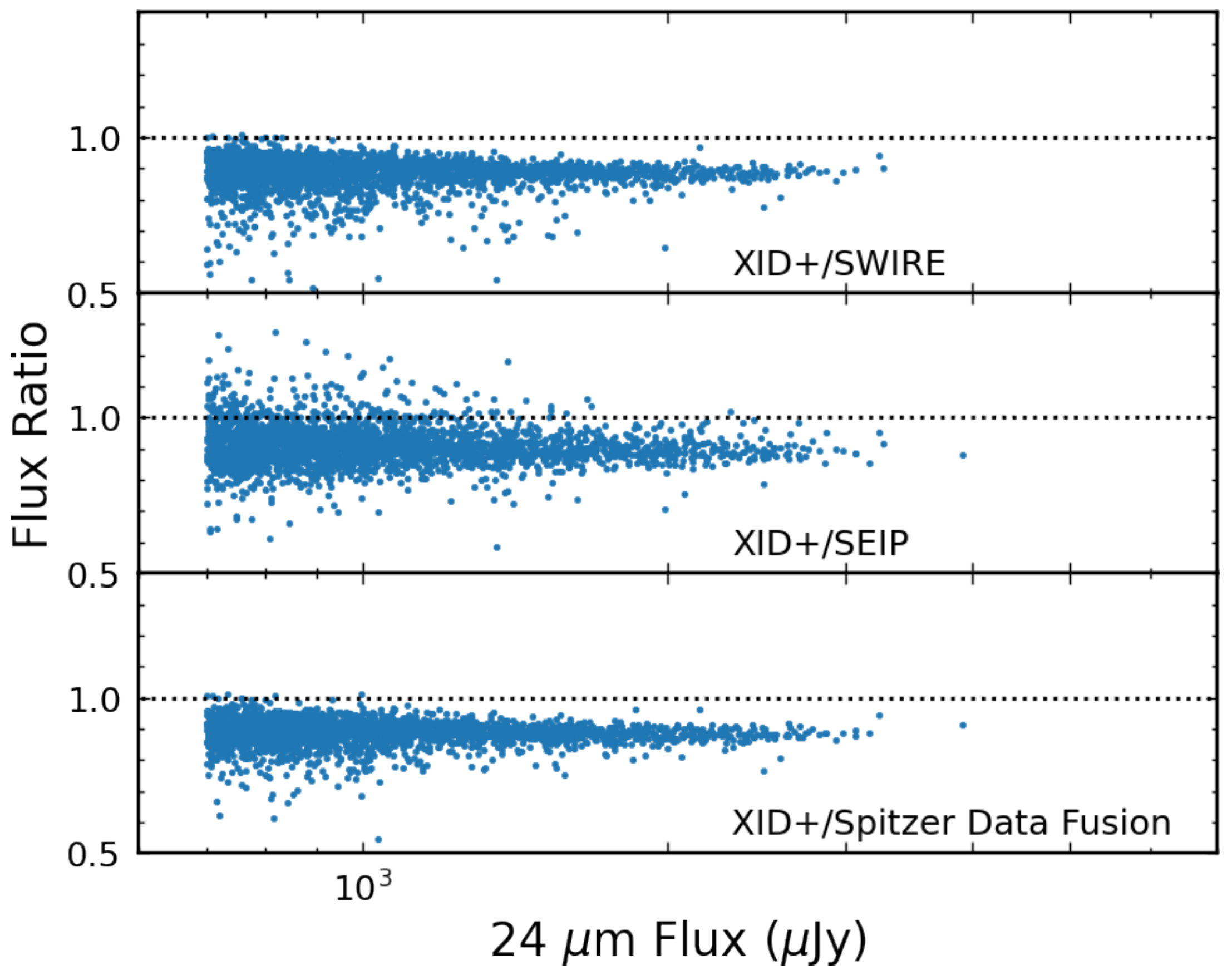}
\caption{The flux ratios of XID+ MIPS 24$\mu$m to those of the SWIRE, SEIP, and Spitzer Data Fusion in the W-CDF-S field, 
    from top to bottom. The dotted line in each panel represents unity flux ratio.}
\label{fig:xid_m24}
\end{figure}

We mainly use the XID+ deblended fluxes from the HELP project (\citealt{shirley2021}) at MIPS 24 $\mu$m through SPIRE 500 $\mu$m.
However, we found that the 24 $\mu$m fluxes downloaded from the HELP project are not always in line with 
those of other MIPS 24 $\mu$m catalogs.
In Fig.~\ref{fig:xid_m24}, we show the ratio of the XID+ fluxes to those from
SWIRE (\citealt{surace2005}),
Spitzer Enhanced Imaging Products (SEIP; \citealt{seipProj}),
and Spitzer Data Fusion (\citealt{vaccari2015}) in the W-CDF-S field.
Note that the reference spectrum of the fluxes from these comparison catalogs in Fig.~\ref{fig:xid_m24} has been 
converted to a 10$^4$ K blackbody.
The median of the flux ratios are 0.907, 0.905, and 0.890 from top to bottom.
The mean of these values is 0.901, and we correct the XID+ fluxes with a factor of $1/0.901=1.11$.
The flux ratios vary with sky field,
and the corresponding correction factors for the ELAIS-S1 and XMM-LSS fields are 0.792 and 1.01, respectively.
Then, a color-correction factor of $1/0.96$ is multiplied to the XID+ 24~$\mu$m fluxes
to convert the reference spectrum to a flat spectrum, $f_\lambda\propto\lambda^0$ (\citealt{mipsHandBook}).
The flux errors produced by the XID+ algorithm are not symmetric, and we conservatively keep the larger one.
However, below the detection level, the flux error might still be underestimated. 
Therefore, we use the flux error of objects with $3\sigma$ detection in the Spitzer Data Fusion.
Furthermore, if the Bayesian P-value residual statistic is larger than 0.5, the XID+ fit shows large residuals 
and the flux is unreliable (\citealt{shirley2021}).
In such cases, we use the MIPS 24 $\mu$m fluxes from the Spitzer Data Fusion.

Similarly, we compared the PACS 100 $\mu$m and 160 $\mu$m fluxes with those from the Herschel/PACS Point Source Catalogs (\citealt{psc2020})
and found median flux ratios of 2.29/1.11/1.49 and 1.17/1.22/0.97 in the W-CDF-S/ELAIS-S1/XMM-LSS fields, respectively.
The XID+ PACS fluxes are then corrected by dividing these factors.
We do not make corrections to XID+ fluxes at the three {\it Herschel}/SPIRE bands either because the two catalogs are consistent 
or the flux uncertainties are too large to find a correction factor.

\bsp    
\label{lastpage}
\end{document}